\documentclass[journal]{IEEEtran}

\IEEEoverridecommandlockouts   

\ifCLASSINFOpdf
   \usepackage[pdftex]{graphicx}
  \graphicspath{{figures/}}
  \usepackage{epstopdf}
  \DeclareGraphicsExtensions{.pdf,.jpeg,.png}
\else
   \usepackage[dvips]{graphicx}
   \graphicspath{{figures/}}
   \DeclareGraphicsExtensions{.eps}
\fi

\usepackage{pgfplots}
\pgfplotsset{compat=newest} 

\usetikzlibrary{plotmarks}
\usetikzlibrary{arrows.meta}
\usepgfplotslibrary{patchplots}
\usepackage{grffile}

\pgfplotsset{plot coordinates/math parser=false}
\newlength\figureheight
\newlength\figurewidth

\usepackage{pdfpages} 

\usepackage{stfloats}
\usepackage{nicefrac}
\usepackage{amsthm}
\usepackage{amssymb}
\usepackage{pifont}

\usepackage{latexsym}
\usepackage{times}

\usepackage{bm} 

\usepackage{array}
\usepackage[cmex10]{amsmath}

\usepackage{mathtools, cuted} 

\usepackage{multicol}
\usepackage{urwchancal}

\usepackage{tikz}
\usepackage{textcomp}
\usepackage{hyperref}
\usepackage{lipsum}

\newcommand\copyrighttext{%
  \footnotesize \textcopyright 2022 IEEE. Personal use of this material is permitted.
  Permission from IEEE must be obtained for all other uses, in any current or future
  media, including reprinting/republishing this material for advertising or promotional
  purposes, creating new collective works, for resale or redistribution to servers or
  lists, or reuse of any copyrighted component of this work in other works.
  DOI: 10.1109/TWC.2022.3177136
  }
\newcommand\copyrightnotice{%
\begin{tikzpicture}[remember picture,overlay]
\node[anchor=south,yshift=10pt] at (current page.south) {\fbox{\parbox{\dimexpr\textwidth-\fboxsep-\fboxrule\relax}{\copyrighttext}}};
\end{tikzpicture}%
}

\usepackage{support-caption}
\usepackage[subrefformat=parens,labelformat=parens]{subcaption}
\usepackage{multirow}
\usepackage{xcolor}

\usepackage[nolist,printonlyused]{acronym}      

\usepackage{optidef} 
\usepackage{algpseudocode, algorithm}

\makeatletter
\algrenewcommand\ALG@beginalgorithmic{\footnotesize}
\makeatother
\algrenewcommand\algorithmiccomment[2][\normalsize]{{#1\hfill\(\triangleright\) #2}}
\usepackage{comment}
\usepackage{cite}

\usepackage{titlecaps}

\usepackage{setspace}

\usepackage{xfrac} 

\newcommand{\altfrac}[2]{\ifmmode\def\tmp{$}\else\def\tmp{}\fi\mbox{%
    {\raisebox{.24\ht\strutbox}{\tmp#1\tmp}}%
    \kern-2.2pt\scalebox{1.6}[1.5]{/}\kern-1.8pt%
    {\tmp#2\tmp}%
    }}

\usepackage{textcomp}

\newcommand\midtilde{\raisebox{0.5ex}{\texttildelow}}

\usepackage{hyperref}

\newtheorem{theorem}{Theorem} 
\newtheorem{definition}{Definition}
\newtheorem{prop}{Proposition}

\newtheorem{observation}{Observation}

\usepackage{color}

\usepackage{xpatch}
\iftrue 
\makeatletter
\xpatchcmd{\proof}{\@addpunct{.}}{\normalfont\,\@addpunct{:}}{}{}
\makeatother
\fi

\DeclareMathAlphabet\mathbfcal{OMS}{cmsy}{b}{n}
\begin{acronym}[LTE-Advanced]
  \acro{2G}{Second Generation}
  \acro{3G}{3$^\text{rd}$~Generation}
  \acro{3GPP}{3$^\text{rd}$~Generation Partnership Project}
  \acro{4G}{4$^\text{th}$~Generation}
  \acro{5G}{5$^\text{th}$~Generation}
  \acro{5GPPP}{5G Infrastructure Public Private Partnership}
  \acro{ACLR}{adjacent channel leakage ratio}
  \acro{ADMM}{alternating direction method of multipliers}
  \acro{ARQ}{automatic repeat request}
  \acro{BADMM}{Bregman \ac{ADMM}} 
  \acro{BER}{bit error rate}
  \acro{BLER}{block error rate}
  \acro{BPC}{binary power control}
  \acro{BPSK}{binary phase-shift keying}
  \acro{BRA}{balanced random allocation}
  \acro{BS}{base station}
  \acro{BW}{bandwidth}
  \acro{CADMM}{\ac{CSI}-aware \ac{ADMM}}
  \acro{CAP}{combinatorial allocation problem}
  \acro{CAPEX}{capital expenditure}
  \acro{CBF}{coordinated beamforming}
  \acro{CCDF}{complementary cumulative distribution function}
  \acro{CFR}{crest-factor reduction}
  \acro{CP}{cyclic prefix}
  \acro{CS}{coordinated scheduling}
  \acro{CPOCS}{\ac{CSI}-aware \ac{POCS}}
  \acro{CRUSADE}{convex suppression of unwanted emission and distortion elimination}
  \acro{CRUISE}{convex reduction of unwanted spectral emission}
  \acro{CRUISE-DAYS}{Convex RedUctIon of unwanted Spectral Emission with Davis And Yin Splitting}
  \acro{cruise}{convex reduction of unwanted spectral emission}
  \acro{CRUISE-SSP}{\ac{CRUISE}-\ac{SSP}}
  \acro{CSI}{channel state information}
  \acro{CSIT}{channel state information at the transmitter}
  \acro{D2D}{device-to-device}
  \acro{DCA}{dynamic channel allocation}
  \acro{DCI}{downlink control information}
  \acro{DE}{differential evolution}
  \acro{DFT}{discrete Fourier transform}
  \acro{DIST}{Distance}
  \acro{DL}{downlink}
  \acro{DMA}{double moving average}
  \acro{DMRS}{demodulation reference signal}
  \acro{D2DM}{D2D mode}
  \acro{DMS}{D2D mode selection}
  \acro{DROPE}{Douglas-Rachford-based \ac{OOBE} power reduction with \ac{EVM} constraint}
  \acro{DRS}{Douglas-Rachford splitting}
  \acro{DYS}{Davis-Yin splitting}
  \acro{DPC}{dirty paper coding}
  \acro{DR}{Douglas-Rachford}
  \acro{DRA}{dynamic resource assignment}
  \acro{DSA}{dynamic spectrum access}
  \acro{eMBB}{enhanced mobile broadband}
  \acro{eV2X}{enhanced vehicle-to-everything}
  \acro{EADMM}{\ac{EVM}-constrained \ac{ADMM}}
  \acro{EIRP}{equivalent isotropically radiated power}
  \acro{EPOCS}{\ac{EVM}-constrained \ac{POCS}}
  \acro{ETSI}{European telecommunications standards institute}
  \acro{EVM}{error vector magnitude}
  \acro{EMSP}{EVM-constrained \ac{MSP}}
  \acro{ENSP}{EVM-constrained \ac{NSP}}
  \acro{ESP}{EVM-constrained spectral precoding}
  \acro{ESSP}{EVM-constrained \ac{SSP}}
  \acro{FD}{fully digital}
  \acro{FDD}{frequency division duplexing}
  \acro{FFT}{(fast) discrete Fourier Transform}
  \acro{GFBS}{generalized forward-backward splitting}
  \acro{HARQ}{hybrid automatic repeat request}
  \acro{HB}{hybrid beamforming}
  \acro{KKT}{Karush-Kuhn-Tucker}
  \acro{KPI}{key performance indicator}
  \acro{IPAPR}{instantaneous \ac{PAPR}}
  \acro{ICF}{iterative clipping and filtering}
  \acro{IDFT}{Inverse \ac{DFT}}
  \acro{IFFT}{Inverse \ac{FFT}}
  \acro{InC}{in-coverage}
  \acro{IoT}{internet of things}
  \acro{ITS}{intelligent transportation systems}
  \acro{L-BFGS}{limited-memory Broyden-Fletcher-Goldfarb-Shanno}
  \acro{LDPC}{low-density parity-check code}
  \acro{LMMSE}{linear minimum mean squared error}
  \acro{LOS}{line-of-sight}
  \acro{LS}{large-scale}
  \acro{LS-MIMO-OFDM}{\ac{LS}-\ac{MIMO}-\ac{OFDM}}
  \acro{LS-MSP}{large-scale MSP}
  \acro{LTE}{long-term evolution}
  \acro{MAC}{medium access control}
  \acro{MADMM}{modified consensus \ac{ADMM}}
  \acro{MCS}{modulation and coding scheme}
  \acro{METIS}{Mobile Enablers for the Twenty-Twenty Information Society}
  \acro{MIMO}{multiple-input multiple-output}
  \acro{MIMO-OFDM}{\ac{MIMO}-\ac{OFDM}}
  \acro{MISO}{multiple-input single-output}
  \acro{MRC}{maximum ratio combining}
  \acro{MS}{mobile station}
  \acro{MSE}{mean squared error}
  \acro{MSP}{mask-compliant spectral precoder}
  \acro{MMSE}{minimum mean squared error}
  \acro{MTC}{machine type communications}
  \acro{MU-MIMO}{multi-user \ac{MIMO}}
  \acro{MU-MIMO-OFDM}{\ac{MU-MIMO}-\ac{OFDM}}
  \acro{mMTC}{massive machine type communications}
  \acro{cMTC}{critical machine type communications}
  \acro{NR}{new radio}
  \acro{NSP}{notching spectral precoder}
  \acro{NSPS}{national security and public safety}
  \acro{NWC}{network coding}
  \acro{OFDM}{orthogonal frequency division multiplexing}
  \acro{OOB}{out-of-band}
  \acro{OOBE}{out-of-band emissions}
  \acro{OoC}{out-of-coverage}
  \acro{PA}{power amplifier}
  \acro{PAPR}{peak-to-average power ratio}
  \acro{PDCCH}{physical downlink control channel}
  \acro{PDSCH}{physical downlink shared channel}
  \acro{POCS}{projection on convex sets}
  \acro{PPG}{proximal-proximal-gradient}
  \acro{PRB}{physical resource block}
  \acro{PRBs}{physical resource block}
  \acro{PRG}{precoding resource block group}
  \acro{PSBCH}{physical sidelink broadcast channel}
  \acro{PSFCH}{physical sidelink feedback channel}
  \acro{PSCCH}{physical sidelink control channel}
  \acro{PSP}{proximal splitting based mask compliant spectral precoding}
  \acro{PSSCH}{physical sidelink shared channel}
  \acro{PHY}{physical}
  \acro{PLNC}{physical layer network coding}
  \acro{PRB}{physical resource block}
  \acro{PSD}{power spectral density}
  \acro{QAM}{quadrature amplitude modulation}
  \acro{QCQP}{quadratic constrained quadratic programming}
  \acro{QoS}{quality of service}
  \acro{QPSK}{quadrature-phase shift keying}
  \acro{PaC}{partial coverage}
  \acro{RAISES}{reallocation-based assignment for improved spectral efficiency and satisfaction}
  \acro{RAN}{radio access network}
  \acro{RA}{resource allocation}
  \acro{RAT}{radio access technology}
  \acro{RB}{resource block}
  \acro{RF}{radio frequency}
  \acro{RSRP}{reference signal received power}
  \acro{Rx}{receive}
  \acro{RX}{receive}
  \acro{RxEVM}{receive-EVM}
  \acro{Rx-TxEVM}{received \ac{TxEVM}}
  \acro{RZF}{regularized zero forcing}
  \acro{SC-FDM}{single carrier frequency division modulation}
  \acro{SEM}{spectral emission mask}
  \acro{SFBC}{space-frequency block coding}
  \acro{SCI}{sidelink control information}
  \acro{SIC}{successive interference cancellation}
  \acro{SINR}{signal-to-interference-plus-noise ratio}
  \acro{SISO}{single-input single-output}
  \acro{SL}{sidelink}
  \acro{SP}{Spectral precoding}
  \acro{SNR}{signal-to-noise ratio}
  \acro{SSP}{semi-analytical spectral precoding}
  \acro{STC}{space-time coding}
  \acro{SU}{single-user}
  \acro{SU-MIMO}{single-user \ac{MIMO}}
  \acro{SU-MIMO-OFDM}{\ac{SU-MIMO}-\ac{OFDM}}
  \acro{SVM}{support vector machine}
  \acro{TDD}{time division duplexing}
  \acro{TDL}{tapped delay line}
  \acro{TOP-ADMM}{three-operator \ac{ADMM}}
  \acro{Tx}{transmit}
  \acro{TX}{transmit}
  \acro{TxEVM}{transmit-EVM}
  \acro{UE}{user equipment}
  \acro{UL}{uplink}
  \acro{UP}{uniform power}
  \acro{URLLC}{ultra-reliable and low latency communications}
  \acro{VUE}{vehicular user equipment}
  \acro{V2N}{vehicular-to-network}
  \acro{V2X}{vehicle-to-everything}
  \acro{V2V}{vehicle-to-vehicle}
  \acro{V2P}{vehicle-to-pedestrian}
  \acro{WMMSE}{weighted minimum mean squared error}
  \acro{wMMSE}{weighted minimum mean squared error}
  \acro{ZF}{zero forcing}
  \acro{ZMCSCG}{zero mean circularly symmetric complex Gaussian}
\end{acronym}

\usepackage{etoolbox}

\usepackage{setspace}

\usepackage{centernot}

\usepackage{mfirstuc} 

\usepackage{enumerate}

\usepackage{footmisc}

\allowdisplaybreaks

\makeatletter
\newcommand\makebig[2]{%
  \@xp\newcommand\@xp*\csname#1\endcsname{\bBigg@{#2}}%
  \@xp\newcommand\@xp*\csname#1l\endcsname{\@xp\mathopen\csname#1\endcsname}%
  \@xp\newcommand\@xp*\csname#1r\endcsname{\@xp\mathclose\csname#1\endcsname}%
}
\makeatother

\makebig{biggg} {3.0}
\makebig{Biggg} {3.5}
\makebig{bigggg}{4.0}
\makebig{Bigggg}{4.5}
\makebig{biggggg}{5.0}
\makebig{Biggggg}{5.5}
\makebig{bigggggg}{6.0}
\makebig{Bigggggg}{6.5}

\begin{document}

\title{ \color{black}{EVM Mitigation with PAPR and ACLR Constraints in Large-Scale MIMO-OFDM Using TOP-ADMM}}
\iftrue

\author{
\iftrue
Shashi~Kant,~\IEEEmembership{Graduate Student Member,~IEEE,}
Mats~Bengtsson,~\IEEEmembership{Senior Member,~IEEE,} 
Gabor~Fodor,~\IEEEmembership{Senior Member,~IEEE,}
Bo~G\"oransson,~\IEEEmembership{Member,~IEEE,} 
and 
Carlo~Fischione,~\IEEEmembership{Senior Member,~IEEE}
\else
Shashi~Kant \\ 
\vspace{15mm}

\fi

\vspace{-1mm}

\iftrue

\thanks{S.\ Kant, B.\ G\"oransson, and G.\ Fodor are with Ericsson AB and KTH Royal Institute of Technology, Stockholm, Sweden (e-mail: \{shashi.v.kant, bo.goransson, gabor.fodor\}@ericsson.com)}
\thanks{M.\ Bengtsson and C.\ Fischione are with KTH Royal Institute of Technology, Stockholm, Sweden (e-mail: \{matben, carlofi\}@kth.se)}
\thanks{The work of S.\ Kant was supported in part by the Swedish Foundation for Strategic Research under grant ID17-0114.}

\thanks{\color{black}{Parts of this work is presented at IEEE Asilomar 2021~\cite{Kant_et_al__asilomar_2021}.}}

\else

\thanks{
\vspace{15mm}
}

\fi
%
}
\fi

\maketitle
\copyrightnotice

\renewcommand\qedsymbol{$\blacksquare$}

\newcounter{subeqsave}
\newcommand{\savesubeqnumber}{\setcounter{subeqsave}{\value{equation}}%
\typeout{AAA\theequation.\theparentequation}}
\newcommand{\recallsubeqnumber}{%
  \setcounter{equation}{\value{subeqsave}}\stepcounter{equation}}

\iftrue
\renewcommand{\vec}[1]{\ensuremath{\boldsymbol{#1}}}
\newcommand{\mat}[1]{\ensuremath{\boldsymbol{#1}}}
\newcommand{\herm}{{\rm H}}
\newcommand{\tran}{{\rm T}}
\newcommand{\trans}{{\rm T}}
\newcommand{\trace}{{\rm Tr}}
\newcommand{\diag}{{\rm diag}}
\newcommand{\Diag}{{\rm Diag}}
\newcommand{\sign}{{\rm sign}}
\newcommand{\rank}{{\rm rank}}
\newcommand{\EVM}{{\rm EVM}}
\newcommand{\SNR}{{\rm SNR}}
\newcommand{\SINR}{{\rm SINR}}
\newcommand{\expect}{\mathbb{E}}
\newcommand{\Cm}{\mathbb{C}}
\newcommand{\Rm}{\mathbb{R}}
\newcommand{\CN}{\mathcal{CN}}
\newcommand{\be}{\begin{equation}}
\newcommand{\ee}{\end{equation}}
\newcommand{\pdf}{\mathcal{P}}
\newcommand{\prox}{\ensuremath{\boldsymbol{\rm{prox}}}}
\newcommand{\proj}{\ensuremath{\boldsymbol{\rm{proj}}}}
\newcommand{\dom}{\rm{dom}}
\newcommand{\epi}{\rm{epi}}
\newcommand{\ie}{\textit{i.e.}}
\newcommand{\eg}{\textit{e.g.}}
\newcommand{\cf}{\textit{cf.}}
\newcommand{\etc}{etc.} 
\newcommand{\avgTxEVM}{\ensuremath{\epsilon_{\rm avg}}}
\newcommand{\TxEVM}{\ensuremath{\bm{\epsilon}}}
\newcommand{\Ind}{{\delta}} 

\newcommand{\vecOp}{\rm{vec}}
\newcommand{\unvecOp}{\rm{unvec}}

\newcommand{\NR}{\ensuremath{N_{\rm R}}}
\newcommand{\NT}{\ensuremath{N_{\rm T}}}
\newcommand{\NL}{\ensuremath{N_{\rm L}}}
\newcommand{\NU}{\ensuremath{N_{\rm U}}}
\newcommand{\Ncp}{\ensuremath{N_{\rm CP}}}
\newcommand{\Nsc}{\ensuremath{N_{\rm SC}}}

\newcommand{\Kcp}{\ensuremath{K_{\rm CP}}}

\newcommand{\RxEVM}{\ensuremath{\bm{\varsigma}}}

\newcommand{\A}{\ensuremath{\boldsymbol{A}}}
\renewcommand{\a}{\ensuremath{\boldsymbol{a}}}
\newcommand{\B}{\ensuremath{\boldsymbol{B}}}
\renewcommand{\b}{\ensuremath{\boldsymbol{b}}}
\newcommand{\C}{\ensuremath{\boldsymbol{C}}}

\newcommand{\D}{\ensuremath{\boldsymbol{D}}}
\renewcommand{\d}{\ensuremath{\boldsymbol{d}}}
\newcommand{\E}{\ensuremath{\boldsymbol{E}}}
\newcommand{\F}{\ensuremath{\boldsymbol{F}}}

\newcommand{\K}{\ensuremath{\boldsymbol{K}}}

\newcommand{\M}{{\mathbf{M}}}
\newcommand{\I}{\ensuremath{\boldsymbol{I}}}

\newcommand{\R}{\ensuremath{\boldsymbol{R}}}

\renewcommand{\S}{\ensuremath{\boldsymbol{S}}}

\newcommand{\T}{\ensuremath{\boldsymbol{T}}}
\renewcommand{\t}{\ensuremath{\boldsymbol{t}}}

\newcommand{\U}{\ensuremath{\boldsymbol{U}}}
\renewcommand{\u}{\ensuremath{\boldsymbol{u}}}

\newcommand{\V}{\ensuremath{\boldsymbol{V}}}
\renewcommand{\v}{\ensuremath{\boldsymbol{v}}}

\newcommand{\X}{\ensuremath{\boldsymbol{X}}}

\newcommand{\Y}{\ensuremath{\boldsymbol{Y}}}

\newcommand{\0}{\ensuremath{\boldsymbol{0}}}
\fi

\iffalse
\colorlet{blueorange}{blue!15!orange!85!}
\definecolor{ao(english)}{rgb}{0.0, 0.5, 0.0}
\def\baselinestretch{1}
\def\sk#1{\textcolor{black}{#1}}
\def\skblue#1{\textcolor{blue}{#1}}
\def\skred#1{\textcolor{red}{#1}}
\def\skgreen#1{\textcolor{ao(english)}{#1}}
\def\skteal#1{\textcolor{teal}{#1}}
\def\gf#1{\textcolor{cyan}{#1}}
\def\bg#1{\textcolor{green}{#1}}
\def\mb#1{\textcolor{magenta}{#1}}
\def\cf#1{\textcolor{teal}{#1}}
\else 
\colorlet{blueorange}{blue!15!orange!85!}
\def\baselinestretch{1}
\def\sk#1{\textcolor{black}{#1}}
\def\skblue#1{\textcolor{black}{#1}}
\def\skred#1{\textcolor{black}{#1}}
\def\skgreen#1{\textcolor{black}{#1}}
\def\skteal#1{\textcolor{black}{#1}}
\def\gf#1{\textcolor{black}{#1}}
\def\bg#1{\textcolor{black}{#1}}
\def\mb#1{\textcolor{black}{#1}}
\def\cf#1{\textcolor{black}{#1}}
\fi

\newcommand{\mx}[1]{\mathbf{#1}}

\DeclareRobustCommand{\BigOh}{%
  \text{\usefont{OMS}{cmsy}{m}{n}O}%
}

\let\oldemptyset\emptyset
\let\emptyset\varnothing

\algblock{ParFor}{EndParFor}
\algnewcommand\algorithmicparfor{\textbf{parfor}}
\algnewcommand\algorithmicpardo{\textbf{do}}
\algnewcommand\algorithmicendparfor{\textbf{end\ parfor}}
\algrenewtext{ParFor}[1]{\algorithmicparfor\ #1\ \algorithmicpardo}
\algrenewtext{EndParFor}{\algorithmicendparfor}

\begin{abstract}
Although signal distortion-based peak-to-average power ratio (PAPR) reduction is a feasible candidate for orthogonal frequency division multiplexing (OFDM) to meet standard/regulatory requirements, the error vector magnitude (EVM) stemming from the PAPR reduction has a deleterious impact on the performance of high data-rate achieving multiple-input multiple-output (MIMO) systems. Moreover, these systems must constrain the adjacent channel leakage ratio (ACLR) to comply with regulatory requirements. Several recent works have investigated the mitigation of the EVM seen at the receivers by capitalizing on the excess spatial dimensions inherent in the large-scale MIMO that assume the availability of perfect channel state information (CSI) with spatially uncorrelated wireless channels. Unfortunately, practical systems operate with erroneous CSI and spatially correlated channels. Additionally, most standards support user-specific/CSI-aware beamformed and cell-specific/non-CSI-aware broadcasting channels. Hence, we formulate a robust EVM mitigation problem under channel uncertainty with nonconvex PAPR and ACLR constraints catering to beamforming/broadcasting. To solve this formidable problem, we develop an efficient scheme using our recently proposed three-operator alternating direction method of multipliers (TOP-ADMM) algorithm and benchmark it against two three-operator algorithms previously presented for machine learning purposes. Numerical results show the efficacy of the proposed algorithm under imperfect CSI and spatially correlated channels.

\end{abstract}


\IEEEpeerreviewmaketitle

\section{Introduction}
Modern mobile and wireless communication systems, such as 5G \ac{NR} and beyond, adopt orthogonal frequency division multiplexing \acused{OFDM}(\ac{OFDM}) with \sk{a} cyclic prefix~\cite{Dahlman5GNR2018}. \sk{This is because of both its backward compatibility with 4G \ac{LTE} as well as the fact that \ac{OFDM} has several appealing attributes,} such as robustness against the adverse effects of time and/or frequency selective channels, simplicity in terms of single-tap (frequency-domain) equalization, and resilience in terms of supporting both low and high symbol rates---thereby supporting \sk{various} quality of service requirements.

It is well-known that \ac{OFDM} suffers from high \ac{PAPR} because of the linear combination of the subcarriers carrying (pseudo-random) $M$-ary quadrature amplitude modulation symbols in the inverse discrete Fourier transform (\acused{IDFT}\ac{IDFT}) operation~\cite{Prasad:04}. 
\skblue{
With the trend of an increasing number of transmit antennas and support of wider carrier bandwidths with the higher spectrum and spectral efficiencies, including support of higher frequency bands, the implementation of radio base station faces many challenges, notably in terms of power consumption and size---primarily dominated by \acp{PA}. Thus, to reduce the power consumption of radio base stations or \acp{PA}, one must reduce the \ac{PAPR} of the transmit signals for each transmit antenna. Additionally, the challenges are to support multiple radio access technologies, \eg, 5G NR, 4G \ac{LTE}, and narrowband internet of things (NB-IoT), by the same radio base station---where NB-IoT carriers can be placed in the guard bands of NR/\ac{LTE}. For some classes of base stations, high average transmit output power of the signals is required for better cell coverage. Therefore, if \ac{PAPR} is high and the required average transmit signal power is high, smaller size and efficient \ac{PA} can cause non-linear distortions rendering both inband and \ac{OOBE}---which may fail the base station to comply with \ac{3GPP} standard and regulatory requirements. Consequently, if \ac{PAPR} is kept high, an inefficient and bigger size \ac{PA} will be employed such that the signals can operate in a linear region to avoid non-linear distortions. Hence, it is paramount to minimize {the} \ac{PAPR} of \ac{OFDM} waveform in large-scale \ac{MIMO} systems. Furthermore, low \ac{PAPR} of the signals with a digital predistorter or a \ac{PA} linearizer can improve the \ac{PA} efficiency by letting \ac{PA} operate close to its bias point. 
}

\skblue{Numerous techniques to reduce \ac{PAPR}, also referred to as crest factor reduction, have been proposed in the literature {over the past decades}.  These can be grouped as 1) distortion-based approaches, such as iterative clipping and filtering~\cite{Armstrong__icf__2002, Wang_and_Luo_ICF_papr__2011}, peak cancellation~\cite{Per_borjesson_etal_peak_cancellation__1999,Kageyama_etal_peak_cancellation__2020}, \sk{and} non-linear companding~\cite{Wang_etal_companding_1999}\sk{;} 2) coding-based methods~\cite{Davis_and_Jedwab__coding_papr_1999}\sk{;} and 3) multiple signaling-based approaches, such as selected mapping, partial transmit sequences~\cite{Bauml_etal__selected_mapping_papr_1996}, \skblue{and tone reservation~\cite{Tellado_and_Cioffi__TR__1198}}  \sk{(see also review papers~\cite{Han_and_Lee__overview_papr__2005, Rahmatallah_Mohan_papr_survey:2013, Wunder__papr__2013}})}. Distortion-based techniques, particularly clipping and filtering, are the most popular methods in practice as they are transparent to the standards and to the receiver, see~\cite{Rahmatallah_Mohan_papr_survey:2013}. However, they cause significant inband distortion\sk{---quantified in terms of error vector magnitude (EVM)\acused{EVM}---and may increase \ac{OOBE}}. \sk{Basically, these} distortion-based techniques \sk{manage the inherent} trade-off between \ac{EVM} and achievable \ac{PAPR}. More precisely, if \sk{the} desired \ac{PAPR} threshold is low\sk{,} then \sk{the} inband transmit \ac{EVM} is high, which consequently penalizes the throughput performance at the receiver. 

\skblue{
\section{Related Works and Contributions}\label{sec:related_works_and_contribution} 
}

In this section, we first give an overview of some prior works on \ac{PAPR} reduction methods with some practical challenges for 5G NR and beyond to motivate distortion-based \ac{PAPR} techniques. Most of the principled approaches for distortion-based \ac{PAPR} reductions are posed as optimization problems, whose (near-)optimal solutions cannot be obtained in closed-form and require efficient iterative optimization techniques to tackle the problem. Thus, we briefly describe some powerful and efficient first-order optimization techniques to tackle the \ac{PAPR} reduction problems for large-scale \ac{MIMO}-\ac{OFDM}-based systems.  
Finally, we highlight our original contribution with mathematical notations and paper organization.

\vspace{1.5mm}
{
\subsection{Practical Research Challenges for PAPR Reduction Techniques in 5G NR and Beyond} 
As mentioned above, the base station supports multiple standards from 4G \ac{LTE} to current 5G NR releases, including cellular \ac{IoT}, possibly in the guard bands of \ac{LTE}/NR. It supports wide bandwidths and multiple bands with growing demand for energy-efficient and lightweight radio heads of the base stations equipped with many transmit antennas. Clearly, \ac{PAPR} of the multiple carriers of the signals must be reduced, which complements \ac{PA} linearization such as digital predistortion to improve \ac{PA} efficiency---targeting energy-efficient and lightweight base stations. The challenges for the \ac{PAPR} reduction techniques are that 1) one cannot exploit guard bands to deploy many non-useful subcarriers for \ac{PAPR} reduction such as in tone reservation~\cite{Tellado_and_Cioffi__TR__1198}, and 2) one cannot send extra non-standard side information to the receiver(s) because of compliance with the stringent \ac{3GPP} standard and regulatory requirements.   
}

\vspace{-1.5mm}
{
\subsection{Related Works for PAPR Reduction Techniques} 
We briefly summarize some of the related works for \ac{PAPR} reduction in \ac{OFDM}-based systems, including \ac{MIMO}, wherever applicable. Moreover, we refer interested readers to overview papers~\cite{Han_and_Lee__overview_papr__2005, Rahmatallah_Mohan_papr_survey:2013, Wunder__papr__2013}.
}

\vspace{-1.5mm}
\skblue{
\subsubsection{Multiple Signaling and Probabilistic Methods}
These techniques reduce \ac{PAPR} by either generating multiple permutations, introducing phase shifts or adding peak reducing subcarriers of \ac{OFDM} signal.
}

\skblue{
Selective mapping~\cite{Mestdagh_and_Spruyt__SelectiveMapping_1996, Han_and_Lee__overview_papr__2005, Rahmatallah_Mohan_papr_survey:2013, Wunder__papr__2013} generates multiple statistically independent input data sequences representing the same useful information, where the generated signal with the lowest \ac{PAPR} is selected for the transmission. However, the receiver must know the chosen mapping at the transmitter for successful demodulation of the signal via side information. Hence, unfortunately, selective mapping cannot be employed in \ac{3GPP}-compliant cellular base stations. There are some related methods to selective mapping, called partial transmit sequences---see~\cite{Han_and_Lee__overview_papr__2005, Rahmatallah_Mohan_papr_survey:2013, Wunder__papr__2013} and references therein. 
}

\skblue{
Tone reservation~\cite{Tellado_and_Cioffi__TR__1198,Wang_et_al__FISTA_TR__2018,Han_and_Lee__overview_papr__2005, Rahmatallah_Mohan_papr_survey:2013, Wunder__papr__2013} is one of the signal distortion-less techniques for \ac{PAPR} reduction, which utilizes unused/inactive subcarriers---that is, adds peak reducing subcarriers---typically in the vicinity of used/active subcarriers carrying useful information. On the one hand, one of the strengths of tone reservation is that it does not induce any signal distortions on the useful data-carrying subcarriers. 
On the other hand, although tone reservation offers low complexity, as discussed above, unfortunately, the limited number of unused/inactive subcarriers can be exploited because of relatively high spectrum efficiency of 5G NR compared to \ac{LTE} and nearly no effective guard bands are available in 5G NR (or even \ac{LTE}) as {NB-IoT} carriers can also be deployed in the guard bands of NR/\ac{LTE}. Nevertheless, ignoring the guard band usage restrictions of NR/LTE, in Section~\ref{subsec:simulation_results}, we have evaluated and compared the performance of two variants of tone reservation schemes against our proposed method---see particularly Fig.~\ref{fig:cmp__tr_unconstrained__constrained__icf__topadmm}. 
}

\vspace{-1.5mm}
\skblue{
\subsubsection{Coding-based PAPR Reduction}
These coding-based techniques, \eg, linear block codes, Golay complementary sequences, and turbo coding---see~\cite{Han_and_Lee__overview_papr__2005, Rahmatallah_Mohan_papr_survey:2013, Wunder__papr__2013} and references therein---reduce the \ac{PAPR} by generating encoded data inputs that produce an \ac{OFDM} symbol with desired \ac{PAPR}. However, these techniques typically require an exhaustive search for a good code, which becomes intractable for an \ac{OFDM}-based system with more than a dozen subcarriers. Unfortunately, such techniques are practically impossible with a 5G NR-like system.
}

\vspace{-1.5mm}
\skblue{
\subsubsection{Distortion-based PAPR Reduction}
These distortion-based methods essentially reduce the \ac{PAPR} by (non-linear) distortion of the useful transmit signal, where the signal quality degradation is measured in terms of \ac{EVM}. Moreover, these distortion-based methods render \ac{OOBE}, which must be handled via appropriate filtering in either time or frequency domain.
}

\skblue{
Active constellation or active set extension~\cite{Krongold_and_Jones__ACE__2003} reduces \ac{PAPR} by modifying or (pre)distorting the modulation constellation carried on the useful/active subcarriers with a minimal penalty in terms of \ac{EVM}. Although no side information needs to be sent to the receiver, this method has, unfortunately, limited application to large modulation alphabet, \eg, alphabet size more than 64QAM~\cite{Han_and_Lee__overview_papr__2005, Rahmatallah_Mohan_papr_survey:2013}.  Hence, active constellation extension is not feasible for a 5G NR-like system where more than 64QAM, \eg, 256/1024QAM, modulation alphabet can be supported.
}

\skblue{
Iterative clipping and filtering~\cite{Armstrong__icf__2002, Wang_and_Luo_ICF_papr__2011} is one of the simplest approaches, which clips the high amplitudes of the signal if above a given peak threshold while retaining the phase of the signals. This peak clipping causes unwanted distortions inband and out-of-band---\ie, \skblue{\ac{EVM} and} \ac{OOBE}. This induced \ac{OOBE} needs to be filtered out either in frequency or time domain to maintain desired \ac{3GPP} \ac{ACLR} requirements. Consequently, filtering of unwanted distortions causes regrowth of the signal peaks. However, after a few repetitions/iterations of clipping and filtering reduces \ac{PAPR} but at the cost of increased signal distortion (\ac{EVM}) depending on the chosen peak threshold. If the signal peaks are clipped at a lower threshold, signal distortion---\ac{EVM}---increases significantly, penalizing the throughput~\cite{Studer_papr_2013}.
}

Recently, for reciprocity-based \skblue{or \ac{CSI}-aware} massive/large-scale \ac{MIMO}, distortion-based \ac{PAPR} reduction techniques, see, e.g.,~\cite{Studer_papr_2013,Bao_papr_admm__2018, Yao__papr_sdp__2019, Zayani_2019, taner2021ellpellqnorm}, are proposed, \sk{which} not only reduce the \ac{PAPR} at the transmitter but also mitigate the incurred signal distortion (\ac{EVM}) seen at the receiver by exploiting downlink channel knowledge available at the base station. \sk{\sk{Most of these distortion-based techniques} relax the nonconvex \ac{PAPR} problem by employing either \sk{minimization of peaks}, see, \eg,~\cite{Studer_papr_2013}, which \sk{renders a suboptimal solution, or semidefinite relaxation, see, \eg,~\cite{Yao__papr_sdp__2019}, which can enable a near-optimal solution at the expense of prohibitive computational complexity. \sk{Another drawback is that}} many of these methods are not robust when the downlink channel knowledge acquired at the transmitter is either imperfect or incomplete \sk{\eg,~\cite{Bao_papr_admm__2018, Zayani_2019, taner2021ellpellqnorm}}.  
Several of these previously proposed methods lack the flexibility {to adapt the desired \ac{PAPR} \sk{target and \ac{OOBE}---typically quantified in terms of \ac{ACLR}---}according to the regulatory requirements.} 
\sk{Furthermore, in realistic systems}, the channel estimates at the transmitter and receiver \sk{are} imperfect and/or incomplete.  
Consequently, under imperfect and/or incomplete channel estimates,  
these previously proposed non-robust schemes}
cannot achieve \sk{(near)} zero EVM at the receiver \sk{(as shown in Section~\ref{sec:simulation_results})}. \sk{Additionally, these {non-robust} prior arts {do not support} the incomplete channel knowledge case, that is, \ac{OFDM} symbols carrying broadcast channels (such as cell-specific control channels), since there is no downlink channel knowledge as these common channels naturally do not target {a specific subset of the served users.}} 
Although~\cite{Yao__papr_sdp__2019} offers various robust methods for reciprocity-based \ac{PAPR} reduction techniques, \sk{in these schemes} there is no constraint on \ac{OOBE} or \ac{ACLR}, which are important for \ac{PAPR} reduction schemes. Additionally and regrettably, semidefinite programming-based algorithms are typically solved by interior-point-based methods~\cite{Boyd2004ConvexOptimization}, which have prohibitive complexity for state-of-the-art base station radio hardware systems. 

\vspace{-3.5mm}
\skblue{
\subsection{Motivation for Distortion and Optimization Based PAPR Reduction} \label{sec:motivation_for_distortion_optimization_based_papr_reduction}
}

\skblue{
Considering practical constraints/restrictions in 5G NR and beyond, an interesting candidate is a distortion-based \ac{PAPR} reduction scheme, which utilizes mainly useful/active subcarriers to reduce the \ac{PAPR} of the \ac{OFDM} signal. Concretely, we seek a one-size-fits-all \ac{PAPR} reduction scheme that can naturally work for base stations equipped with single or multiple transmit antennas and adaptively caters to non-\ac{CSI}-aware, \eg, broadcast channels, and \ac{CSI}-aware if possible. \skblue{In large-scale \ac{MIMO}}, the base station has a large number of antennas compared to the total number of served/scheduled spatial layers for the transmission---applicable to both \ac{SU-MIMO} or \ac{MU-MIMO}---and additionally has some good enough quality of downlink channel estimates of all the served users. \skblue{Consequently,} one could \skblue{potentially relax the \ac{EVM} requirements at the transmitter stipulated by the standards, such as \ac{3GPP} NR~\cite{3GPPTS38.1042018NRReception}, by mitigating/reducing the \ac{EVM}} 
seen at the receivers of the users accordingly while meeting the desired \ac{PAPR} and \ac{ACLR} requirements at the transmitter of a base station. Moreover, if the base station has no downlink channel knowledge of the users for instance supporting broadcast or control channels in some \ac{OFDM} symbols over all the active subcarriers within a bandwidth, we propose minimizing the signal distortion energy subject to the desired \ac{PAPR} and \ac{ACLR} constraints. Additionally, we seek a principled approach to describe the distortion-based \ac{PAPR} reduction scheme \skblue{using an optimization formulation} and subsequently develop a (near-)optimal and computationally-efficient solution. \skblue{In practice, we do not require highly accurate solution, especially for the problems of interest in this work. Therefore, the} subsequent section introduces first-order optimization schemes\skblue{, which offer a good trade-off between computational complexity and accuracy. Hence, these first-order methods can} tackle complicated composite problems and yield implementation-friendly algorithms suitable for large-scale \ac{MIMO}-\ac{OFDM} systems. 
}

\vspace{1.0mm}
\subsection{\skblue{Related Works on} Operator/Proximal Splitting for Signal Processing in Communications} \label{sec:related_works_on_proximal_splitting}
\sk{
In the optimization problems mentioned above, proximal/operator splitting techniques---typically first-order optimization algorithms---\sk{have become} popular {thanks to their} low computational complexity. This \sk{can be seen} not only in wireless communications but also in machine learning applications. Operator splitting methods employ a divide-and-conquer approach, which essentially breaks the large problem into smaller, easy-to-solve subproblems.
}
\sk{However, first-order methods can be very slow to converge \sk{to solutions having high accuracy}. Nonetheless, modest accuracy can be sufficient for many practical applications, such as \sk{in this work}.} There is a plethora of proximal/operator splitting \skblue{methods~\cite{Combettes2011, Boyd2011, Parikh2013, Wang_Banerjee__Bregman_ADMM__2014, Komodakis_and_Pesquet:2015, Glowinski2016, Beck2017, Davis2017, Yan_PD3O:2018, Ryu_Yin_ls_book_2022_draft}  
to} tackle {the} generic composite {problems}. The \sk{operator splitting algorithms can be {roughly grouped} into two classes of algorithms~\cite{Yan_PD3O:2018}, namely primal-dual-type\sk{---\sk{\eg, Chambolle-Pock-type}~\cite{Komodakis_and_Pesquet:2015}---}and primal-/dual-type---\sk{\eg,} \ac{ADMM}-type~\cite{Ryu_Yin_ls_book_2022_draft} \sk{algorithms}.}

Operator splitting for \skblue{two operators} (or, loosely speaking, \sk{optimization problems \sk{in which the} objective function is given by the} \skblue{sum of two functions}) has been widely studied \cite{Combettes2011, Boyd2011, Parikh2013, Komodakis_and_Pesquet:2015, Glowinski2016, Beck2017, Ryu_Yin_ls_book_2022_draft}. {Operator splitting} with more than two composite terms in the objective has been an active research problem without resorting to problem or product space reformulations~\cite{Davis2017} because these \sk{techniques} are either not straightforward or \sk{converge slowly} \cite{Davis2017, Yan_PD3O:2018}. In other words, if the {optimization} problem comprises more than two functions, then \sk{under certain separability and smoothness assumptions,} a typical approach is to {re}formulate the problem by modeling {it} as a block-wise  
\sk{two-operator} problem such that one can exploit the separability {of grouped functions} within each block. However, as \sk{shown} in~\cite{Davis2017,Yan_PD3O:2018, Kant_journal_msp_top_admm:2020}, \sk{this can lead to slow convergence} or/and~{exhibit memory inefficiency}. Therefore, novel algorithms are  
researched and developed  
to solve composite problems comprising more than two operators.

\textit{Related Works on Three-Operator Splitting}: 
Recently, some authors have extended both primal-/dual-type (including \ac{ADMM}-type) \sk{algorithms} and primal-dual \sk{classes} of splitting algorithms from two operators to three operators \cite{Wang_Banerjee__Bregman_ADMM__2014, Davis2017, Yan_PD3O:2018, Banert2020, Kant_journal_msp_top_admm:2020, Combettes_data_science__2021, Ryu_Yin_ls_book_2022_draft}. 
{
\sk{\ac{ADMM} (primal-/dual-type splitting) algorithms for three operators have gained interest in the literature. }
There are many varieties of \ac{ADMM} in the literature, see, \eg, \cite{Glowinski2016}. One of the popular variants is \sk{the} so-called proximal \ac{ADMM}\sk{, which} essentially regularizes each subproblem with a quadratic proximal regularization (or weighted norm). \sk{A} special case of proximal \ac{ADMM} is linearized proximal \ac{ADMM}, which basically cancels out the inconvenient quadratic terms in the subproblem.  
\sk{However}, ``linearization" in \sk{the} \ac{ADMM}-type literature \sk{can be ambiguous}~\cite{Ryu_Yin_ls_book_2022_draft}. Besides the aforementioned linearization of the subproblem, there are also techniques that linearize the $L$-smooth{\footnote{\skblue{See Definition~\ref{def:definition_of_lipschitz_continous_gradient} of $L$-smooth function.}}} \sk{objective} function within \sk{the} \ac{ADMM}-type class of algorithms designed to solve \sk{the} three-operator problem \cite{Wang_Banerjee__Bregman_ADMM__2014, Davis2017, Banert2020, Kant_journal_msp_top_admm:2020}, which \sk{we collectively refer} to as \ac{TOP-ADMM}-type \sk{algorithms} in this work.
}

\subsection{Contribution}

In this paper, we develop scalable, flexible, and computationally efficient \sk{optimization methods} to tackle \sk{a} robust and nonconvex \ac{PAPR} reduction problem while catering \sk{to} both beamforming and broadcasting in standard-compliant wireless communication systems such as 5G NR. 
{Our contributions are as follows:}
\begin{itemize}
    \item We formulate \sk{a} \ac{PAPR} reduction \sk{optimization} problem amenable to large-scale \ac{MIMO}-\ac{OFDM}, where the transmit distortion (or \ac{EVM}) stemming from the \ac{PAPR} reduction subproblem is mitigated at the receiver without imposing any additional signal processing \sk{requirement} at the receiver.
    \item \sk{Additionally, the proposed problem formulation offers flexibility to adapt the desired \ac{PAPR} and \ac{ACLR} targets, and can seamlessly support both beamforming and broadcasting carrying resources}. \sk{Note that user-specific and common resources may exist in the same \ac{OFDM} symbol.} {\sk{One could argue that it is not non-trivial to extend the previously proposed methods supporting both beamforming and broadcasting, particularly a special case of our proposed formulation. 
    }} 
    \item We introduce a powerful yet simple \sk{optimization solution algorithm based on our recently proposed \ac{TOP-ADMM} method~\cite{Kant_journal_msp_top_admm:2020}, which is intended for a class of three-operator proximal splitting schemes. 
   To benchmark it against state-of-the-art \sk{algorithms}, we also provide details on how to apply \ac{BADMM}~\cite{Kant_et_al__asilomar_2021,Wang_Banerjee__Bregman_ADMM__2014} and \ac{DYS}~\cite{Davis2017} to the problem at hand.
    } 
    \item We present simulation results for downlink 5G NR-like \sk{systems} to substantiate the capability of the proposed \ac{TOP-ADMM}-based algorithm. Additionally, we analyze the performance considering \ac{PAPR} and \ac{EVM} metrics under  imperfect \ac{CSI} \sk{available} at the transmitter \sk{and considering spatially-correlated wireless channels. 
    } 
\end{itemize}
 
\subsection{Notation}
Let the set of complex  and real numbers be denoted by $\mathbb{C}$ and $\mathbb{R}$, respectively. $\Re\{x\}$ denotes the real part of a complex number $x \!\in\!\Cm$. 
The $i$-th element of a vector $\vec{a} \! \in \! \mathbb{C}^{m \times 1}$ and $j$-th column vector of a matrix $\mat{A} \! \in\! \mathbb{C}^{m \times n}$ are denoted by ${a}[i] \!\coloneqq \left(\vec{a}\right)_i \!\! \in \! \mathbb{C}$ 
and  $\mat{A}\left[:, j\right] \! \in \! \mathbb{C}^{m \times 1}$, respectively. An $i$-th element of three-order tensor is denoted in matrix form as $\mat{X}[i] \! \in \! \mathbb{C}^{m \times n}$. \skblue{We form a matrix by stacking the set of higher order vectors $\left\{ \vec{a}[n]\! \in \! \mathbb{C}^{M \times 1} \right\}_{n=1}^N$ and $\left\{ \vec{b}[m] \! \in \! \mathbb{C}^{1 \times N} \right\}_{m=1}^M$ column-wise and row-wise as {$\mat{A} \!= \! \left[ \vec{a}[1],\ldots,\vec{a}[N] \right] \! \in \! \mathbb{C}^{M \times N}$ and $\mat{B} \! = \! \left[ \vec{b}[1];\ldots;\vec{b}[M] \right] \! \in \! \mathbb{C}^{M \times N} $}, respectively.} 
The transpose and conjugate transpose of a vector or matrix are denoted by $\left(\cdot\right)^{\rm T}$ and $\left(\cdot\right)^{\herm}$, respectively. The complex conjugate is represented by $\left(\cdot\right)^*$. The $K  \times  K$ identity matrix is written as $\vec{I}_K$. \skblue{The expectation operator is denoted by $\expect\{\cdot\}$.} 
An $i$-th iterative update is denoted by $(\cdot)^{(i)}$; $\vec{e}_k \! \in \! \Rm^{N \times 1}$ is a standard basis vector. \sk{$\mat{F}$ refers to \sk{the} \ac{DFT} matrix; \skblue{$\mat{A}^{\nicefrac{1}{2}}$ denotes the square root of $\mat{A}$.} 
}

\vspace{-5.5mm}
\skblue{
\subsection{Paper Organization:}
}

The remainder of the paper is organized as follows. The next section briefly introduces the system and \ac{EVM} model with useful performance and algorithm design metrics. Section \ref{sec:robust_papr_reduction_algorithm} formulates \ac{EVM} mitigation with \ac{PAPR} and \ac{ACLR} constraints suitable for any practical 5G NR-like system, \ie, supporting both user-specific and cell-specific signals. Moreover, in Section \ref{sec:robust_papr_reduction_algorithm}, we \sk{first introduce our novel \ac{TOP-ADMM} algorithm as a generic splitting method and thereafter present an algorithm tackling the proposed large-scale problem using \ac{TOP-ADMM}. Then, we also present the benchmark algorithms tackling the same proposed problem using \ac{BADMM}~\cite{Kant_et_al__asilomar_2021,Wang_Banerjee__Bregman_ADMM__2014} and \ac{DYS}~\cite{Davis2017}. Moreover, we \sk{characterize the} computational complexities of these \ac{TOP-ADMM}-type algorithms \sk{by} solving the proposed problem and compare their complexities with two prior arts. } 
Section~\ref{sec:simulation_results} \sk{gives}  a rich set of numerical results, and Section \ref{sec:conclusion_future_work} concludes with a summary. 

\vspace{+.5mm}
\section{Preliminaries} \label{sec:prelim}
This section briefly introduces the downlink \ac{MIMO}-\ac{OFDM} system model followed by a short description of the performance metrics \sk{that are} useful \sk{for} designing \skblue{distortion-based} \ac{PAPR} reduction algorithms.

\vspace{-1.0mm}
\subsection{System Model \skblue{for Distortion-Based PAPR Reduction in Frequency-Domain}} \label{sec:system_model}

\begin{figure*}[tp!]
    \centering
    \scalebox{0.74}{\includegraphics[trim=0mm 0mm 0mm 0.1mm,clip]{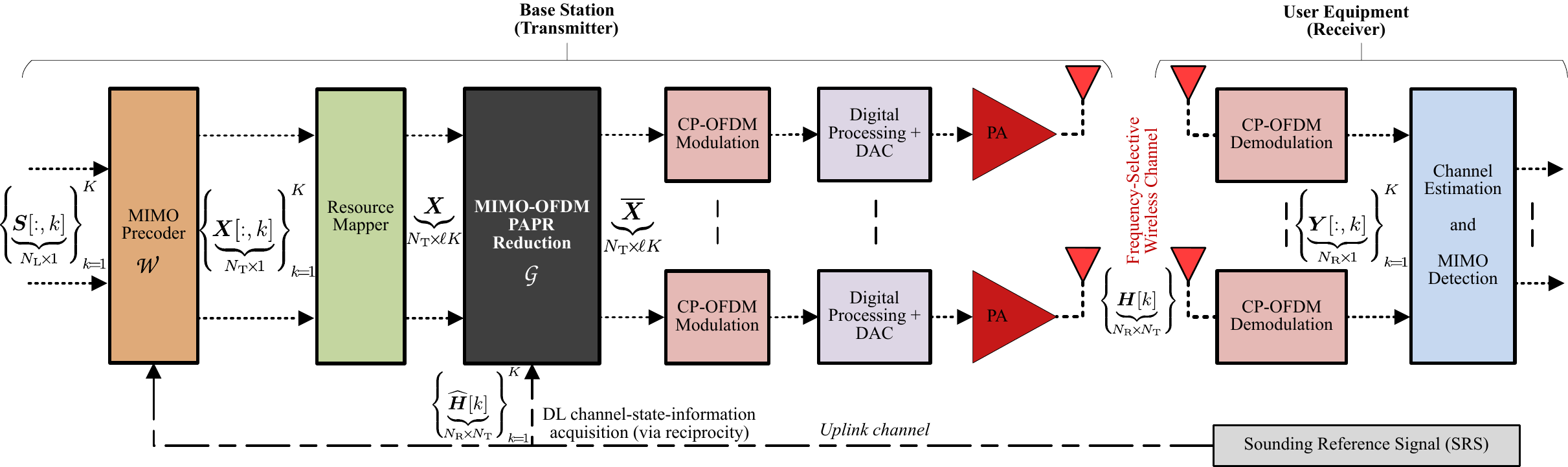}}
    \vspace{-4mm}
    \caption{\footnotesize Simplified block diagram of a single-user MIMO-OFDM transceiver with \ac{PAPR} reduction scheme.}
    \label{fig:block_diag_mimo_ofdm_tx_rx}
    \vspace{-3mm}
\end{figure*}

We consider \sk{an} \ac{OFDM}-based \ac{MIMO} downlink---as depicted in Fig. \ref{fig:block_diag_mimo_ofdm_tx_rx}---where the base station is equipped with $\NT$ \ac{Tx} antennas, and the \ac{UE} is equipped with $\NR$ \ac{Rx} antennas. Moreover, we assume \sk{a} spatial multiplexing transmission scheme with $\NL \!\leq\! \min\left\{\NT,\NR\right\}$ spatial layers. \skblue{For \ac{MU-MIMO}, $\NU$ represent total number of scheduled users sharing all the resources for the data transmission, where each $\mu$-th user can have $N_{{\rm L}_{\mu}} \!\leq \! \NL $ spatial layers such that $\NL \!=\! \sum_{\mu=1}^{\NU} N_{{\rm L}_{\mu}}$.} 

\skblue{
A (generalized) spatially precoded (or beamformed) 
symbol matrix in frequency-domain input to an $\ell$-times oversampled $K$-point \ac{IDFT} matrix is denoted by ${\mat{X}} \! \in \! \Cm^{\NT \times \ell K}$ such that at the $k$-th (useful) subcarrier,
\iffalse
\vspace{-2mm}
\begin{align*} 
    {\Cm^{\NT \times 1} \! \ni\! \mat{X}}[:,k]\, = \mathcal{W} \left( \vec{S}[:,k] \right),
    \vspace{-2mm}
\end{align*}
\else
${\Cm^{\NT \times 1} \! \ni\! \mat{X}}[:,k]\, = \mathcal{W} \left( \vec{S}[:,k] \right)$,
\fi
where non-precoded symbol vector, for all the spatial layers of \ac{SU-MIMO}/\ac{MU-MIMO}, $\mat{S}[:,k] \! \in \! \mathcal{S}^{\NL\times 1}$  belongs to a complex-valued finite-alphabet set $\mathcal{S}$, \eg, corresponding to a $2^Q$-\ac{QAM} constellation with $Q\!\in\!\{2,4,6,8\}$. The spatial {(possibly nonlinear)} precoding $\mathcal{W}\!: \! \mathcal{S}^{\NL \times 1} \! \rightarrow \! \Cm^{\NT \times 1}$ maps the non-precoded symbol vector $\mat{S}[:,k] \!\in\! \mathcal{S}^{\NL \times 1}$ to spatially precoded symbol vector $\vec{X}[:,k] \! \in \! \Cm^{\NT\times 1}$ catering either \ac{SU-MIMO} or \ac{MU-MIMO}. For \ac{MU-MIMO}, $\mat{S}[:,k] \!=\! \left[\mat{S}_{1}[:,k]; \ldots; \mat{S}_{\NU}[:,k]  \right]$, where $\mat{S}_{\mu}[:,k] \! \in \! \mathcal{S}^{N_{{\rm L}_{\mu}}\times 1}$ represent $\mu$-th user's data vector.
}

\subsubsection{\skblue{Transmit SU/MU-MIMO Signal Model}}
\skblue{
Inspired by, \eg, \cite{Kant_journal_emsp:2019, Kant_journal_msp_top_admm:2020}, we model a frequency-domain data matrix $\overline{\mat{X}}$ having reduced \ac{PAPR}  as $\Cm^{\NT \! \times\! \ell K} \! \ni \!  \overline{\mat{X}}  \! \coloneqq \!  \mathcal{G}\left({\mat{X}}\right)$, where a distortion-based  \ac{PAPR} reduction function ${\mathcal{G}}: \! \mat{X} \!\in\! \Cm^{\NT \! \times\! \ell K} \!  \! \rightarrow \! \Cm^{\NT \! \times\! \ell K} \!\ni\! \overline{\mat{X}}$ manipulates the data symbol matrix $\mat{X}$. Consequently, this perturbation signal model capable of reducing \ac{PAPR} at a given $k$-th subcarrier can be expressed as,
\begin{equation} \label{eqn:tx_signal_freq_domain_kth_subcarrer__after_papr__additive_model}
\overline{\mat{X}}\left[:,k\right] \!=\! \alpha[k] {\mat{X}}\left[:,k\right] \!+\! \underbrace{\TxEVM[k]}_{\rm Tx \ distortion} \! \approx \! {\mat{X}}\left[:,k\right] + \TxEVM[k]
\end{equation}
where $\alpha[k] \! \in \! \Cm$ {is} a deterministic scalar and $\TxEVM[k]$ is an instantaneous distortion that {is} uncorrelated with $\mat{X}[:,k]$ but statistically dependent. However, appealing to the works in~\cite{Moghadam:18, Larsson:18, Moghadam_distortion_corr_mimo:12, Sienkiewicz_spatially_dependent:14, Kant_journal_msp_top_admm:2020}, $\TxEVM[k]$ stemming from any (possibly non-linear) source, \eg, clipping, spectral precoding, is correlated across antennas and consequently beamformed in the similar direction as the signal $\mat{X}[:,k]$, for instance, depending on the ratio of transmit spatial layers and the transmit antenna branches---\ie, rank of the channel, frequency-granularity of the spatial precoding---\ie, channel coherence bandwidth. 
Observe that {although the scaling factor $\alpha[k]$ may vary across subcarriers, on average, the scaling is assumed to be real-valued and nearly unity---our extensive simulations in  Section~\ref{sec:simulation_results} show that the approximate additive \ac{EVM} model \eqref{eqn:tx_signal_freq_domain_kth_subcarrer__after_papr__additive_model} is accurate and ignoring $\alpha[k]$ has nearly negligible impact on the key performance metrics}.
}

\subsubsection{\skblue{Received SU/MU-MIMO Signal Model}}
\skblue{
For \ac{MU-MIMO}, let $N_{{\rm R}_{\mu}}$ denote the number of receive antennas of user $\mu$ such that the total number of receive antennas for all the users is $\NR \!=\! \sum_{\mu=1}^{\NU} N_{{\rm R}_{\mu}}$. Additionally, let $\mat{H}_{\mu}\left[k\right] \!\in\!\Cm^{N_{{\rm R}_{\mu}} \times \NT}$ describe the channel matrix between user $\mu$ and base station such that the composite channel matrix over all the receive antennas of users is $\mat{H}\left[k\right] \!=\! \left[\mat{H}_{1}\left[k\right]; \ldots; \mat{H}_{\NU}\left[k\right] \right]$. Therefore, the received signal model of user $\mu$ can be expressed as
\vspace{1.5mm}
\begin{subequations} \label{eqn:rx_mu_su_signal_freq_domain_kth_subcarrer__with_unprecoded_pertubation}
\begin{align} 
{\Cm^{N_{{\rm R}_{\mu}} \times 1} \!\ni} \mat{Y}_{\mu}[:,k]  
=&  \mat{H}_{\mu}[k]\overline{\mat{X}}\left[:,k\right] \! +\! \vec{n}_{\mu}[k] \\ 
=& \underbrace{\mat{S}_{\mu}\left[:,k\right]}_{\rm desired} \nonumber \\ 
&+\! \underbrace{\left(\mat{H}_{\mu}\left[k\right]\overline{\mat{X}}\left[:,k\right] \!-\! \mat{S}_{\mu}\left[:,k\right] \right)}_{\rm interuser \, interference \, and \, received \, Tx \, distortion } \! \nonumber \\ 
&+ \vec{n}_{\mu}[k],
\end{align}
\end{subequations}
\begin{subequations} \label{eqn:rx_mu_su_signal_freq_domain_kth_subcarrer__with_desired_and_interlayer_interference}
\begin{align} 
{\phantom{{\Cm^{N_{{\rm R}_{\mu}} \times 1} \!\ni} \mat{Y}_{\mu}[:,k] }} 
\approx&  \mat{H}_{\mu}\left[k\right]{\mat{X}}\left[:,k\right] \!+\!  {\mat{H}_{\mu}\left[k\right]\TxEVM\left[k\right]} \! + \! \vec{n}_{\mu}[k]   \\ 
=&\! \underbrace{\mat{S}_{\mu}\left[:,k\right]}_{\rm desired} \nonumber \\ 
&+\! \underbrace{\left(\mat{H}_{\mu}\left[k\right]{\mat{X}}\left[:,k\right] \!-\! \mat{S}_{\mu}\left[:,k\right] \right)}_{\rm interuser \ interference} \nonumber \\
&+\! \underbrace{\mat{H}_{\mu}\left[k\right]\TxEVM\left[k\right]}_{{\rm received    \, {Tx} \, {distortion}}} \! + \vec{n}_{\mu}[k],
\end{align}
\end{subequations}
where $\vec{n}_{\mu}[k]$ is effective noise at user $\mu$, which can be modeled as a zero-mean complex symmetric white Gaussian noise with $\mathcal{CN}\left(\vec{0}, N_0\vec{I}_{N_{{\rm R}_{\mu}}}\right)$, where $N_0$ is the noise variance at the receiver. 
The transmit distortion on the receiver side of user $\mu$ at subcarrier $k$ can be expressed by $\mat{H}_{\mu}\left[k\right] \TxEVM\left[ k\right]$, which unfortunately correlates the total noise spatially at the receiver since the effective noise covariance matrix seen at the user side---ignoring the $k$-th index for brevity---$\mat{R}_{\mu} \!= \! \mat{H}_{\mu} \mat{R}_{\TxEVM\TxEVM} \mat{H}_{\mu}^\herm \! + \! N_0 \vec{I}_{N_{{\rm R}_{\mu}}}$ where $\mat{R}_{\TxEVM\TxEVM} \! = \!\expect\left\{ \TxEVM\TxEVM^\herm \right\}$ is the transmit distortion covariance. Therefore, the impact of transmit distortion has unfortunately a {well-known} pernicious effect on the system-wide throughput, see, \eg,~\cite{Studer_papr_2013, Bao_papr_admm__2018, Yao__papr_sdp__2019, Kant_journal_msp_top_admm:2020} and their references. To minimize the impact on the system throughput due to transmit distortion, we focus on reducing the transmitted $\TxEVM$ or/and received $\mat{H}_{\mu}\TxEVM$ distortion energy. 
For \ac{SU-MIMO} received signal model, all the $\NL$ spatial layers target a single user $\mu\!=\!1$ having $\NR$ receive antennas. Thus, the received signal model for both \ac{SU-MIMO} and over all the receive antennas of $\NU$ users in \ac{MU-MIMO} is exactly same as $\Cm^{\NR \times 1}\! \ni\! \mat{Y}[:,k] \!=\! \left[\mat{Y}_{1}[:,k]; \ldots; \mat{Y}_{\NU}[:,k] \right] \!\approx\! \mat{H}\left[k\right]{\mat{X}}\left[:,k\right] \!+\!  {\mat{H}\left[k\right]\TxEVM\left[k\right]} \! + \! \vec{n}[k]$. Hence, we will use the overall received signal model that is valid for both \ac{SU-MIMO} and \ac{MU-MIMO}. 
}

\subsection{Performance Metrics} \label{sec:performance_metrics}
 
We utilize three \sk{figures of merit: 1) \ac{PAPR} and \ac{IPAPR}, 2) inband distortions, and 3) \ac{OOBE} for the \ac{PAPR} reduction algorithm design and its performance.}

\subsubsection{PAPR and \sk{IPAPR}} \label{sec:papr_intro}
\sk{The \ac{PAPR} and \sk{\ac{IPAPR}} of \sk{the} \ac{MIMO}-\ac{OFDM} signal in \sk{the} time domain, per \sk{the} $j$-th transmit antenna branch, \sk{are} defined as 
\sk{$\mathrm{PAPR}_{j} \! \coloneqq \! {\left\|\mat{F}^\herm \left(\mat{X}[j,:]\right)^\trans\right\|_{\infty}^2}/{\left\|\mat{F}^\herm \left(\mat{X}[j,:]\right)^\trans\right\|_2^2}$ and  $\mathrm{IPAPR}_{j}\!\left[n\right] \! \coloneqq \! {\left|\!\left(\!\mat{F}^\herm \left(\mat{X}[j,:]\right)^\trans\!\right)_n\right|^2}/{\left\|\mat{F}^\herm \left(\mat{X}[j,:]\right)^\trans\right\|_2^2}$}, respectively, where $\mat{F}^\herm \! \in \! \Cm^{\ell K \times \ell K}$ is the~\ac{IDFT}~matrix.}

\subsubsection{Inband distortions}\label{sec:performance_metrics__inband_distortions}
The inband distortions are evaluated in terms of three \ac{EVM} metrics\footnote{\sk{The} \ac{EVM} metric evaluation provides insightful and useful information on the link quality in terms of \ac{SNR} seen at the receiver due to the aggregated digital and analogue hardware imperfections~\cite{Hassun:1997, Mahmoud:2009}. Hence, in this study, we evaluate the inband performance in terms of proposed \ac{EVM} metrics.}, namely i) (unequalized) \ac{EVM} at the transmitter, ii) \skblue{predicted} received \ac{EVM}, and iii) \skblue{estimated} received \ac{EVM}. Using \eqref{eqn:rx_mu_su_signal_freq_domain_kth_subcarrer__with_unprecoded_pertubation} and  \eqref{eqn:rx_mu_su_signal_freq_domain_kth_subcarrer__with_desired_and_interlayer_interference}, we define these three \skblue{instantaneous}  performance and algorithm design metrics at subcarrier~$k$ for a given \ac{MIMO}-\ac{OFDM} symbol.
\begin{enumerate}[(i)]
    \item \skblue{Transmit} (unequalized) \ac{EVM}: 
    $ 
    {\rm EVM}_{\rm Tx}\left[k\right] 
    \!\coloneqq\! {{\left\| \left(\! \overline{\mat{X}}\!\left[:,k\right] \!-\! \mat{X}\!\left[:,k\right] \right)  \right\|_2}/{\left\|  \mat{X}\!\left[:,k\right]  \right\|_2}} \skblue{ \approx \! \left\| \TxEVM\!\left[k\right]\right\|_2/{\left\|  \mat{X}\!\left[:,k\right]  \right\|_2}}
    $
    represents the transmit signal distortion energy. For the performance evaluation, we assume that the average transmit signal power for each $j$-th transmit antenna branch $\expect{\left\{ \left\| \mat{X}[j,:] \right\|_2^2 \right\}}$ is fixed, and the total signal energy per \ac{MIMO}-\ac{OFDM} symbol is \sk{the} same before and after \ac{PAPR} reduction. 
    
    \item \skblue{Predicted} received \ac{EVM}: 
    $
    {\rm EVM}_{\rm pred}\left[k\right] 
    \!\coloneqq\! \left\| \mat{H}\!\left[k\right]\! \left(\! \overline{\mat{X}}\!\left[:,k\right] \!-\! \mat{X}\!\left[:,k\right] \right)  \right\|_2\!/\!\left(\!\left\| \mat{H}\!\left[k\right]\!  \mat{X}\!\left[:,k\right]  \right\|_2\!\right)$ 
    where $\left\{\mat{H}\!\left[k\right]\right\}$ is \ac{CSI} acquired at the transmitter via \sk{the} uplink channel \sk{for the user-specific transmission}.   
    
    \item \skblue{Estimated} received \ac{EVM} per spatial layer after \ac{MIMO} detection (\sk{with} channel equalization):
    \sk{$ \allowbreak 
    {\rm EVM}_{\rm est}\!\left[l,k\right]  
    \!\coloneqq\! {\left|\widehat{\overline{\mat{S}}}\!\left[l,k\right] \!-\! \mat{S}\!\left[l,k\right]  \right|}/{\left| \mat{S}\!\left[l,k\right] \right|},
    $} 
    where $\widehat{\overline{\mat{S}}}\!\left[l,k\right] \!\coloneqq\!\mat{G}\!\left[k\right]\!\mat{H}\!\left[k\right]\!\overline{\mat{X}}\!\left[:,k\right]$ corresponds to {the} estimated symbol for spatial layer $l$ and subcarrier $k$ with $\mat{G}\left[k\right] \!\in\! \Cm^{\NL \times \NT}$ equalizer at the receiver.
\end{enumerate}

\subsubsection{Out-of-band emissions}
In this work, the \ac{OOBE} requirements are simply characterized by the amount of distortion \sk{energy} on the unused subcarriers $k \!\in\!\mathcal{T}^{\perp}$ with respect to the useful data-carrying subcarriers $k \!\in\!\mathcal{T}$, which is referred to as the \ac{ACLR} \sk{per $j$-th antenna} and can mathematically be described as \sk{$\mathrm{ACLR}_{j} \! \coloneqq \! { \left\|\overline{\mat{X}}\left[j,\mathcal{T}^{\perp}\right] \right\|_2^2}/{ \left\|\overline{\mat{X}}\left[j,\mathcal{T}\right]\right\|_2^2 }$}. \sk{Note our \ac{ACLR} definition in the description of optimization problem---posed in the subsequent section---can be seen as reciprocal to the 3GPP \ac{ACLR} definition~\cite{3GPPTS38.1042018NRReception}. However, we interchangeably use both definitions but with the appropriate sign of the \ac{ACLR} value.}  

\section{Proposed EVM Mitigation with PAPR and ACLR Constraints \sk{IN} 
MIMO-OFDM} \label{sec:robust_papr_reduction_algorithm}

\skblue{
This section introduces proximal/operator splitting techniques, particularly, \ac{TOP-ADMM} proposed recently in~\cite{Kant_journal_msp_top_admm:2020}, that can tackle an appropriate optimization problem potentially yield computationally efficient algorithm for the large-scale optimization problem. In the sequel, we formulate {a} robust \ac{PAPR} reduction problem that mitigates the predicted received \ac{EVM} under channel estimates' uncertainty while targeting desired \ac{PAPR} and \ac{ACLR} levels. Subsequently, we formulate a robust problem amenable to the \ac{TOP-ADMM}-type algorithms. Finally, we propose a computationally efficient \ac{TOP-ADMM}-based algorithm and develop algorithms utilizing two popular \ac{BADMM}~\cite{Wang_Banerjee__Bregman_ADMM__2014} and \ac{DYS}~\cite{Davis2017} techniques. 
}

\subsection{\skblue{Efficient Proximal/Operator Splitting Algorithms}} \label{sec:top_admm_algorithm}
\skblue{
As discussed in Section~\ref{sec:related_works_and_contribution}, we present some proximal/operator splitting techniques, notably \ac{TOP-ADMM}, \ac{DYS}, and \ac{BADMM} that can render computationally efficient algorithms, typically using proximal operators, to solve a large-scale problem in a principled manner. We consider a generic problem,
\begin{equation}
\label{eqn:general__form}
\underset{ \vec{x} \in \mathcal{X}}{\operatorname{minimize}}  \quad \sum_{m=1}^M \! f_m \! \left( {\vec{x}}\right) \!+\! g\left( {{\vec{x}}} \right) \!+\! \beta h\!\left( {{\vec{x}}} \right), 
\end{equation}
for some $M \! \geq \! 1$, where $\mathcal{X}$ is a real/complex Euclidean space, $\left\{f_m(\cdot) \right\}$, $g(\cdot)$ and $h(\cdot)$ are closed, convex, and proper functions, and $h$ is $L$-smooth (see Definition~\ref{def:definition_of_lipschitz_continous_gradient}) with some scaling $\beta \!\in\!\Rm_{\geq0}$.
Foremost, we layout two useful definitions in the sequel to introduce proximal/operator splitting algorithms.
}

\skblue{
\vspace{-2mm}  
\begin{definition}[$L$-smooth function~{\cite{Bauschke:2011, Beck2017}}]  \label{def:definition_of_lipschitz_continous_gradient}
A differentiable function $f\!\left(\!\vec{z}\right) \! \in\! \Rm$, where $\vec{z} \!\in\! \Cm^n$, \skblue{is $L$-smooth, \ie,} has $L$-Lipschitz continuous gradient (for $L \!>\! 0$) if
$\left\| \nabla \!f\!\left(\vec{z}_1\right)  \!-\! \nabla\! f\!\left(\vec{z}_2\right) \right\| \!\leq\! L \!\left\|\! \vec{z}_1 \!-\! \vec{z}_2 \!\right\| \: \forall \vec{z}_1,\! \vec{z}_2 \!\in\! \Cm^n$.
\end{definition}
}

\skblue{
\vspace{-2mm}
\begin{definition}[Proximal mapping~\cite{Parikh2013}\cite{Beck2017}] \label{definition:prox_operator}
Given a proper closed convex function $f: \dom_f \mapsto \left(-\infty\right., \left. +\infty \right]$, then the proximal mapping of $f$ is the operator given by 
\begin{align}
    \prox_{\lambda f}\left( \mat{X}\right) \!= \!\arg\min_{\mat{Z} \in  \dom_f} \left\{ f\left(\mat{Z}\right) \! + \! \frac{1}{\beta \lambda} \left\|\mat{X} \! - \! \mat{Z} \right\|_F^2\right\} 
\end{align}
for any $\mat{X} \in  \dom_f$, where $\dom_f$ corresponds to the domain of a function $f$ and $\lambda > 0$. If $\mat{Z}$ is complex-valued or real-valued, $\beta = 1$ or $\beta=2$, respectively. 
\end{definition}
}

\skblue{
\begin{definition}[Proximal mapping of the indicator function~\cite{Parikh2013}\cite{Beck2017}] \label{definition:prox_operator_on_indicator_function}
Let $f \!\coloneqq\! \Ind_{C}: \dom_f \!\mapsto\! \left(-\infty\right., \left. +\infty \right]$ be an indicator function (also known as characteristic function),
\begin{align} \label{eqn:definition_of_characteristic_function}
	f(\vec{z}) = \Ind_{\mathcal{C}}\left(\vec{z}\right) \coloneqq
	\left\{ 
		 \begin{matrix}
		  0,        &  \vec{z} \in    \mathcal{C}  \\
		  +\infty,  &  \vec{z} \notin \mathcal{C},
		\end{matrix} 
	\right.
\end{align} 
then the proximal mapping of a given set $\mathcal{C} \!\neq\!\emptyset$ is {the} orthogonal projection operator onto $\mathcal{C}$, \ie, 
\begin{align*}
    \prox_{\lambda \Ind_{\mathcal{C}}}\left( \vec{z}\right) 
    &= \arg\min_{\vec{y} \in  \dom_f} \left\{ \Ind_{\mathcal{C}}\left(\vec{y}\right) + \frac{1}{\beta \lambda} \left\|\vec{z} - \vec{y} \right\|_2^2\right\}  \\
    &\equiv \arg\min_{\vec{y} \in  \mathcal{C}} \left\{ \left\|\vec{z} - \vec{y} \right\|_2^2\right\} \!=\! \proj_{\mathcal{C}}\left( \vec{z}\right).
\end{align*}
\end{definition}
}

\skblue{
In the subsequent sections, we present our proposed \ac{TOP-ADMM}~\cite{Kant_journal_msp_top_admm:2020} generic algorithm, and briefly introduce \ac{DYS}~\cite{Davis2017} and \ac{BADMM}~\cite{Wang_Banerjee__Bregman_ADMM__2014} algorithms.
}

\skblue{
\subsubsection[TOP-ADMM Algorithm]{TOP-ADMM Algorithm~{\cite{Kant_journal_msp_top_admm:2020}}}
It is one of the generalized algorithms for the classical consensus \ac{ADMM}---see details of \ac{ADMM} and its applications in, \eg, review paper~\cite{Boyd2011}. This \ac{TOP-ADMM} can be employed to solve many optimization problems\footnote{\skblue{For instance, see~\cite{Kant_journal_msp_top_admm:2020}, where we applied \ac{TOP-ADMM} to tackle a large-scale mask-compliant spectral precoding problem in the MIMO-OFDM-based wireless communication systems, particularly 5G NR.}} that can be expressed as a composite problem such as~\eqref{eqn:general__form}---which utilizes the gradient of $L$-smooth function $h$, unlike classical \ac{ADMM}. The \ac{TOP-ADMM} algorithm can be classified as a divide-and-conquer method, which decomposes a big optimization problem---difficult to solve in a composite form~\eqref{eqn:general_consensus_top_admm__generic_form}---into smaller subproblems that are easy to solve. 
}

\skblue{
Let us reformulate problem~\eqref{eqn:general__form}, without loss of generality, which has at least one solution, as
\begin{alignat}{2}\label{eqn:general_consensus_top_admm__generic_form}
    &\underset{ \left\{{\vec{x}}_m \in \Cm^n \right\}\!,  {\vec{z}} \in \Cm^n}{\operatorname{minimize}}&  \sum_{m=1}^M \! f_m \! \left( {{\vec{x}}_m} \right)  \!+\! g\left( {{\vec{z}}} \right) \!+\! \beta h\!\left( {{\vec{z}}} \right) \nonumber \\ &\operatorname{subject \, to}& \  {{\vec{x}}_m} - {\vec{z}} = {\vec{0}}, \forall m\!=\!1,\ldots,M.
\end{alignat}
}

\skblue{
We lay out the general definition of \sk{our} \ac{TOP-ADMM} method, which was introduced in~\cite{Kant_journal_msp_top_admm:2020}. 
}
\begin{theorem}[TOP-ADMM] \label{thm:definition_of_topadmm_algorith__feasible_problem}
\skblue{Consider a problem given in~\eqref{eqn:general_consensus_top_admm__generic_form} with at least one solution and a suitable step-size $\tau \! \in \! \Rm_{\geq 0}$.  Assume subproblems~\eqref{eqn:update_xm__step1_parallel__general_top_admm__prox__for_convergence} and \eqref{eqn:update_z__step2__general_top_admm__prox__for_convergence} have solutions, and consider a relaxation/penalty parameter $\rho \! \in \! \Rm_{> 0}$ with some arbitrary initial $ \left(\left\{\vec{x}_m^{(0)}\right\}, {\vec{z}^{(0)}}, \left\{\vec{y}_m^{(0)}\right\}\right)$. Then} 
\begin{subequations} \label{eqn:generalized_top_admm_algorithm_iterates__for_convergence}
\begin{align}
    \label{eqn:update_xm__step1_parallel__general_top_admm__prox__for_convergence}
    {\vec{x}}_m^{\left(i+1\right)} \! 
    =& \arg\min_{{\vec{x}}_m}  f_m\left( {\vec{x}}_m \right) \!+ \! \rho \! \left\|\!  {{\vec{x}}_m} \! - \! {\vec{z}^{\left(i\right)}} \! +\! \frac{{\vec{y}}_m^{\left(i\right)}}{\rho} \right\|_2^2 \; \forall m \\ 
    \label{eqn:update_z__step2__general_top_admm__prox__for_convergence}
    {\vec{z}}^{\left(i+1\right)} \! 
    =& \arg\min_{{{\vec{z}}}}  g\left( {\vec{z}} \right)  \nonumber \\ 
    &+\!   \sum_{m=1}^M  \rho \left\|  {{\vec{x}}_m^{\left(i+1\right)}} \! - \! {\vec{z}} \! {- \tau \nabla h \! \left( \! {\vec{z}}^{\left( i \right)} \right)} \! + \!  \frac{{\vec{y}}_m^{\left(i\right)}}{\rho} \right\|_2^2 \\
    \label{eqn:update_dual_ym__step3_parallel__general_top_admm}
    {\vec{y}}_m^{\left(i+1\right)} =& {\vec{y}}_m^{\left(i\right)} + \rho \left( {{\vec{x}}_m^{\left(i+1\right)}} - {\vec{z}^{\left(i+1\right)}} \right) \quad \forall m=1,\ldots,M,
\end{align}
\end{subequations}
at any limit point, converges to a \ac{KKT} stationary point of~\eqref{eqn:general_consensus_top_admm__generic_form}.
\vspace{-2.5mm}
\begin{proof}
\skblue{We prove dual residual $\lim_{i  \rightarrow +\infty} \! \left(\! \vec{z}^{\left( i+1 \right)} \! - \! \vec{z}^{\left( i \right)} \! \right) \! = \! 0$ and primal residual $\lim_{i \rightarrow +\infty} \! \left(\! \vec{x}_m^{\left( i+1 \right)} \! - \! \vec{z}^{\left( i+1 \right)} \!\right) \! = \! 0$, $\forall m \! = \! 1,\ldots,M$, hold under mild assumptions. Consequently, we establish the global convergence~\cite{Kant_etal__intro_to_TOPADMM_2022}.}
\end{proof}
\end{theorem}

\skblue{
Please observe that the classical consensus \ac{ADMM} algorithm is a special case of our proposed \ac{TOP-ADMM} algorithm when $h\!=\!0$ in~\eqref{eqn:general_consensus_top_admm__generic_form} or $\nabla h\!=\!0$ in~\eqref{eqn:generalized_top_admm_algorithm_iterates__for_convergence}. Although classical consensus \ac{ADMM} can solve many problems in signal processing for communications, it does not necessarily yield an implementation-friendly algorithm, particularly, if the proximal operator of $L$-smooth function $h$ is inefficient to compute.  
}

\subsubsection[DYS Algorithm]{\skblue{DYS Algorithm~\cite{Davis2017}}}\label{sec:intro_to_DYS}
\skblue{
It tackles the subform of~\eqref{eqn:general__form}, \ie,
\begin{equation}\label{eqn:problem_setup_for_dys_algorithm}
\operatorname{minimize}_{\vec{x} \in \mathcal{X}} \quad f_1(\vec{x}) \!+\! g(\vec{x}) \!+\! h(\vec{x}).   
\end{equation}
\begin{theorem}[DYS] \label{thm:definition_of_dys_algorithm}
Consider a step-size $\tau \!\coloneqq\! \left(0, 2/L \right)$ with some arbitrary initial $ \vec{z}^{(0)} \!\in\! \mathcal{X}$. Set $\varrho\!=\!2\!-\!(\tau L)/2$. Let $\left(\rho^{(i)}\right)_{i \in \mathbb{N}}$ be a sequence in $\left[0, \varrho\right]$ such that $\sum_{i \in \mathbb{N}} \rho^{(i)} \left(\varrho \!-\! \rho^{(i)}\right) \!=\! +\infty$. Then, the sequence $\left\{\vec{x}^{\left(i+1\right)}\right\}$ generated by the following iterative scheme
\begin{subequations} \label{eqn:dys_algorithm}
\begin{align}
    \label{eqn:update_x__step1__dys}
    \vec{x}^{\left(i+1\right)} \! 
    =& {\prox}_{\tau f_1} \left(  \vec{z}^{\left(i\right)}  \right) \\ 
    \label{eqn:update_z__step2__dys}
    \vec{z}^{\left(i+1\right)} \! 
    =& \vec{z}^{\left(i\right)} \!+\! \rho^{(i)} \Bigl[\!{\prox}_{\tau g}\!\bigl( 2\vec{x}^{\left(i+1\right)} \nonumber \\ 
    &\hspace{5mm}-\! \vec{z}^{\left(i\right)} \!-\! \tau \nabla h\!\left( \vec{x}^{\left(i+1\right)}\right) \bigr) \!-\! \vec{x}^{\left(i+1\right)}\!\Bigr],
\end{align}
\end{subequations}
converges to a solution of~\eqref{eqn:problem_setup_for_dys_algorithm}.
\end{theorem}
}

\skblue{
\subsubsection[BADMM Algorithm]{BADMM Algorithm~\cite{Wang_Banerjee__Bregman_ADMM__2014}} \label{sec:intro_to_BADMM} Although this algorithm may seem to solve a two-operator problem, it utilizes the Bregman distance to linearize the $L$-smooth function $h$, and consequently solve the three-operator problem~\eqref{eqn:problem_setup_for_dys_algorithm}. We omit the details, but we refer the interested reader to~\cite[Section~2.1]{Wang_Banerjee__Bregman_ADMM__2014}. 
}

\skblue{
To this end, we formulate the problem that is amenable to the \ac{TOP-ADMM}, \ac{DYS}, and \ac{BADMM}.
}

\subsection{Proposed Problem Formulations}

\skblue{We propose a principled approach to mitigate (in \ac{CSI}-aware mode under channel uncertainty) and/or minimize (in non-\ac{CSI}-aware mode) signal distortion according to the appropriate conditions and/or requirements subject to the \ac{PAPR} and \ac{ACLR} constraints.
Consequently, we first pose an optimization} problem~\texttt{P0} that essentially minimizes the total \skblue{transmit} \ac{EVM} over all the allocated subcarriers and transmit antennas, \skblue{see~\eqref{eqn:problem_p0__tx_evm}}, subject to the \sk{three performance metrics introduced in Section~\ref{sec:performance_metrics}, namely  nonconvex $\left\{\mathrm{PAPR}_{j}\right\}$\skblue{, see~\eqref{eqn:problem_p0__papr_constraint}}, frequency-selective {\skblue{predicted}} received $\left\{{\rm EVM}_{\rm pred}\left[k\right]\right\}$\skblue{, see~\eqref{eqn:rx_tx_evm_per_subcarrier__ce_unaware}}, and nonconvex $\left\{\mathrm{ACLR}_{j}\right\}$\skblue{, see~\eqref{eqn:aclr__ce_unaware},} constraints}.  
Therefore, the optimization problem can mathematically be \mbox{described~as:} 
\begin{subequations} \label{eqn:problem_p0__ce_unaware}
\begin{align} \label{eqn:problem_p0__tx_evm}
(\texttt{P0}) \quad & \underset{ \overline{\mat{X}}  \in \Cm^{ \NT \times \ell K }}{\operatorname{minimize}} 
& &  \left\| \overline{\mat{X}} - \mat{X} \right\|_F^2   \\  
\label{eqn:problem_p0__papr_constraint}
& \operatorname{subject \, to}
& & \! \frac{\left\| \mat{F}^\herm \overline{\mat{X}}^\trans \right\|_{\infty}^2}{\left\| \mat{F}^\herm \overline{\mat{X}}^\trans \right\|_2^2} \preceq \!   \bm{\gamma}_{\rm par} \\ 
\label{eqn:rx_tx_evm_per_subcarrier__ce_unaware}
& & & \left\| {\mat{H}}\!\left[k\right]\left( \overline{\mat{X}} \! - \! \mat{X} \right) \vec{e}_k \right\|_2 \! \leq \!  \epsilon_{\rm evm}\left[k\right] \ \forall k \! \in \! \mathcal{T}   \\
\label{eqn:aclr__ce_unaware}
& & & \sk{\frac{ \left\|\overline{\mat{X}}\left[:,\mathcal{T}^{\perp}\right] \right\|_2^2}{ \left\|\overline{\mat{X}}\left[:,\mathcal{T}\right]\right\|_2^2 } \preceq  \bm{\psi}_{\rm aclr}},
\end{align}
\end{subequations} 
where $\preceq$ denotes element-wise inequality, and \sk{$\bm{\gamma}_{\rm par}\!\in\!\Rm^{\NT\!\times\!1}$, $\epsilon_{\rm evm}\!\left[k\right]\!\in\!\Rm $, and $\bm{\psi}_{\rm aclr} \!\in\!\Rm^{\NT\!\times\!1}$ are predefined thresholds for the desired \ac{PAPR}, the received \ac{EVM} per $k$-th useful data-carrying subcarrier, and \ac{ACLR} constraint}, respectively. These predefined thresholds can be scalar, \ie, the same values are used for all the antennas. \sk{Observe that this ACLR constraint is not shaping the spectrum such as the methods described in~\cite{Kant_journal_emsp:2019, Kant_journal_msp_top_admm:2020}. However, this constraint could ensure \sk{that the} minimum 3GPP requirement of 45~dB \cite{3GPPTS38.1042018NRReception} \sk{is met} after some spectrum shaping, \eg, ${\psi}_{\rm aclr}\!=\!-50$~dB, \sk{see} Section~\ref{sec:simulation_results}}. \skblue{Furthermore, although in~\eqref{eqn:rx_tx_evm_per_subcarrier__ce_unaware} we have used a perfect channel estimate notation, subsequently, we develop a robust formulation that utilizes imperfect and/or incomplete channel estimates. Nevertheless, constraint~\eqref{eqn:rx_tx_evm_per_subcarrier__ce_unaware} can essentially exploit the excess spatial degrees-of-freedom  to mitigate the transmit distortion at the receiver(s)---see Section~\ref{subsec:simulation_results}, particularly, Fig.~\ref{fig:Average_Estimated_Rxd_EVM_over_Mantennas__Mx2_or_Mx4} that shows that the (predicted/estimated) received \ac{EVM} decreases with the increasing spatial degrees-of-freedom or number of base station transmit antennas. Notice that instead of constraining predicted received \ac{EVM} in~\eqref{eqn:rx_tx_evm_per_subcarrier__ce_unaware}, it is straightforward to constrain interlayer interference and the received distortion energy---cf.~\eqref{eqn:rx_mu_su_signal_freq_domain_kth_subcarrer__with_unprecoded_pertubation}---that is, replace the left hand of~\eqref{eqn:rx_tx_evm_per_subcarrier__ce_unaware} with $\left\|\left({\mat{H}}\!\left[k\right]\overline{\mat{X}}\left[:,k\right]  \! - \! \mat{S}\left[:,k\right]\right) \right\|_2$. On the contrary, if the base station has no downlink channel knowledge, \eg, catering to non-CSI-aware mode for the broadcast/control channel, then constraint~\eqref{eqn:rx_tx_evm_per_subcarrier__ce_unaware} can be omitted. In the sequel, we will reformulate this problem, which can easily adapt to CSI-aware and non-\ac{CSI}-aware modes via suitable parameter settings.} 

\skblue{
Clearly, this problem~\texttt{P0} is nonconvex, in particular, due to {the} \ac{PAPR} constraint \eqref{eqn:problem_p0__papr_constraint} and {the} \ac{ACLR} constraint \eqref{eqn:aclr__ce_unaware}, whereas objective~\eqref{eqn:problem_p0__tx_evm} and constraint~\eqref{eqn:rx_tx_evm_per_subcarrier__ce_unaware} are convex. One could convexify the \ac{PAPR} constraint by resorting to peak clipping, \ie, replacing the denominator in~\eqref{eqn:problem_p0__papr_constraint}, \skblue{\ie, $\left\| \mat{F}^\herm \overline{\mat{X}}^\trans \right\|_2^2$,} with a fixed energy of the \ac{OFDM} symbol, \ie, $\expect\left\{ \left\| \mat{F}^\herm {\mat{X}}^\trans \right\|_2^2 \right\}\!$, see, \eg,~\cite{Studer_papr_2013}. However, such an approach would render {a} suboptimal solution. Therefore, one could employ a semidefinite relaxation to the \ac{PAPR} constraint, see, \eg,~\cite{Yao__papr_sdp__2019} to improve the accuracy of the solution at the cost of prohibitive computational complexity $\BigOh\left( \left(\ell K \NT \right)^{4.5}\right)$. Similarly, one could convexify the \ac{ACLR} constraint~\eqref{eqn:aclr__ce_unaware} by replacing the denominator $\left\|\overline{\mat{X}}\left[:,\mathcal{T}\right]\right\|_2^2$ with $\left\|{\mat{X}}\left[:,\mathcal{T}\right]\right\|_2^2$. Thus, even if one convexifies problem~\texttt{P0} by employing fixed values to the denominators of \ac{PAPR} and \ac{ACLR} constraints, computing an optimal solution in real-time using off-the-shelf convex optimization solvers, typically employs interior-point methods, such as CVX~\cite{Grant2014CVX:Beta}  will result in prohibitively high computational complexity $\BigOh\left( \left(\ell K \NT \right)^{4.5}\right)$. Thus, an interior-point-based algorithm is infeasible to deploy on current state-of-the-art radio base stations. Hence, we seek a computationally efficient algorithm tackling the proposed problem that can be employed in realistic radio systems. 
}

\skblue{
Toward the goal to develop efficient algorithms, unfortunately proposed problem~\texttt{P0} is not amenable to first-order optimization methods, notably operator/proximal splitting---see Section~\ref{sec:related_works_on_proximal_splitting} and Section~\ref{sec:top_admm_algorithm}. Hence, we must reformulate the problem~\texttt{P0}. For the problem reformulation, we now define two sets:
\iffalse
, 1)~$
    \mathbfcal{P} \!\coloneqq \! \left\{ \overline{\mat{T}} \bm{:} \left\|\! \overline{\mat{T}}^\trans \right\|_{\infty} \!\preceq\! \sqrt{\bm{\gamma}_{\rm par}} \left\| \! \left(\overline{\mat{T}}\right)^\trans \right\|_2 \right\}
$ corresponding to the nonconvex \ac{PAPR} constraint~\eqref{eqn:problem_p0__papr_constraint}, and 2)~$
    \mathbfcal{U} \!\coloneqq \! \left\{ \overline{\mat{X}} \bm{:} \left\| \overline{\mat{X}}\left[:,\mathcal{T}^{\perp}\right] \right\|_{2} \!\preceq\! \sqrt{\bm{\psi}_{\rm aclr}} \left\|  \overline{\mat{X}} \left[:,\mathcal{T}\right]  \right\|_2 \right\}
$ corresponding to the {nonconvex} \ac{ACLR} constraint~\eqref{eqn:problem_p3__oobe_constraint}.
\else
\begin{align} \label{eqn:nonconvex_papr_constraint_set}
    \mathbfcal{P} 
    &\coloneqq \! \left\{\! \overline{\mat{X}} \! \bm{:} \!\left\| \overline{\mat{T}} \right\|_{\!\infty} \!\preceq\! \bm{\gamma}_{\rm par}^{\frac{1}{2}} \! \left\|  \overline{\mat{T}} \right\|_2;  \overline{\mat{T}} \!=\! \mat{F}^\herm \overline{\mat{X}}^\trans \!\right\} \nonumber \\ 
    &\equiv \! \left\{ \! \overline{\mat{X}} \!\bm{:} \! \left\|\! \overline{\mat{T}} \! \vec{e}_j \!\right\|_{\infty}^2 \!\leq\! {\gamma}_{\rm par} \! \left\|\!  \overline{\mat{T}} \!\vec{e}_j \!\right\|_2^2;  \overline{\mat{T}} \!=\! \mat{F}^\herm \!\overline{\mat{X}}^\trans;  j\!=\!1,\!\ldots\!,\NT \! \right\}  
\end{align}
\begin{align}
    \label{eqn:nonconvex_aclr_constraint_set}
    \mathbfcal{U} 
    &\coloneqq \! \left\{\! \overline{\mat{X}} \! \bm{:} \!\left\| \overline{\mat{X}}\!\left[:,\mathcal{T}^{\perp}\!\right]\! \right\|_{2} \!\preceq\! \bm{\psi}_{\rm aclr}^{\frac{1}{2}} \! \left\|  \overline{\mat{X}}\! \left[:,\mathcal{T}\right] \!  \right\|_2 \right\} \nonumber \\ 
    &\equiv \! \left\{ \! \overline{\mat{X}} \bm{:} \! \left\| \!\left(\overline{\mat{X}}\!\left[:,\mathcal{T}^{\perp}\right]\right) \! \vec{e}_j^{\trans} \right\|_{2}^2 \!\leq\! {\psi}_{\rm aclr} \! \left\| \!  \left(\overline{\mat{X}}\!\left[:,\mathcal{T}\right]\right) \!\vec{e}_j^{\trans} \right\|_2^2\!;  \forall j \right\} 
\end{align}
corresponding to the nonconvex \ac{PAPR}~\eqref{eqn:problem_p0__papr_constraint} and \ac{ACLR}~\eqref{eqn:aclr__ce_unaware} constraints, respectively.
\fi
 Subsequently, we use an indicator function to these sets $\mathbfcal{P}$ and $\mathbfcal{U}$, \ie, $\Ind_{\mathbfcal{P}}\left(\overline{\mat{X}} \right)$ and $\Ind_{\mathbfcal{U}}\left( \overline{\mat{X}} \right)$, where the indicator function definition to any set $\mathcal{C}$ is given in~\eqref{eqn:definition_of_characteristic_function}.
 Furthermore, we define $h_{\rm TxEVM}\left( \overline{\mat{X}}  \right) \!\coloneqq\!  \left\| \overline{\mat{X}} \!-\! \mat{X} \!\right\|_F^2$. Consequently, using these definitions, we can reformulate an equivalent problem to \texttt{P0}, without loss of generality, as
\begin{subequations} \label{eqn:problem_p3__ce_unaware__with_sets}
\begin{alignat}{3} \label{eqn:problem_p3__tx_evm__with_sets}
(\texttt{P1}) \quad & \underset{ \overline{\mat{X}}  \in \Cm^{ \NT \times \ell K }}{\operatorname{minimize}} 
& & \quad  h_{\rm TxEVM}\left( \overline{\mat{X}}  \right) \! +\! \Ind_{\mathbfcal{P}}\left(\overline{\mat{X}} \right)  \! +\! \Ind_{\mathbfcal{U}}\left( \overline{\mat{X}} \right) \\  
\label{eqn:rx_tx_evm_per_subcarrier__ce_unaware__with_sets}
\quad & \operatorname{subject \, to}
& & \quad  \left\| {\mat{H}}\!\left[k\right]\left( \overline{\mat{X}} \! - \! \mat{X} \right) \vec{e}_k \right\|_2^2 \! \leq \!  \epsilon_{\rm evm}^2\left[k\right] \ \forall k \! \in \! \mathcal{T} .
\end{alignat}
\end{subequations} 
Clearly, problem \texttt{P1} is equally hard to solve as \texttt{P0} because of nonconvexity of indicator functions to nonconvex sets $\mathbfcal{P}$ and $\mathbfcal{U}$. Question remains whether there are any possible (approximate) techniques to solve \texttt{P1}. Unfortunately, the answer is still negative unless there is some reformulation. 
}

\skblue{
Before we progress further on tackling nonconvex problem \texttt{P0}/\texttt{P1}, let us express convexified problem \texttt{P0}/\texttt{P1} in an unconstrained form: 
\begin{align} \label{eqn:problem_p2__convex}
&(\texttt{Convexified: P2}) \nonumber \\  
&\underset{ \overline{\mat{X}}  \in \Cm^{ \NT \times \ell K }}{\operatorname{minimize}} 
\   h_{\rm TxEVM}\left( \overline{\mat{X}}  \right) \!+\! \Ind_{\mathbfcal{P}^{\left(0\right)}}\left(\overline{\mat{X}} \right)  \! +\! \Ind_{\mathbfcal{U}^{\left(0\right)}}\left( \overline{\mat{X}} \right) \! +\! \Ind_{\mathbfcal{E}}\left( \overline{\mat{X}} \right),
\end{align}
where convexified \ac{PAPR} constraint set reads 
\begin{equation*}
\mathbfcal{P}^{(0)} \!\coloneqq \! \left\{\! \overline{\mat{X}} \bm{:} \left\| \overline{\mat{T}} \right\|_{\infty} \!\preceq\! \sqrt{\bm{\gamma}_{\rm par}} \left\|  {\mat{T}} \right\|_2;  \overline{\mat{T}} \!=\! \mat{F}^\herm \overline{\mat{X}}^\trans; {\mat{T}} \!=\! \mat{F}^\herm \!{\mat{X}}^\trans \!\right\}
\end{equation*}
and convexified \ac{ACLR} constraint set is 
\begin{equation*}
\mathbfcal{U}^{(0)} \!\coloneqq \! \left\{ \overline{\mat{X}} \bm{:} \left\| \overline{\mat{X}}\left[:,\mathcal{T}^{\perp}\right] \right\|_{2} \!\preceq\! \sqrt{\bm{\psi}_{\rm aclr}} \left\|  {\mat{X}} \left[:,\mathcal{T}\right]  \right\|_2 \right\},    
\end{equation*}
\ie, the respective denominators are fixed as discussed previously. The convex set corresponding to the predicted received \ac{EVM} can be deonoted by $\mathbfcal{E} \!\coloneqq \! \left\{ \overline{\mat{X}} \bm{:} \left\| {\mat{H}}\!\left[k\right]\left( \overline{\mat{X}} \! - \! \mat{X} \right) \vec{e}_k \right\|_2^2 \! \leq \!  \epsilon_{\rm evm}^2\left[k\right] \ \forall k \! \in \! \mathcal{T} \right\}$.
}

\skblue{
Although problem~\texttt{P2} is convex, employing off-the-shelf solvers like CVX~\cite{Grant2014CVX:Beta} is not easible for a realistic radio systems due to prohibitive high computational complexity. Hence, developing a computationally efficient algorithm is challenging using any (first-order) proximal splitting algorithm for problem~\texttt{P1}/\texttt{P2}, notably and unfortunately due to the computation of proximal operator for $\Ind_{\mathbfcal{E}}\left( \overline{\mat{X}} \right)$, \ie, an orthogonal projection operator on an ellipsoid having $\operatorname{rank}\!\geq\!1$ as it involves eigenvalue decomposition---see, \eg~\cite[Theorem~4]{Kant_journal_msp_top_admm:2020} and their references. In other words, the orthogonal projection operator of such an ellipsoid for the predicted received \ac{EVM} is computationally expensive. Although the orthogonal projection operators corresponding to the nonconvex \ac{ACLR} and \ac{PAPR} constraints are still notoriously hard, we will propose later in the subsequent sections how to possibly tackle these nonconvex constraints. Before proceeding further on attacking predicted received \ac{EVM} constraint, let us formulate a robust version of predicted received \ac{EVM} against channel uncertainty if channel estimates are available.
}

\noindent\textit{\skblue{Problem \texttt{P1} Reformulation for Robust Mitigation of Distortion}}:
Dropping subcarrier index~$k$, let the channel \sk{estimation} error model be described by $\mat{H} \! \coloneqq \! \widehat{\mat{H}} \!+\! \Delta \mat{H}$, where $\mat{H}$, $\widehat{\mat{H}}$, and $\Delta \mat{H}$ denote true, estimated, and error, respectively. Additionally, let the \sk{error} at subcarrier $k$ be denoted by $\Delta {\mat{X}} \!\coloneqq\!  \left( \overline{\mat{X}}  \! \left[:, k\right] \! - \!   {\mat{X}}  \! \left[:, k\right] \right)$. Furthermore, for simplicity, we assume that $\Delta \mat{H}$ takes values from the bounded set $\left\{ \left\|\Delta \mat{H} \right\|_F^2 \leq \sigma^2_{\rm ce} \right\}$, where $\sigma^2_{\rm ce} \!>\! 0$ describes the channel uncertainty that is assumed to be known to the transmitter. Applying the triangle inequality followed by the Cauchy-Schwarz inequality to \eqref{eqn:rx_tx_evm_per_subcarrier__ce_unaware}/\eqref{eqn:rx_tx_evm_per_subcarrier__ce_unaware__with_sets}, we have 
\begin{align} \label{eqn:rx_tx_evm_per_subcarrier__ce_aware}
\Bigl\| {\mat{H}} \Delta\!{\mat{X}} \! \Bigr\|_2^2 &=\! \Bigl\| \left(\widehat{\mat{H}} \!+\! \Delta\!\mat{H} \right)  \Delta {\mat{X}}   \Bigr\|_2^2 \nonumber \\
&\leq \left\| \widehat{\mat{H}} \Delta {\mat{X}}   \right\|_2^2  \! + \! \left\| \Delta \mat{H}    \Delta {\mat{X}}   \right\|_2^2 \nonumber \\
&\leq \Delta\! {\mat{X}}^\herm \!\left( \widehat{\mat{H}}^\herm \widehat{\mat{H}} \! + \! \sigma^2_{\rm ce} \I \!\right) \Delta\!{\mat{X}}.
\end{align}

Using the inequality \eqref{eqn:rx_tx_evm_per_subcarrier__ce_aware} in the constraint \eqref{eqn:rx_tx_evm_per_subcarrier__ce_unaware}/\eqref{eqn:rx_tx_evm_per_subcarrier__ce_unaware__with_sets}, the channel estimation error-aware constraint {is guaranteed if} $ \left\| \mat{Q}^{\nicefrac{1}{2}}\left[k\right]\left( \overline{\mat{X}} \! - \! \mat{X} \right) \vec{e}_k \right\|_2 \! \leq \!  \epsilon_{\rm evm}[k]$, where matrix {$\mat{Q}$ is defined~as}
\begin{equation}\label{eqn:def_Qk_matrix}
    \sk{\Cm^{\NT \times \NT} \ni} \mat{Q}[k] \! \coloneqq \! \left( \widehat{\mat{H}}[k]^\herm \widehat{\mat{H}}[k] \! + \! \skblue{\nu} \I_{\NT} \! \right),
\end{equation}
\skblue{where $\nu$ is a user-defined parameter to regularize the Gram matrix under channel estimation error. However, the $\nu$ parameter does not have to be the same as the channel estimation error variance $\sigma^2_{\rm ce}$---see Fig.~\ref{fig:64x2__Avg_EstRxdEVM__vs__channel_est_error_var__TOPADMM}, which shows the estimated received \ac{EVM} metric against channel estimation error variance for a fixed $\nu$.}

\skblue{
To achieve our computationally efficient algorithms for our proposed problem, let us further reformulate the problem~\texttt{P1} under channel uncertainty by forming the Lagrangian such that we have
\begin{align*}
    L_{\texttt{P1}} \!\left( \overline{\mat{X}}, \left\{ \lambda_{k} \right\} \!\right)  
    \!\coloneqq&\!   h_{\rm TxEVM}\left( \overline{\mat{X}}  \right) \! +\! \Ind_{\mathbfcal{P}}\left(\overline{\mat{X}} \right)  \! +\! \Ind_{\mathbfcal{U}}\left( \overline{\mat{X}} \right) \nonumber \\ &\!+\! \sum_{k \in \mathcal{T}} \!\lambda_{k} \!\left( \! \left\| {\mat{Q}}^{\nicefrac{1}{2}}\!\left[k\right]\left( \overline{\mat{X}} \! - \! \mat{X} \!\right) \!\vec{e}_k \right\|_2^2 \!-\! \epsilon_{\rm evm}^2\left[k\right] \!\right)\!. 
\end{align*}
As we know from the Lagrange duality~\cite{Boyd2004ConvexOptimization}, the primal problem \texttt{P1} can equivalently be expressed as
\begin{equation*}
    \underset{\overline{\mat{X}}}{\operatorname{minimize}} \ \sup_{\left\{ \lambda_{k} \right\}} L_{\texttt{P1}} \left( \overline{\mat{X}}, \left\{ \lambda_{k} \right\} \right).
\end{equation*}
It is well-known that the dual function $d(\left\{ \lambda_{k} \right\})\!\coloneqq\! \inf_{\overline{\mat{X}}} L_{\texttt{P1}} \left( \overline{\mat{X}}, \left\{ \lambda_{k} \right\} \right)$ is a concave function due to pointwise infimum of affine functions of $\lambda_{k}$, even if the primal problem is nonconvex. Moreover, for the inequality constraint, the Lagrange multipliers are nonnegative $\lambda_{k} \! \in \! \Rm_{\geq 0}$. Unfortunately, computing a dual function for problem~\texttt{P1} is not trivial, which yields a lower bound on the optimal primal value. Therefore, toward the goal to develop implementation-friendly algorithms, let us assume that we have access to near/sub-optimal Lagrange multipliers $\lambda_{k}^{\star} \! \in \! \Rm_{\geq 0}$. 
Hence, we now reformulate problem~\texttt{P1} that is robust against channel uncertainty and avoids costly orthogonal projections onto ellipsoids corresponding to predicted received \ac{EVM} even though the problem is still nonconvex: 
\begin{align} \label{eqn:problem_p3_four_operators}
\underset{ \overline{\mat{X}}  \in \Cm^{ \NT \times \ell K }}{\operatorname{minimize}} \! \Bigl\{ \! L_{\texttt{P1}} & \! \left( \overline{\mat{X}}, \left\{ \lambda_{k}^\star \right\} \right) \nonumber \\ 
=&  h_{\rm TxEVM}\left( \overline{\mat{X}}  \right) \!+\! \sum_{k \in \mathcal{T}} \!\lambda_{k}^{\star} \!\left\| {\mat{Q}}^{\nicefrac{1}{2}}\!\left[k\right]\left( \overline{\mat{X}} \! - \! \mat{X} \right) \vec{e}_k \right\|_2^2 \nonumber \\ 
&+\! \Ind_{\mathbfcal{P}}\left( \overline{\mat{X}} \right)  \! +\! \Ind_{\mathbfcal{U}}\left( \overline{\mat{X}} \right) \Bigr\}.
\end{align}
We now absorb $\lambda_k\!=\!\lambda_{k}^{\star}$ in $\mat{Q}\!\left[k\right]$ and offer a trade-off between $h_{\rm TxEVM}$ and the predicted received \ac{EVM} with a user-defined parameter $\zeta \!\in \!\Rm_{\geq 0}$. Consequently, we succinctly represent the above problem as
\begin{equation} \label{eqn:problem_p3}
(\texttt{P3}) \quad \underset{ \overline{\mat{X}}  \in \Cm^{ \NT \times \ell K }}{\operatorname{minimize}} \  \Ind_{\mathbfcal{P}}\left( \overline{\mat{X}} \right)  \! +\! \Ind_{\mathbfcal{U}}\left( \overline{\mat{X}} \right) \!+\! h\left( \overline{\mat{X}}  \right),
\end{equation}
where 
\begin{equation}
    h\left( \overline{\mat{X}}  \right) \!\coloneqq\! \sum_{k \in \mathcal{T}} \left\| {\mat{Q}}^{\nicefrac{1}{2}}\!\left[k\right]\left( \overline{\mat{X}} \! - \! \mat{X} \right) \vec{e}_k \right\|_2^2 \!+\! \zeta h_{\rm TxEVM}\left( \overline{\mat{X}}  \right)
\end{equation}
is a differentiable function, the gradient of which is
\begin{equation} \label{eqn:grad_h__ver1}
    \nabla h\left( \overline{\mat{X}}  \right) 
    = \sum_{k \in \mathcal{T}} \mat{Q}\!\left[k\right] \left( \overline{\mat{X}} \!-\! \mat{X} \! \right)\! \vec{e}_k \vec{e}_k^\trans \!+\! \zeta\!\left( \overline{\mat{X}} \!-\! \mat{X} \! \right). 
\end{equation}
If $\zeta$ is a large value, then \ac{EVM} at the transmitter will be minimized more than the predicted received \ac{EVM}. Observe that when the base station has no downlink channel knowledge, \eg, for broadcast channels, then one possibility is $\mat{Q}[k] \!=\! \lambda_k\I_{\NT}$ with $\lambda_k\!\in\!\Rm_{\geq0}$ and other possibilities {can be explored in} future work. Observe that one could argue that it is not easy to find optimal dual variables $\lambda_{k}^{\star}$. Therefore, alternatively, one can envisage to minimize the maximum over all the predicted received \ac{EVM}, \ie, $\max\left\{  \left\| {\mat{Q}}^{\nicefrac{1}{2}}\!\left[k\right]\left( \overline{\mat{X}} \! - \! \mat{X} \right) \vec{e}_k \right\|_2^2 \right\}$---reminiscent of infinite norm. Unfortunately, this approach is equally implementation unfriendly as minimizing peaks (or convexified \ac{PAPR}), \ie, minimizing the infinite norm. Thus, we do not pursue this approach further in this work.
}

\skblue{
Before we attack on the nonconvexity of problem~\texttt{P3}, we convexify~\texttt{P3} similar to~\eqref{eqn:problem_p2__convex}: 
\begin{align} \label{eqn:problem_p4__convex}
&(\texttt{Convexified: P4}) \nonumber \\  
&\underset{ \overline{\mat{X}}  \in \Cm^{ \NT \times \ell K }}{\operatorname{minimize}} 
 \quad  \Ind_{\mathbfcal{P}^{\left(0\right)}}\left(\overline{\mat{X}} \right)  \! +\! \Ind_{\mathbfcal{U}^{\left(0\right)}}\left( \overline{\mat{X}} \right) \! +\! h\left( \overline{\mat{X}}  \right).
\end{align}
Reiterating, although this convex problem~\texttt{P4} can be solved by any off-the-shelf solvers such as CVX~\cite{Grant2014CVX:Beta}, clearly, a general purpose solver is infeasible in realistic radio systems. Hence, we forage for computationally efficient algorithms to solve large-scale problem~\texttt{P3} and~\texttt{P4}.
}

\subsection{\skblue{Efficient Solution of \texttt{P3} Utilizing TOP-ADMM-type Algorithms}}\label{sec:formulation_of_problem_p2}
\skblue{In this section, we develop three implementation-friendly algorithms utilizing \ac{TOP-ADMM}, \ac{DYS}, and \ac{BADMM} techniques described in Section~\ref{sec:top_admm_algorithm}, which solve problem~\texttt{P3} by invoking our proposed conjecture. Naturally, these designed algorithms inherently tackle problem~\texttt{P4} as a special case.   
}

\skblue{
In general, the considered proximal/operator splitting techniques, described in Section~\ref{sec:top_admm_algorithm}, need to compute the proximal operators of indicator functions $\Ind_{\mathbfcal{P}}$  and $\Ind_{\mathbfcal{U}}$ to nonconvex sets, \ie, orthogonal projections onto the respective nonconvex sets---see Definition~\ref{definition:prox_operator_on_indicator_function}. Unfortunately, no (efficient) projection operators ${\proj}_{\mathbfcal{P}}$ and ${\proj}_{\mathbfcal{U}}$ are known for the nonconvex \ac{PAPR} and \ac{ACLR} sets, \ie, $\mathbfcal{P}$ and $\mathbfcal{U}$, respectively. On the contrary, for the convexified problem~\texttt{P4}, the orthogonal projection operator ${\proj}_{\mathbfcal{P}^{(0)}}$ for the convexified \ac{PAPR} constraint set $\mathbfcal{P}^{(0)}$ is $l_{\infty}$-norm ball for each transmit antenna, which is provided below in Proposition~\ref{prop:proj_papr_constraint}. Furthermore, the orthogonal projection operator  ${\proj}_{\mathbfcal{U}^{(0)}}$ for the convexified \ac{ACLR} constraint set $\mathbfcal{U}^{(0)}$ is $l_{2}$-norm ball for each transmit antenna, which is given below in Proposition~\ref{prop:l2_ball_proj_operator}.  
}

\skblue{
\begin{prop}[\texttt{Proj} onto the $l_{\infty}$-norm ball] \label{prop:proj_papr_constraint}
Let 
$\mathcal{C} \!=\! \left\{\vec{x} \!\in\!  \Cm^{K \times 1} \!  \bm{:} \! \left\|\vec{x}\right\|_{\infty} \!-\! r \!\leq\!  0 \right\}$, where 
$r \!\in\! \Rm_{\geq 0}$, then the orthogonal projection operator is
$ \proj_{\mathcal{C}} \left( \vec{x} \right)  
    \!=\! 
    \left( \left(\frac{x\left[ k\right]}{\left| x\left[ k\right] \right|} \right) \min \Bigl\{ \bigl| x\left[ k\right] \bigl|, r \Bigr\} \right)_{1 \leq k \leq K}. $
\begin{proof}
Invoking the Moreau decomposition theorem~\cite[Section~6.6]{Beck2017} and using the $l_1$-norm proximal operator result~\cite[Example~6.8]{Beck2017}, the proof follows after the rearrangement.
\end{proof}
\end{prop}
}

\skblue{
\begin{prop}[\texttt{Proj} onto the $l_2$-norm ball {\cite[Lemma~6.26]{Beck2017}}] \label{prop:l2_ball_proj_operator}
    Let $E \subseteq \Cm^{p \times 1}$ and $E \neq \emptyset$ be given by 
    $E \!\coloneqq \! \mathcal{B}_{\|\cdot\|_2}\left[\vec{c},  r \right] \!=\! \left\{ \vec{x} \!\in\!  \Cm^{p \times 1}: \left\| \vec{x} \! -\! \vec{c} \right\|_2 \!\leq\! r \right\}$,
    then the orthogonal projection operator for the $l_2$-norm ball, \ie, $\mathcal{B}_{\|\cdot\|_2}\!\left[\vec{c},  r \right]$ with a given center $\vec{c}$ and the radius~$r$,~is
    $
        \proj_{ \mathcal{B}_{\|\cdot\|_2}\left[\vec{c},  r \right]}  \left( \vec{x} \right) 
        \!= \! \vec{c} \! + \!  \left( \! \frac{r}{\max\left\{  \left\| \vec{x} \!  - \!  \vec{c} \right\|_2, r \right\}} \! \right) \!  \left(  \vec{x} \!  - \!  \vec{c}  \right).
    $
\end{prop}
}

\skblue{
Toward the goal to develop efficient algorithms \skblue{utilizing \ac{TOP-ADMM}-type techniques} for nonconvex problem~\texttt{P3} (and inherently for~\texttt{P4} as a special case)\skblue{, we solve the appropriate subproblems \skblue{at \mbox{iteration~$(i\!+\!1)$}} by replacing the nonconvex constraint sets~\eqref{eqn:nonconvex_papr_constraint_set} and~\eqref{eqn:nonconvex_aclr_constraint_set} with the following convex but employing generated sequence $\overline{\mat{X}}^{\left( i\right)}$ of previous iteration $i$ for \ac{PAPR} and \ac{ACLR} constraint sets:}
\begin{align*}
    \mathbfcal{P}^{\left( i\right)} &\coloneqq \! \Biggl\{\! \overline{\mat{X}} \bm{:} \left\| \overline{\mat{T}} \right\|_{\infty} \!\preceq\! \sqrt{\bm{\gamma}_{\rm par}} \left\|  \overline{\mat{T}}^{\left( i\right)} \right\|_2; \nonumber \\  
    &\hspace{10mm} \overline{\mat{T}} \!=\! \mat{F}^\herm \overline{\mat{X}}^\trans; \, \overline{\mat{T}}^{\left( i\right)} \!=\! \mat{F}^\herm \left(\overline{\mat{X}}^{\left( i\right)}\right)^{\!\trans} \!\Biggr\}, \\
    \mathbfcal{U}^{\left( i\right)} &\coloneqq \! \left\{ \overline{\mat{X}} \bm{:} \!\left\| \overline{\mat{X}}\!\left[:,\mathcal{T}^{\perp}\!\right]\! \right\|_{2} \!\preceq\! \bm{\psi}_{\rm aclr}^{\frac{1}{2}} \! \left\|  \overline{\mat{X}}^{\left( i\right)}\! \left[:,\mathcal{T}\right] \!  \right\|_2 \right\}. 
\end{align*}
Theorem~\ref{thm:definition_of_topadmm_algorith__feasible_problem} establishes the global convergence of \ac{TOP-ADMM} for the general convex problem. \skblue{However, based on extensive numerical evidence,} we conjecture that the \skblue{asymptotically} vanishing property of the (dual/primal) residual errors of \ac{TOP-ADMM} holds for the nonconvex problem~\texttt{P3} with iteration-dependent sets---see Fig.~\ref{fig:convergence_behaviour__topadmm}--Fig.~\ref{fig:convergence_behaviour__topadmm_two_different_psi}. Consequently, under this conjecture, the proposed \ac{TOP-ADMM} algorithm satisfies the \ac{KKT} optimality conditions of nonconvex problem~\texttt{P3}---see~Appendix~\ref{appendix:optimality_condition_of_top_admm_for_p3_or_p4}. 
Additionally, this optimality condition implies that $\lim_{i \rightarrow \infty} \mathbfcal{P}^{\left( i\right)} \! = \! \mathbfcal{P}$ and $\lim_{i \rightarrow \infty} \mathbfcal{U}^{\left( i\right)} \! = \! \mathbfcal{U}$. 
Therefore, we argue that a \ac{KKT} point for nonconvex~\texttt{P3} is not worse than a \ac{KKT} point of convex~\texttt{P4} by applying \ac{TOP-ADMM}-type techniques. 
Noticeably, in Fig.~\ref{fig:cmp_ccdf__P3__P4__non_csi_aware}~and Fig.~\ref{fig:imperfect_csi_aware__cmp_p3_and_p4}, the iteration-dependent set renders a better solution compared to the fixed set for the considered test cases. However, proving the conjecture\footnote{\skblue{During the \skblue{revision} process, we came across~\cite{Ying_Sun__MM_principle_survey__2017} and its references, which evidently employs \skblue{similar techniques} within the ``majorization-minimization" framework to approximate the nonconvex constraint set for a difference of convex functions, \ie, $f(x) \!-\! g(x) \!\leq\!0$.}} is the topic of future research. 
}

\setlength{\textfloatsep}{3mm}
\begin{algorithm} [t]
\caption{{TOP-ADMM-based PAPR Reduction \skblue{for \texttt{P3}/\texttt{P4}}}} \label{algo:generalized_top_admm_algorithm_iterates} 
 \begin{algorithmic}[1] 
  \algrenewcommand\algorithmicrequire{\textbf{Initialization:}}
    \algrenewcommand\algorithmicensure{\textbf{Output(s):}}
  \Require $\overline{\mat{X}}^{\left(0\right)}, \overline{\mat{Z}}^{\left(0\right)}\!, {{\bm{\Lambda}}^{\left(0\right)}} \! \in \! \mathbb{C}^{\NT \times \ell K}$, {$\tau\!\in\! \mathbb{R}$}; $\left\{\mat{Q}\left[k\right] \! \in \! \mathbb{C}^{\NT \times \NT}\right\}$; {$\rho \!\in\! \mathbb{R}_{>0} $}
  \Ensure  $\overline{\mat{X}}^{\left(\cdot\right)} \in \mathbb{C}^{\NT \times \ell K} $ or $\overline{\mat{Z}}^{\left(\cdot\right)} \in \mathbb{C}^{\NT \times \ell K} $ 
  \For {$i = 0,1,\ldots$ } 
  \Statex
  \vspace{-3mm}
  \begin{subequations} 
   \begin{flalign}
   \label{eqn:update_Y__step2__general_top_admm__prox}
    \overline{\mat{X}}^{\left(i+1\right)} \! 
     &\coloneqq 
    \! \proj_{\mathbfcal{U}^{\left( i\right)}}\left( \! {\overline{\mat{Z}}}^{\left(i\right)} \!-\! \frac{{\bm{\Lambda}}^{\left(i\right)}}{\rho} \right)  \\
   \label{eqn:update_Z__step1__general_top_admm__prox}
   \overline{\mat{Z}}^{\left(i+1\right)} \! 
    &\coloneqq 
    \! \proj_{\mathbfcal{P}^{\left( i\right)}}\! \left(\!  \overline{\mat{X}}^{\left(i+1\right)} \! -\! \tau \nabla h \! \left( \! \overline{\mat{Z}}^{\left( i \right)} \right) \! + \!  \frac{{\bm{\Lambda}}^{\left(i\right)}}{\rho} \!\right) \nonumber \\ 
    &\equiv \! \left[\!\mat{F}\proj_{\mathbfcal{P}_T^{\left( i\right)}}\!\left( \!\mat{F}^\herm \! \left(\!  \overline{\mat{X}}^{\left(i+1\right)} \! -\! \tau \nabla h \! \left( \! \overline{\mat{Z}}^{\left( i \right)} \right) \! + \!  \frac{{\bm{\Lambda}}^{\left(i\right)}}{\rho} \!\right)^{\!\trans}\! \right)\!\right]^\trans \\
   \label{eqn:update_Lambda__step1__general_top_admm__prox}   
    {\bm{\Lambda}}^{\left(i+1\right)} 
    &\coloneqq  {\bm{\Lambda}}^{\left(i\right)} \!+\! \rho \left( \overline{\mat{X}}^{\left(i+1\right)} \!-\! \overline{\mat{Z}}^{\left(i+1\right)} \right) 
  \end{flalign} 
  \end{subequations}
  \EndFor
\end{algorithmic} 
\end{algorithm}

\skblue{
\subsubsection{TOP-ADMM-Based Algorithm for \texttt{P3}/\texttt{P4}}
For the brevity of the algorithm description, we redefine the \ac{PAPR} set ${\mathbfcal{P}_{T}}^{\left( i\right)} \!\coloneqq \! \left\{\! \overline{\mat{T}} \bm{:} \left\| \overline{\mat{T}} \right\|_{\infty} \!\preceq\! \sqrt{\bm{\gamma}_{\rm par}} \left\|  \overline{\mat{T}}^{\left( i\right)} \right\|_2 \!\right\}$ and apply \ac{TOP-ADMM} method~\eqref{eqn:generalized_top_admm_algorithm_iterates__for_convergence} to tackle \texttt{P3}/\texttt{P4} by \skblue{setting $M\!=\!1$, $f_1\!=\!\Ind_{\mathbfcal{U}^{(i)}}$, $g\!=\!\Ind_{\mathbfcal{P}^{(i)}}$, $\overline{\mat{X}} \!=\!\vec{x}_1$, $\overline{\mat{Z}} \!=\!\vec{z}$, $\bm{\Lambda} \!=\!\vec{y}_1$}. The resulting \ac{TOP-ADMM}-based algorithm for \texttt{P3}/\texttt{P4} is given in Algorithm~\ref{algo:generalized_top_admm_algorithm_iterates}.
The gradient $\nabla h$ is given by \eqref{eqn:grad_h__ver1}. Observe that both primal updates, \ie, \eqref{eqn:update_Y__step2__general_top_admm__prox} and \eqref{eqn:update_Z__step1__general_top_admm__prox}, are easy to solve and have closed-form solutions. For the $\overline{\mat{X}}^{\left(i+1\right)}$ update in \eqref{eqn:update_Y__step2__general_top_admm__prox}---corresponding to $\proj_{\mathbfcal{U}^{\left( i\right)}}$---we apply Proposition~\ref{prop:l2_ball_proj_operator} with center $\mat{C}\!=\!\mat{0}$ and radius $r\!=\!\sqrt{\bm{\psi}_{\rm aclr}} \left\|  \overline{\mat{X}}^{\left(i\right)} \left[:,\mathcal{T}\right]  \right\|_2$. Subsequently, we invoke Proposition~\ref{prop:separable_prox_operators} for the \sk{$\overline{\mat{X}}^{\left(i+1\right)}$ and $\overline{\mat{Z}}^{\left(i+1\right)}$ updates in \eqref{eqn:update_Y__step2__general_top_admm__prox} and \eqref{eqn:update_Z__step1__general_top_admm__prox}, respectively, such that the updates are performed independently for each transmit antenna. Consequently, the orthogonal projection in \eqref{eqn:update_Y__step2__general_top_admm__prox}, \ie, $\proj_{\mathbfcal{U}^{\left( i\right)}}$, corresponding to the \ac{ACLR} constraint for each transmit antenna, is given by Proposition~\ref{prop:l2_ball_proj_operator}. The orthogonal projection in \eqref{eqn:update_Z__step1__general_top_admm__prox}, \ie, $\proj_{\mathbfcal{P}^{\left( i\right)}}$, corresponding to the \ac{PAPR} constraint per transmit antenna branch is given by Proposition~\ref{prop:proj_papr_constraint}.} Moreover, \eqref{eqn:update_Z__step1__general_top_admm__prox} update exploits the proximal operator property given in Proposition~\ref{prop:prox_composition_with_affine}. Notice that to solve \texttt{P4}, the sets for both \ac{PAPR} and \ac{ACLR} are not iteration dependent, \ie, we set ${\proj}_{\mathbfcal{U}^{(i)}}\!=\!{\proj}_{\mathbfcal{U}^{(0)}}$ and ${\proj}_{\mathbfcal{P}^{(i)}}\!=\!{\proj}_{\mathbfcal{P}^{(0)}}$ for all $i\!\geq\!0$. 
}

\skblue{
\begin{prop}[Separation theorem for proximal operators\cite{Beck2017,Parikh2013}] \label{prop:separable_prox_operators}
    If a function $f\left( \mat{X} \right) \!=\! \sum_{k=1}^n f_i\left( \vec{x}_k \right)$ is separable across the variables column-wise $\mat{X} \!=\! \left[\vec{x}_1, \ldots, \vec{x}_n \right]$ {and} row-wise $\mat{X} \! =\! \left[\vec{x}_1; \ldots; \vec{x}_{\NT} \right]$, then the respective $\prox$ operators {are} as
    $\prox_{f}\left( \mat{X} \right) \!=\! \left[ \prox_{f_1} \left( \vec{x}_1 \right), \ldots, \prox_{f_n} \left( \vec{x}_n \right)  \right]$ {and} $\prox_{f}\left( \mat{X} \right) \!=\! \big[ \prox_{f_1} \left( \vec{x}_1 \right); \ldots; \allowbreak \prox_{f_{\NT}} \left( \vec{x}_{\NT} \right)  \big]$, respectively.
\end{prop}
}

\skblue{
\begin{prop}[\texttt{Prox} map for composition with an affine mapping~{\cite[Theorem~6.15]{Beck2017}}] \label{prop:prox_composition_with_affine}
Let $g: \Cm^{m \times 1} \! \mapsto \!  \bigl(-\infty, +\infty \bigr]$ be a closed proper convex, and $f\left(\vec{x}\right) \!=\! g\left( \mathcal{A} \left(\vec{x}\right) \!+\! \vec{b}\right)$,  where $\vec{b} \!\in\! \Cm^{m \times 1}$ and $\mathcal{A}: \mathcal{X} \! \mapsto \! \Cm^{m \times 1}$ satisfying $\mathcal{A} \circ \mathcal{A}^{\herm} \!=\! \nu \I$  where $\nu \!\in \Rm_{>0}$, then for any $\vec{x} \!\in\! \Cm^{n \times 1}$
$
    \prox_{f} \left( \vec{x} \right)  
    \!=\! \vec{x} \!+\! \frac{1}{\lambda}\mathcal{A}^\herm \! \left( \prox_{\lambda g} \!\left( \mathcal{A}\!\left(\vec{x}\right) \!+\! \vec{b} \right) \!-\!  \mathcal{A}\!\left(\vec{x}\right) \!-\! \vec{b} \!\right).
$
\end{prop}
}

\setlength{\textfloatsep}{3mm}
\begin{algorithm} [t]
\caption{{Bregman ADMM-based PAPR Reduction \skblue{for \texttt{P3}/\texttt{P4}}}} \label{algo:generalized_badmm_algorithm_iterates} 
 \begin{algorithmic}[1] 
  \algrenewcommand\algorithmicrequire{\textbf{Initialization:}}
    \algrenewcommand\algorithmicensure{\textbf{Output(s):}}
  \Require $\overline{\mat{X}}^{\left(0\right)}, \overline{\mat{Z}}^{\left(0\right)}\!, {{\bm{\Lambda}}^{\left(0\right)}} \! \in \! \mathbb{C}^{\NT \times \ell K}$; $\left\{\mat{Q}\left[k\right] \! \in \! \mathbb{C}^{\NT \times \NT}\right\}$; {$\rho, \rho_x, \rho_z, \tau \!\in\! \mathbb{R}_{>0} $}
  \Ensure  $\overline{\mat{X}}^{\left(\cdot\right)} \in \mathbb{C}^{\NT \times \ell K} $ or $\overline{\mat{Z}}^{\left(\cdot\right)} \in \mathbb{C}^{\NT \times \ell K} $ 
  \For {$i = 0,1,\ldots$ } 
  \Statex
  \vspace{-3mm}
  \begin{subequations} 
    \begin{flalign}    
    \label{eqn:update_Y__step2__general_badmm__prox___U}
	\mat{U} &\coloneqq  \rho_x \overline{\mat{X}}^{\left(i\right)} \!-\! \nabla h \! \left( \! \overline{\mat{X}}^{\left( i \right)} \right) \! + \! \rho \left( \overline{\mat{Z}}^{\left(i\right)} \!  \!-\! \frac{{\bm{\Lambda}}^{\left(i\right)}}{\rho} \right)  \\
	\label{eqn:update_Y__step2__general_badmm__prox}
    \overline{\mat{X}}^{\left(i+1\right)} \! 
    &\coloneqq 
    \! \proj_{\mathbfcal{U}^{\left( i\right)}}\left( \frac{\mat{U}}{\left(\rho_x\!+\!\rho_z\right)} \!  \right)  \\ 
    \label{eqn:update_Z__step1__general_badmm__prox___V}
    \mat{V} &\coloneqq \frac{1}{\left(\rho_z\!+\!\rho\right)}\!\left(\!\rho_z \overline{\mat{Z}}^{\left(i\right)} \! - \! \nabla h \! \left( \! \overline{\mat{Z}}^{\left( i \right)} \right) \! + \!  \rho\left( \overline{\mat{X}}^{\left(i+1\right)} \!+\! \frac{{\bm{\Lambda}}^{\left(i\right)}}{\rho}\right) \!\right) \\
	\label{eqn:update_Z__step1__general_badmm__prox}
   \overline{\mat{Z}}^{\left(i+1\right)} \! 
    &\coloneqq 
    \! \left[\!\mat{F}\proj_{\mathbfcal{P}_T^{\left( i\right)}}\left( \!\mat{F}^\herm \mat{V}^{\trans}  \right)\!\right]^\trans \\
   \label{eqn:update_lambda__step3__general_badmm__prox}   
    {\bm{\Lambda}}^{\left(i+1\right)} \coloneqq&  {\bm{\Lambda}}^{\left(i\right)} \!+\! \tau \left( \overline{\mat{X}}^{\left(i+1\right)} \!-\! \overline{\mat{Z}}^{\left(i+1\right)} \right) 
  \end{flalign} 
  \end{subequations}
  \EndFor
\end{algorithmic} 
\end{algorithm}

\sk{To benchmark our proposed \sk{Algorithm~\ref{algo:generalized_top_admm_algorithm_iterates} using} \ac{TOP-ADMM} against state-of-the-art algorithms, we next present two algorithms applying \ac{BADMM} and \ac{DYS} methods \skblue{to solve \texttt{P3}/\texttt{P4}. Recall, to solve \texttt{P4}, the sets for both \ac{PAPR} and \ac{ACLR} are iteration independent, \ie, ${\proj}_{\mathbfcal{U}^{(i)}}\!=\!{\proj}_{\mathbfcal{U}^{(0)}}$ and ${\proj}_{\mathbfcal{P}^{(i)}}\!=\!{\proj}_{\mathbfcal{P}^{(0)}}$ for all the iterations $i\!\geq\!0$.}}

\subsection{Bregman ADMM (BADMM) Algorithm}
We develop \sk{first reference algorithm using \ac{BADMM}~\cite{Wang_Banerjee__Bregman_ADMM__2014}, \skblue{cf. Section~\ref{sec:intro_to_BADMM},} for benchmarking purposes} to solve \skblue{\texttt{P3}/\texttt{P4}, whose} 
recipe  
is presented in Algorithm~\ref{algo:generalized_badmm_algorithm_iterates}\sk{---\sk{see~\cite{Kant_et_al__asilomar_2021} for additional algorithmic details}}. The operators $\proj_{\mathbfcal{P}^{\left(i\right)}}$, $\proj_{\mathbfcal{U}^{\left(i\right)}}$, and $\nabla h$ can be computed as described for \sk{the} \ac{TOP-ADMM} algorithm.

\subsection{Davis-Yin Splitting (DYS) Algorithm}
{Lastly, we also designed another reference algorithm for benchmarking purposes using \ac{DYS}~\cite{Davis2017}, \skblue{cf. Section~\ref{sec:intro_to_DYS}} tackling} \skblue{\texttt{P3}/\texttt{P4}}---the recipe is given in Algorithm~\ref{algo:generalized_dys_algorithm_iterates}. Similarly, the projection operators $\proj_{\mathbfcal{P}^{\left(i\right)}}$ and $\proj_{\mathbfcal{U}^{\left(i\right)}}$, and gradient $\nabla h$ are same as described for \sk{the} \ac{TOP-ADMM} algorithm.

\subsection{Computational Complexity Analysis}
We compare the run-time computational complexities of our proposed \ac{TOP-ADMM}-type algorithms with two prior arts, namely semidefinite relaxation-based~\cite{Yao__papr_sdp__2019} and Bao et al.~\cite{Bao_papr_admm__2018}. Let $K_D \!\coloneqq\! \left| \mathcal{T} \right|$ denote the number of useful data-carrying subcarriers. Consequently, let  $K_C \!\coloneqq\! \left| \mathcal{T}^\perp \right|$ such that $K\!=\!K_D \!+\! K_C$. 

\textit{Common processing of \ac{TOP-ADMM}-type algorithms}: the major computational complexity of all the three presented algorithms stem from \ac{IDFT} and \ac{DFT} operations, and gradient $\nabla h$. The cost of \ac{IDFT} or \ac{DFT} is typically $\BigOh\left(\ell K \log\left(\ell K\right)\right)$ per iteration. Additionally, the worst-case cost in computing the gradient $\nabla h$ \eqref{eqn:grad_h__ver1} is $\BigOh\left(K_D\NT^2\right)$. Observe that the computation of $\left\{\mat{Q}[k]\right\}$ matrices is required only once per subcarrier, which is $\BigOh\left(K_D\NR\NT^2\right)$. The computational cost of orthogonal projection operators is relatively low compared to \ac{IDFT}/\ac{DFT} operation. Hence, the overall complexity of the proposed \ac{TOP-ADMM}-type algorithms is $\BigOh\left(\left(\NT\ell K \log\left(\ell K\right)\right)\!+\! K_D\NT^2\right)$. 
Table~\ref{table:complexity_comparison} summarizes and compares the complexity of the proposed schemes with \skblue{three prior arts}.  
Clearly, the semidefinite relaxation-based method~\cite{Yao__papr_sdp__2019} is impractical to deploy on the current state-of-the-art chipsets due to prohibitive complexity cost compared to the proposed first-order-based approaches. Moreover, \sk{another drawback} of the prior art~\cite{Bao_papr_admm__2018}  
is \sk{the need} to resort to iterative numerical optimization, \eg, a bisection method. \sk{This can be used} to solve the proximal operator for the infinite norm---referred to as \texttt{PROXINF} in~\cite{Bao_papr_admm__2018}---\sk{{which has a complexity of}} $\BigOh\left(I^{\prime}\NT \ell K \right)$, \sk{where} $I^{\prime}$ denotes the number of iterations. Nevertheless, the complexity of the prior art, Bao et al.~\cite{Bao_papr_admm__2018}, is similar to our \ac{TOP-ADMM}-type schemes.

\setlength{\textfloatsep}{3mm}
\begin{algorithm} [t]
\caption{ DYS-based PAPR Reduction \skblue{for \texttt{P3}/\texttt{P4}}} \label{algo:generalized_dys_algorithm_iterates} 
 \begin{algorithmic}[1] 
  \algrenewcommand\algorithmicrequire{\textbf{Initialization:}}
    \algrenewcommand\algorithmicensure{\textbf{Output(s):}}
  \Require $\overline{\mat{X}}^{\left(0\right)}, \overline{\mat{Z}}^{\left(0\right)}\!, {{\bm{\Lambda}}^{\left(0\right)}} \! \in \! \mathbb{C}^{\NT \times \ell K}$; $\left\{\mat{Q}\left[k\right] \! \in \! \mathbb{C}^{\NT \times \NT}\right\}$; {$\tau, \mu  \!\in\! \mathbb{R}_{\geq0} $}
  \Ensure  $\overline{\mat{X}}^{\left(\cdot\right)} \in \mathbb{C}^{\NT \times \ell K} $ or $\overline{\mat{Z}}^{\left(\cdot\right)} \in \mathbb{C}^{\NT \times \ell K} $ 
  \For {$i = 0,1,\ldots$ } 
  \Statex
  \vspace{-3mm}
  \begin{subequations} 
   \begin{flalign}
   \overline{\mat{X}}^{\left( i+1 \right) } &\coloneqq \proj_{\mathbfcal{U}^{\left( i\right)}} \!\left( \bm{\Lambda}^{\left( i \right) } \right) \\
    \overline{\mat{Z}}^{\left( i+1 \right) } &\coloneqq  \left[\!\mat{F}\proj_{\mathbfcal{P}_T^{\left(i\right)}}\!\left( \!\mat{F}^{\herm} \!\left(\! 2\overline{\mat{X}}^{\left( i+1 \right) } \!-\! \bm{\Lambda}^{\left( i \right) } \!-\! \tau \nabla h\!\left(\! \overline{\mat{X}}^{\left( i+1 \right) } \right) \!  \right)^{\!\trans} \right) \!\right]^{\!\trans}    \\
   \label{eqn:update_dual_y__step3__general_dys}   
    \bm{\Lambda}^{\left( i+1 \right) } &\coloneqq {\bm{\Lambda}}^{\left( i \right) } + \mu \left(  \overline{\mat{Z}}^{\left( i+1 \right) } - \overline{\mat{X}}^{\left( i+1 \right) }   \right)  
  \end{flalign} 
  \end{subequations}
  \EndFor
\end{algorithmic} 
\end{algorithm}

\vspace*{-2.5mm}
\begin{table*} 
\begin{center}
\caption{Comparison of computational complexities of various PAPR reduction methods }\label{table:complexity_comparison}
\vspace*{-2mm}
{
\scalebox{1}{
\resizebox{\textwidth}{!}{%
	\begin{tabular}{|l||c|c|} \hline
	\textbf{Method}   & \textbf{Complexity per (outer) iteration} \\ \hline
	\skblue{Prior art (non-CSI-aware): Iterative clip \& filter~\cite{Armstrong__icf__2002}}  & \multicolumn{1}{c|}{$\BigOh\left(\left(\NT\ell K \log\left(\ell K\right)\right)\right)$}   \\ \hline
    \skblue{Prior art (CSI-aware)}: Semidefinite relaxation-based~\cite{Yao__papr_sdp__2019}   & \multicolumn{1}{c|}{$\BigOh\left( \left({\ell K \NT}\right)^{4.5}\right)$ }  \\ \hline
    \skblue{Prior art (CSI-aware)}: Bao et al.~\cite{Bao_papr_admm__2018}   & \multicolumn{1}{c|}{ $\BigOh\left(\left(\NT\ell K \left(I^\prime \!+\! \log\left(\ell K\right)\right)\right)\!+\!I_{\rm max} K_D\NT^2\right)$; $I^\prime$ and $I_{\rm max}$ iterations for \texttt{PROXINF} and inner loop, resp.}  \\ \hline
    \ac{TOP-ADMM}-type: \ac{TOP-ADMM}/\ac{BADMM}/\ac{DYS}                            & {$\BigOh\left(\left(\NT\ell K \log\left(\ell K\right)\right)\!+\! K_D\NT^2\right)$}    \\ \hline
	\end{tabular}
}
}
}
\end{center}
\end{table*}

\section{Simulation Results} \label{sec:simulation_results}

We present the performance of the proposed algorithms and benchmark their performance with the prior art~\cite{Bao_papr_admm__2018} utilizing \sk{a} 5G NR-like simulator \sk{for \ac{PDSCH}}. We analyze the \ac{PAPR} reduction performance in terms of three \sk{figures of merit} briefly described in Section~\ref{sec:performance_metrics}. 

\subsection{Simulation Assumptions and Parameters}
The general simulation assumptions and parameters are given in Table~\ref{table:sim_assumptions}, and 3GPP NR wideband (average) \ac{EVM} requirements (after equalization\footnote{\label{footnote:3gpp_evm_limit_comment}\sk{The} EVM measurement procedure as described in~\cite[Annex~B for FR1]{3GPPTS38.1042018NRReception} does not include any wireless channel, but (realistic) impairments stemming from the transmitter such as \sk{clipping} noise and inband distortion due to nonlinear power amplifier. Therefore, for BS type~1-C and FR1, \sk{3GPP \ac{EVM} requirements consider linear equalization of the distortion at the receiver, where the distortion is solely emanating from the non-ideal transmitter. In these measurements for 3GPP requirements, the transmitter and receiver are connected with a wire-like connector. Thus, the 3GPP EVM requirements do not consider any wireless channel equalization.}}) for \sk{each} respective modulation alphabet are shown in Table~\ref{table:nr_wbevm_requirements}. Furthermore, we assume \sk{a} base station supporting sub-6 GHz, \eg, frequency ranges between 410 MHz and 7.125 GHz, referred to as FR1 in 5G NR~\cite[Section~5.1]{3GPPTS38.1042018NRReception}. Note that we do not \sk{model} any additional radio hardware impairments at the transmitter and/or receiver besides the distortion generated at the transmitter by the considered \ac{PAPR} reduction methods, if enabled.

In Fig. \ref{fig:fig1__channel_frequency_responses}, we illustrate channel frequency responses, corresponding to the first receive and transmit antenna branch $|H_{1,1}|$ of one of the realizations of the \ac{TDL} 3GPP NR channel \skblue{models~\cite{3GPPTR38.901_NR_channel_model}}: TDL-D (30 ns, 10 Hz) in Fig.~\ref{fig:fig1__channel_frequency_responses}\subref{fig:FINAL__TDLD30ns10Hz__H11_mag} and \ac{TDL}-A  (300 ns, 100 Hz) in Fig.~\ref{fig:fig1__channel_frequency_responses}\subref{fig:FINAL__TDLA300ns10Hz__H11_mag}, where the considered parameters in terms of root mean square delay spread in nano seconds (ns) and maximum Doppler shift in Hz parameters are shown in the brackets. Notice that the taps of \sk{the} \ac{TDL}-D channel model follow both Ricean and Rayleigh fading, \ie, \ac{LOS} and non-\ac{LOS}, whereas the taps of \sk{the}  \ac{TDL}-A follow only Rayleigh fading, \ie, non-LOS---\skblue{cf.~\cite[Section~B.2.3.2]{3GPPTR38.901_NR_channel_model}}. Moreover, we have also extended \sk{the} 3GPP \ac{MIMO} channel spatial correlation model for \skblue{$\NT \! > \! 8$} transmit (with cross-polarized antennas) and \skblue{two/four} receive antennas---cf.~\cite[Section~B.2.3.2]{3GPPTS38.101-4_NRUEdemod}. The considered parameters of the \ac{MIMO} spatial correlation model are given in Table~\ref{table:sim_assumptions}.

\begin{table*} 
\begin{center}
\begin{minipage}[t]{0.774\linewidth}\centering\vspace{-3mm}
\caption{Simulation Parameters for 5G NR-like PDSCH}\label{table:sim_assumptions}
\vspace{-2.5mm}
\scalebox{1}{
\resizebox{\textwidth}{!}{%
	\begin{tabular}{|l||c|c|} \hline
	\textbf{Parameters}                                       & \textbf{Test $1$}                            & \textbf{Test $2$}        \\ \hline
	\multicolumn{1}{|l||}{Subcarrier spacing}                       & \multicolumn{2}{c|}{15 kHz}                                 \\ \hline
	\multicolumn{1}{|l||}{Carrier bandwidth (PRB alloc{ation})}           & \multicolumn{2}{c|}{$20$ MHz ($100$ PRBs, \ie, {active subcarriers} $\Nsc \!=\! 1200$)}                     \\ \hline
	\multicolumn{1}{|l||}{DL \skblue{SU/MU-MIMO} $\NT, \NR$}            
	& \multicolumn{2}{c|}{ \skblue{$2/4/8/16/32/64/128/256{\rm Tx}, 2/4 {\rm Rx}$; In MU-MIMO, each user with 1 Rx} }                \\ \hline 
	\multicolumn{1}{|l||}{ Spatial layers ($\rank$)  }        & \multicolumn{2}{c|}{Fixed $\rank$ 2 \skblue{when $\NR\!=\!2$ or 4 when $\NR\!=\!4$ }}             \\ \hline
	\multicolumn{1}{|l||}{Spatial precoding}                &  \multicolumn{2}{c|}{RZF precoding with regularization $\alpha\!=\!0.001$}                                                       \\ \hline
	\multicolumn{1}{|l||}{Precoding resource block group (PRG)  size}                & \multicolumn{2}{c|}{2 PRBs granularity}                                                       \\ \hline
	\multicolumn{1}{|l||}{Modulation}                         & \multicolumn{2}{c|}{256QAM}                                              \\ \hline 
	\multicolumn{1}{|l||}{Channel model \cite{3GPPTS38.101-4_NRUEdemod}}              & \multicolumn{1}{c|}{TDL-D ($30$ ns, $10$ Hz, K-factor=$9$) }   & \multicolumn{1}{c|}{TDL-A ($300$ ns, $10$ Hz) }  \\ \hline
	\multicolumn{1}{|l||}{MIMO spatial correlation \cite{3GPPTS38.101-4_NRUEdemod}}         & \multicolumn{1}{c|}{High-like: $\alpha_1, \alpha_2,\beta\!=\!0.9$, $\gamma\!=\!0.3$} & \multicolumn{1}{c|}{Low-like: $\alpha_1, \alpha_2\!=\!0.09$, $\beta\!=\!0.03$, $\gamma\!=\!0.0$ }  \\ \hline
	\multicolumn{1}{|l||}{DL channel knowledge at the BS}     & \multicolumn{2}{c|}{\sk{Noisy channel estimate} with \ac{SNR}=5~dB} \\ \hline
	\multicolumn{1}{|l||}{DL channel knowledge at the UE}     & \multicolumn{2}{c|}{Perfect channel estimate} \\ \hline
    \multicolumn{1}{|l||}{Other information}          & \multicolumn{2}{c|}{\sk{\ac{CSI} delay is 1 ms, no additional impairments}; \skblue{fixed $\nu\!=\!0.001$ for $\left\{ \mat{Q}[k]\right\}$---cf.~\eqref{eqn:def_Qk_matrix} } }   \\ \hline	
	\end{tabular}
	}
}
\end{minipage}\hfill%
\begin{minipage}[t]{0.225\linewidth}\centering\vspace{-2.9mm}
\caption{\ac{EVM} Req.~\cite{3GPPTS38.1042018NRReception}}\label{table:nr_wbevm_requirements}\vspace{-2.5mm}
\scalebox{1}{
\resizebox{\textwidth}{!}{%
	\begin{tabular}{|l||c|} \hline
	\textbf{Modulation Scheme}     & \textbf{EVM Threshold}         \\ \hline
	\multicolumn{1}{|l||}{QPSK}    & 17.5\%                        \\ \hline
	\multicolumn{1}{|l||}{16QAM}   & 12.5\%                        \\ \hline
    \multicolumn{1}{|l||}{64QAM}   & 8.0\%                         \\ \hline
    \multicolumn{1}{|l||}{256QAM}  & 3.5\%                         \\ \hline	
	\end{tabular}
	}
} 
\end{minipage}
\end{center}
\vspace{-3mm}
\end{table*}

\iftrue
\begin{figure*}[tp!]
  \centering
  \begin{subfigure}[t]{0.495\textwidth}
    \centering
    \includegraphics[width=\textwidth,trim=32.5mm 92mm 31mm 92mm,clip]{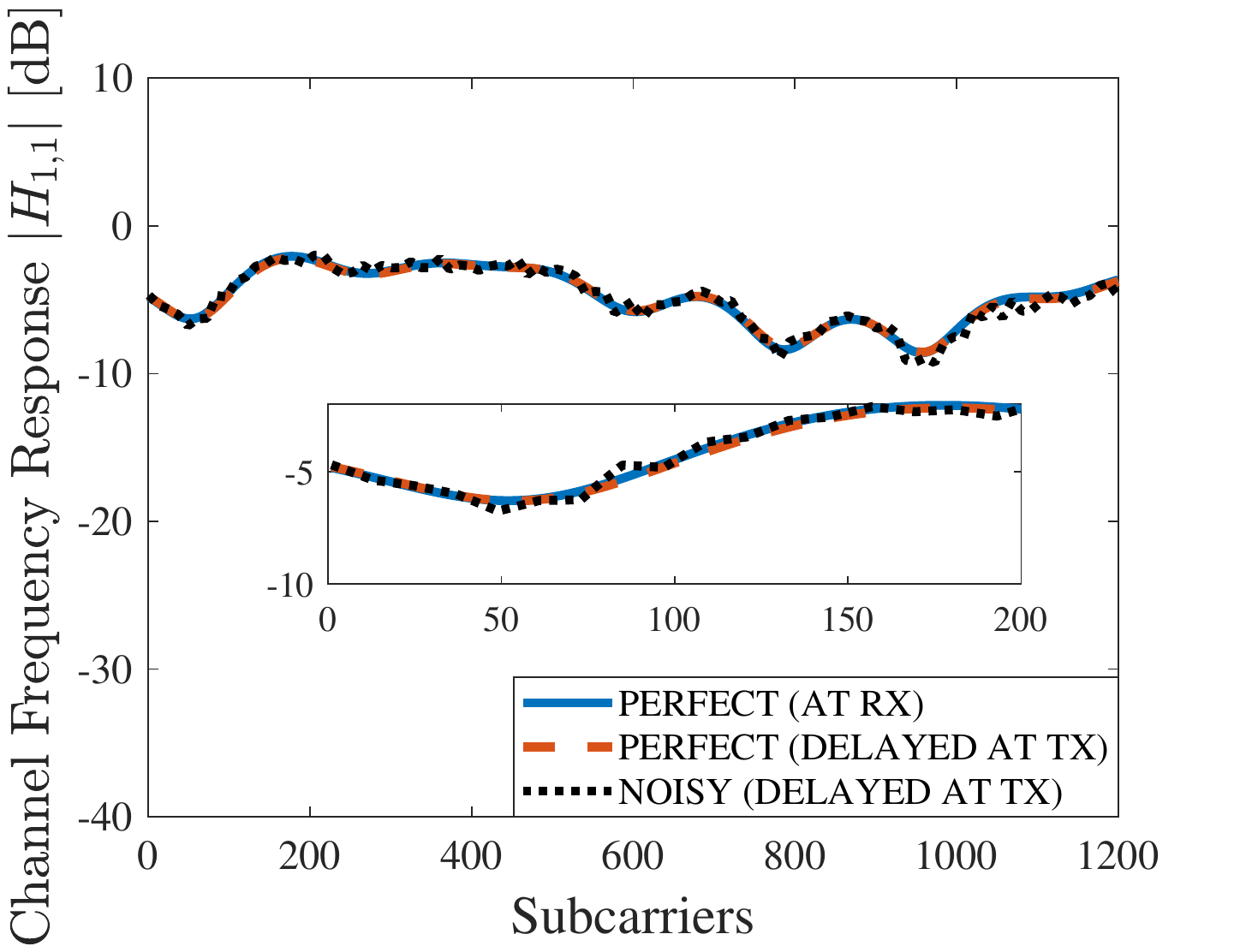}
    \caption{{TDL-D (30 ns, 10 Hz)}}\label{fig:FINAL__TDLD30ns10Hz__H11_mag}
  \end{subfigure}
  \begin{subfigure}[t]{0.495\textwidth}
    \centering
    \includegraphics[width=\textwidth,trim=32.5mm 92mm 31mm 92mm,clip]{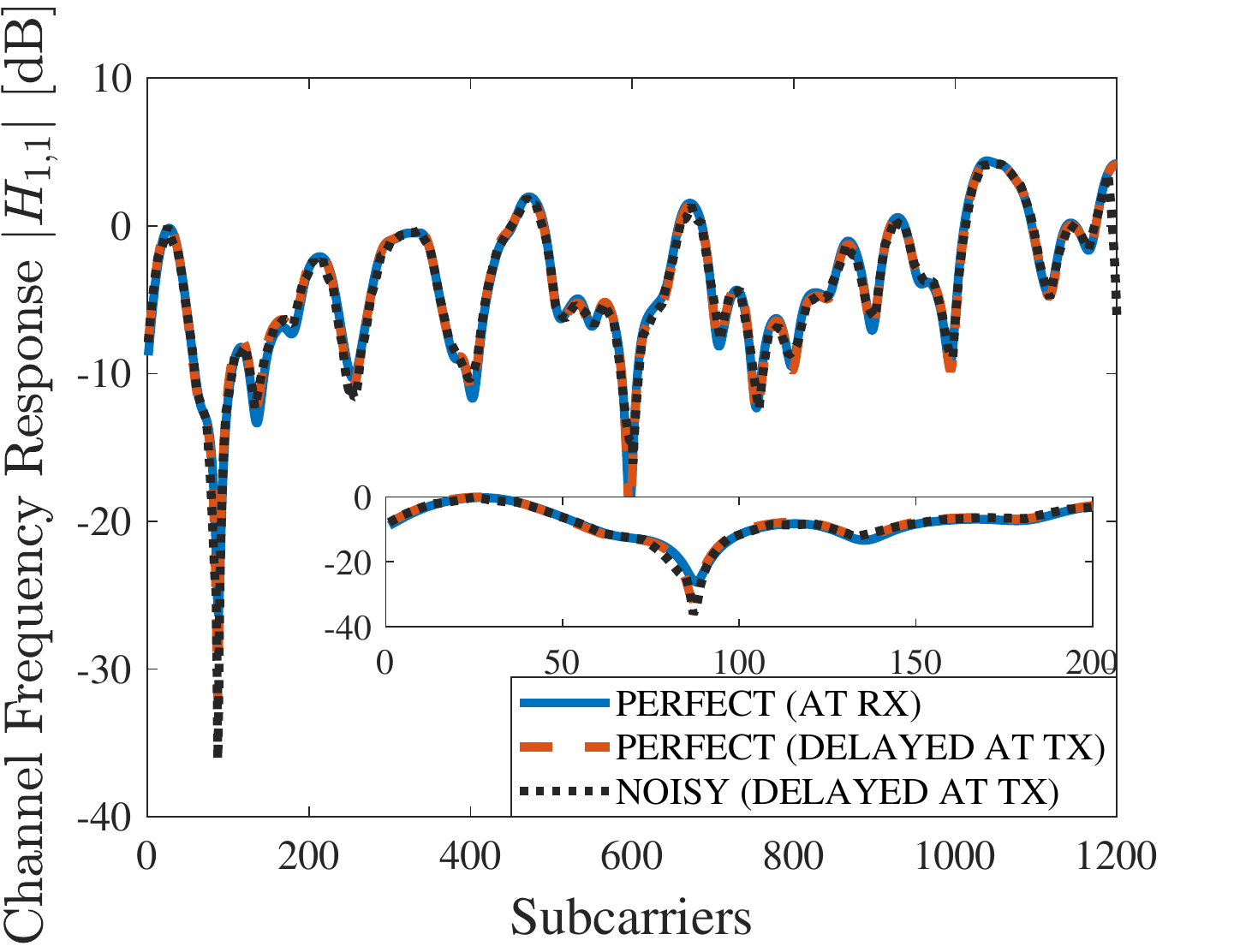}
    \caption{TDL-A (300 ns, 10 Hz)}\label{fig:FINAL__TDLA300ns10Hz__H11_mag}
  \end{subfigure}
  \caption{Magnitude of channel frequency responses $\left| H_{1,1} \right|$ of (a) TDL-D (LOS and non-LOS) and (b) TDL-A (non-LOS) channel models.}\label{fig:fig1__channel_frequency_responses}
  \vspace{-4.5mm}
\end{figure*}

\begin{figure}[!ht]
    \begin{minipage}[t]{0.99\linewidth}
    \centering
    \includegraphics[width=\textwidth,trim=32.5mm 92mm 30mm 92mm,clip]{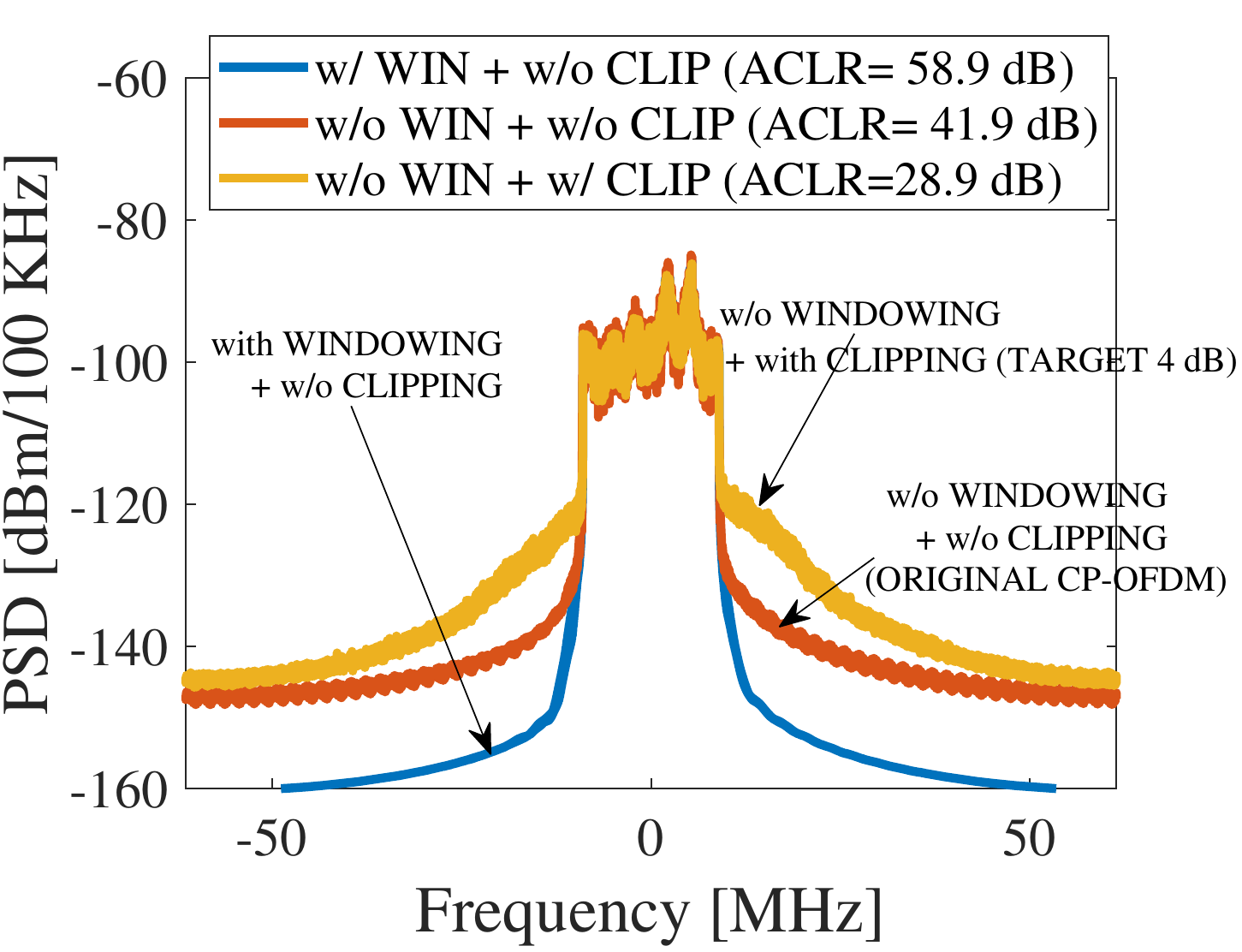}
    \caption{{{PSD} of \ac{OFDM} waveform with and without (raised cosine) transmit windowing. }}\label{fig:psd_of_64x2_tdld}
    \end{minipage}
    \vspace{-1.5mm}
\end{figure}

\else
\begin{figure*}[tp!]
  \begin{minipage}[t]{.63\linewidth}
  \begin{subfigure}[t]{0.49\textwidth}
    \centering
    \includegraphics[width=\textwidth,trim=32.5mm 92mm 31mm 92mm,clip]{figures/FINAL__TDLD30ns10Hz__H11_mag.pdf}
    \caption{TDL-D (30 ns, 10 Hz)}\label{fig:FINAL__TDLD30ns10Hz__H11_mag}
  \end{subfigure}
  \begin{subfigure}[t]{0.49\textwidth}
    \centering
    \includegraphics[width=\textwidth,trim=32.5mm 92mm 31mm 92mm,clip]{figures/FINAL__TDLA300ns10Hz__H11_mag.pdf}
    \caption{TDL-A (300 ns, 10 Hz)}\label{fig:FINAL__TDLA300ns10Hz__H11_mag}
  \end{subfigure}
  \end{minipage}\hfill
  \begin{minipage}[t]{.315\linewidth}
    \centering
    \includegraphics[width=\textwidth,trim=32.5mm 92mm 30mm 92mm,clip]{figures/FINAL_psd_64x2_no_cfr_with_and_without_windowing_avg.pdf}
  \end{minipage}
  \medskip
  
  \begin{minipage}[t]{.63\linewidth}
    \caption{Magnitude of channel frequency responses $\left| H_{1,1} \right|$ of (a) TDL-D (LOS and non-LOS) and (b) TDL-A (non-LOS) channel models.}\label{fig:fig1__channel_frequency_responses}
  \end{minipage}\hfill
  \begin{minipage}[t]{.315\linewidth}
    \caption{{{PSD} of \ac{OFDM} waveform with and without (raised cosine) transmit windowing. }}\label{fig:psd_of_64x2_tdld}
  \end{minipage}
  \vspace{-4.5mm}
\end{figure*}
\fi

We depict three channel frequency responses in Fig.~\ref{fig:fig1__channel_frequency_responses}: one at the \ac{UE}/receiver and two at the \ac{BS}/transmitter. At the \ac{UE}/receiver, perfect channel estimates are utilized for the estimated received \ac{EVM}. However, at the \ac{BS}/transmitter, we have employed \sk{a} channel estimation error model $\mat{H} \!= \! \widehat{\mat{H}} \!+\! \Delta \mat{H}$, where $\Delta \mat{H}$ has entries that are independent and identically distributed and follow zero mean circularly symmetric complex Gaussian distribution with given variance $\widetilde{\sigma}^2_{\rm ce}$. The noisy channel estimates $ \widehat{\mat{H}}$ with \ac{SNR}$\!= -10\log_{10}\left(\widetilde{\sigma}^2_{\rm ce}\right)\!=\!5$~dB are used both for \ac{MIMO} signal precoding and computation of $\mat{Q}$ matrices as defined in \eqref{eqn:def_Qk_matrix}. Both \ac{MIMO} signal precoding $\mat{W}$ and $\mat{Q}$ matrices have two physical resource blocks (PRBs)\acused{PRBs} for grouping---\sk{see} Table~\ref{table:sim_assumptions}---\sk{which} implies that the mean channel estimate within considered \ac{PRG} is used for the computation of $\mat{W}$ and $\mat{Q}$. 

In Fig.~\ref{fig:psd_of_64x2_tdld}, we illustrate three plots of \ac{PSD} of \sk{the} \ac{OFDM} waveform \sk{having} a useful signal bandwidth \sk{of} 20 MHz, see Table~\ref{table:sim_assumptions}, 1) without \sk{{either}} transmit windowing \sk{or} any \ac{PAPR} reduction method, 2) \sk{with {transmit windowing but without any \ac{PAPR} reduction method}}, and 3) with a na\"{i}ve amplitude clipping (\ac{PAPR} target of 4~dB) but without any transmit windowing. The transmit windowing is typically employed to shape the spectrum to meet the minimum 3GPP NR \ac{ACLR} requirement of 45~dB~\cite{3GPPTS38.1042018NRReception}, see, \eg,~\cite{Bala_windowing:2013, Kant_journal_emsp:2019, Kant_journal_msp_top_admm:2020}. In this study, we perform transmit windowing---using raised cosine windowing, see, \eg,~\cite{Bala_windowing:2013}---after the (proposed) \ac{PAPR} reduction if enabled. Although windowing penalizes \sk{\midtilde10\% of the cyclic prefix}, the spectrum-shaped signal has 58.9~dB of \sk{\ac{ACLR}}, which clearly meets \sk{the} minimum 45~dB \ac{ACLR} requirement (with a good implementation margin). However, as we show \sk{in the} \ac{PSD} result \sk{subsequently}, depending on the \ac{ACLR} constraint $\bm{\psi}_{\rm aclr}$, the effective \ac{ACLR} after the (proposed) \ac{PAPR} reduction method(s) can be lower than \midtilde59~dB but must \sk{be more than or equal to} 45~dB (with some margin) to comply with the requirement. Clearly, distortion-based \ac{PAPR} reduction methods, such as clipping, require filtering or windowing to remove unwanted distortion on the unused subcarriers. 

\vspace{-2.5mm}
\subsection{Simulation Results} \label{subsec:simulation_results}
In this section, we present the simulation results for the test cases given in Table~\ref{table:sim_assumptions}. \skblue{We consider \ac{TOP-ADMM}-type performance with iteration-dependent sets targeting problem~\texttt{P3} unless specified explicitly. Additionally, we set $\zeta\!=\!0$ typically for the computation of the gradient of $h$ in~\eqref{eqn:grad_h__ver1} unless stated otherwise. Moreover, the typical \ac{MIMO} test setup is 64 transmit antennas, and two receive antennas unless mentioned otherwise. Since the estimated \ac{EVM} at the receiver is computed for each spatial layer---see Section~\ref{sec:performance_metrics__inband_distortions}---we do not differentiate between \ac{SU-MIMO} and \ac{MU-MIMO} performance evaluation.}
\begin{figure*}[tp!]
  \centering
  \begin{minipage}[t]{.99\linewidth}
  \begin{subfigure}[t]{0.49\textwidth}
    \centering
    \includegraphics[width=\textwidth,trim=32.5mm 92mm 31mm 92mm,clip]{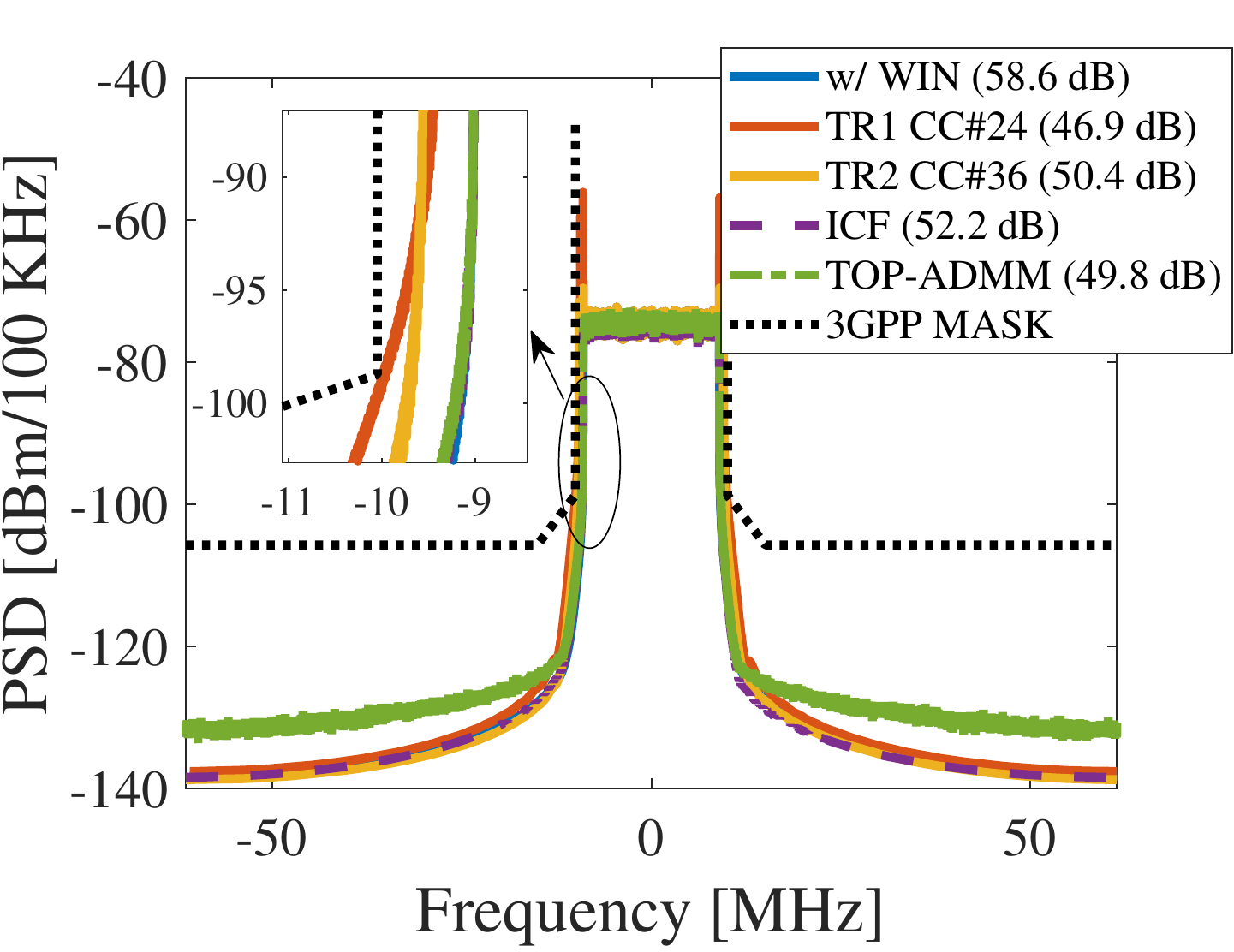}
    \caption{\ac{PSD} with 3GPP spectral emission mask}\label{fig:1x1__TDL-A300ns10Hzlow__tr_unconstrained__constrained__icf__topadmm__psd}
  \end{subfigure}
  \begin{subfigure}[t]{0.49\textwidth}
    \centering
    \includegraphics[width=\textwidth,trim=32.5mm 92mm 31mm 92mm,clip]{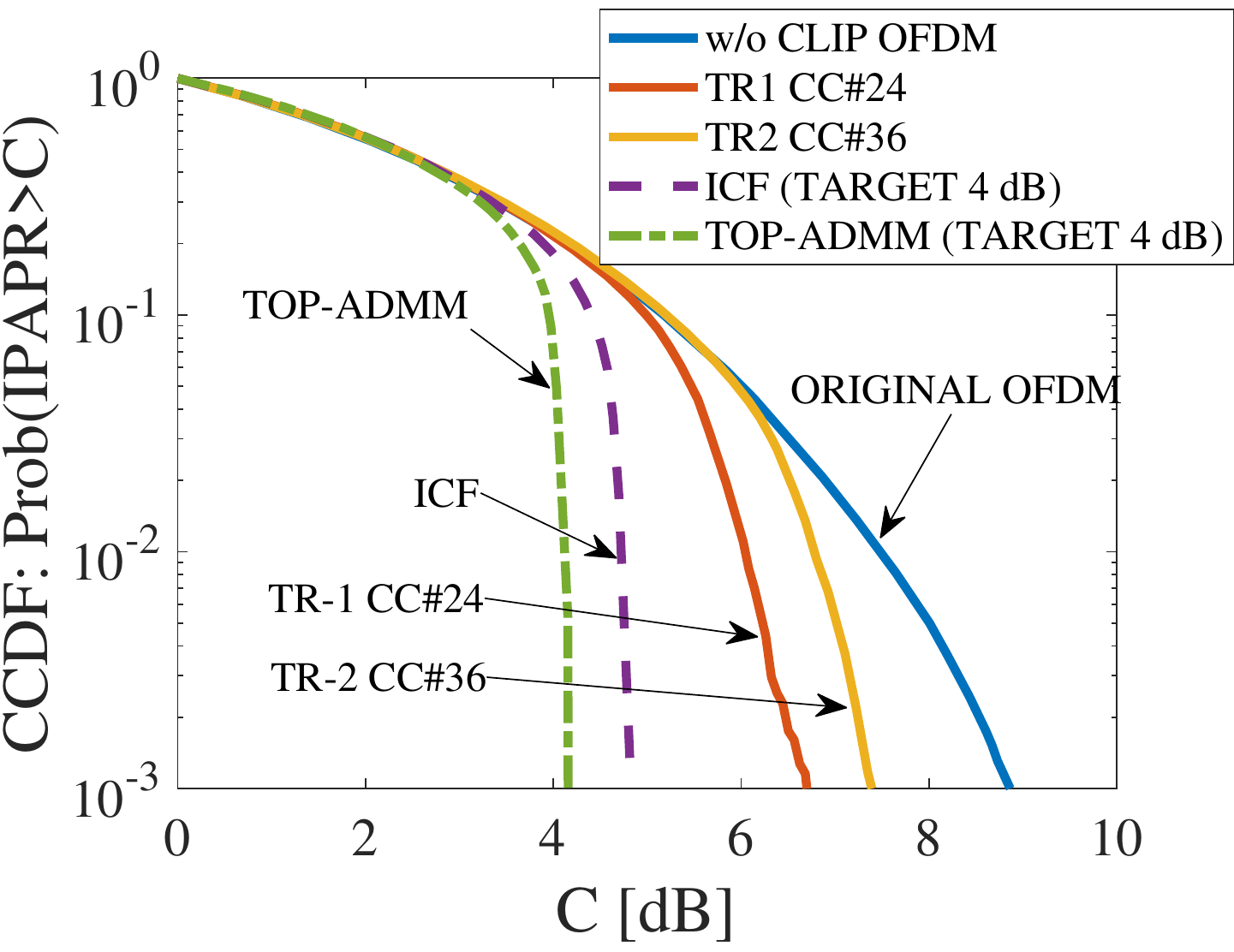}
    \caption{\ac{IPAPR} distribution}\label{fig:1x1__TDL-A300ns10Hzlow__tr_unconstrained__constrained__icf__topadmm__ccdf_iPAPR}
  \end{subfigure}
  \caption{\skblue{Comparison among 1) distortion-based: ICF and (non-CSI-aware) TOP-ADMM, and 2) distortion-less: TR1 and TR2. Noticeably, TR1/TR2 has poor PAPR reduction performance compared to ICF/TOP-ADMM.}}\label{fig:cmp__tr_unconstrained__constrained__icf__topadmm}
  \end{minipage}
  \vspace{-3.5mm}
\end{figure*}

We have employed linear \ac{RZF}-based {\ac{MIMO} (spatial)} precoding in the downlink, 
which can be expressed as $\mat{X}[:,k] \!=\!\mathcal{W}\left(\mat{S}[:,k]\right)\! = \! \mat{W}[k] \mat{S}[:,k]$, where the \ac{MIMO} precoder reads $\mat{W}[k] \!= \!\Xi \odot \widetilde{\mat{W}}[k]$, \skblue{where $\Xi$ normalizes power discussed later in the following text}. The un-normalized \ac{RZF} as given by, see, \eg, \cite{Fatema_dl_precoders_survey:2018}, is
\iffalse
\vspace{-2mm}
\begin{align*}
    \Cm^{\NT \!\times\! \NL} \!\ni\! \widetilde{\mat{W}}[k]  
    &= \!\widehat{\mat{H}}[k]^\herm \! \left(\widehat{\mat{H}}\![k] \widehat{\mat{H}}\![k]^\herm \!+\! \widetilde{\mat{R}} \!+\! \alpha \I_{\NR} \! \right)^{-1} \\ \nonumber
    &\coloneqq \left[ \widetilde{\mat{W}}_1[k],\ldots, \widetilde{\mat{W}}_{\NL}[k]\right].
    \vspace{-1.5mm}
\end{align*}
\else 
$\Cm^{\NT \!\times\! \NL} \!\ni\! \widetilde{\mat{W}}[k]  
    \!= \!\widehat{\mat{H}}[k]^\herm \! \left(\widehat{\mat{H}}\![k] \widehat{\mat{H}}\![k]^\herm \!+\! \widetilde{\mat{R}} \!+\! \alpha \!\I_{\NR} \! \right)^{\!-1} \! \coloneqq \! \left[ \widetilde{\vec{w}}_1\![k],\!\ldots,\! \widetilde{\vec{w}}_{\NL}\![k]\right]$.
\fi
The choice of suitable regularizing parameters, namely Hermitian positive semidefinite $\widetilde{\mat{R}}$ and $\alpha \! \in \! \Rm_{\geq 0}$, are typically tunable to maximize the desired performance metrics of the system, such as the sum-rate, see, \eg, \cite{Fatema_dl_precoders_survey:2018}. Moreover, {the} power normalization matrix can be 
\iffalse
\vspace{-3.5mm}
\begin{align*}
    \Cm^{\NT \times \NL} \ni \Xi = \!\left[ \frac{\sqrt{\eta}_1}{\left\| \widetilde{\vec{w}}_1[k] \right\|_2} \vec{1}_{\NT}, \ldots, \frac{\sqrt{\eta}_{\NL}}{\left\| \widetilde{\vec{w}}_{\NL}[k] \right\|_2} \vec{1}_{\NT} \right],
    \vspace{-1.5mm}
\end{align*}
\else 
$\Cm^{\NT \times \NL} \ni \Xi = \!\left[ \frac{\sqrt{\eta}_1}{\left\| \widetilde{\vec{w}}_1[k] \right\|_2} \vec{1}_{\NT}, \ldots, \frac{\sqrt{\eta}_{\NL}}{\left\| \widetilde{\vec{w}}_{\NL}[k] \right\|_2} \vec{1}_{\NT} \right]$,
\fi
where $\left\{\sqrt{\eta}_{s}\right\}$ is a set of powers that can be assigned to respective layers or users. Consequently, the total power allocated to {the} $k$-th subcarrier must satisfy $\left\|  \mat{W}[k] \right\|_F^2 \! = \! \sum_{s=1}^{\NL} \eta_s$. \skblue{In our simulations, we have simply set $\Xi$ as all-ones matrix.}

\skblue{Most of all, to motivate the signal distortion-based \ac{PAPR} reduction schemes for 5G NR and beyond, in Fig.~\ref{fig:cmp__tr_unconstrained__constrained__icf__topadmm}, we compare the distortion-based \ac{PAPR} reduction techniques, notably, classical iterative clipping and frequency-domain filtering (\acused{ICF}\ac{ICF})~\cite{Armstrong__icf__2002} and our proposed \ac{TOP-ADMM} with two variants of signal distortion-less tone reservation (TR) schemes under a special case with a single transmit and receive antenna. In \ac{ICF}, let $\Cm^{\ell K \times \NT}\!\ni\!{\mat{T}} \!=\! \mat{F}^\herm {\mat{X}}^\trans$ be a time-domain input signal;  $\overline{\mat{T}}^{\left(I\right)}$ denotes a time-domain output from \ac{ICF} after $I$ iterations. 
Then, within one iteration cycle of \ac{ICF}, for each $j$-th transmit antenna and an $i$-th iteration, the amplitude of a discrete time-domain (oversampled) signal is clipped $\overline{\mat{T}}^{\left(i\right)}[n,j]\!=\!A \exp\left\{\iota \phi\left( n\right) \right\}$ if $\left|\overline{\mat{T}}^{\left(i-1\right)}[n,j]\right\| \!>\! A$, where $A\!=\!\sqrt{\gamma_{\rm par} P_{\rm avg}}$, $P_{\rm avg}$ is average transmit power of the signal, and $\phi\left( n\right)$ is the phase of $\overline{\mat{T}}^{\left(i-1\right)}[n,j]$. In the frequency-domain of a clipped signal, the unwanted distortions on the unused/inactive subcarriers are zeroed out. After filtering, the frequency-domain signal is transformed back to the time-domain signal. We repeat this time-domain and frequency-domain filtering for $I$ iterations. For tone reservation schemes, there are at least two popular variants: 1) unconstrained method, referred to as TR1, \ie, $\operatorname{minimize}_{\overline{\mat{C}}} \, \left\| \mat{F}^{\herm} \left( \mat{X} \!+\! \overline{\mat{C}} \right) \right\|_{\infty}$, where $\overline{\mat{C}}$ corresponds to tones/subcarriers in the guard bands, and 2) peak-constrained method, called TR2, \ie, $\left\|\overline{\mat{C}}\right\|_F \!\leq\! {\rm threshold}$. In the unconstrained TR1 scheme, the energy of the subcarriers carrying \ac{PAPR} reduction capable information does not have any constraint/restriction on the energy. Therefore, the unconstrained TR1 method causes peaks, see Fig.~\ref{fig:cmp__tr_unconstrained__constrained__icf__topadmm}\subref{fig:1x1__TDL-A300ns10Hzlow__tr_unconstrained__constrained__icf__topadmm__psd} for 24 tones (CCs) on each edge of guard bands, on the power spectral density, which can cause many radio frequency-related implementation issues, including \ac{PA} linearization. Hence, the peak-constrained TR2 technique can control the peaks but penalizes the capability to reduce the \ac{PAPR} of signal---see Fig.~\ref{fig:cmp__tr_unconstrained__constrained__icf__topadmm}\subref{fig:1x1__TDL-A300ns10Hzlow__tr_unconstrained__constrained__icf__topadmm__ccdf_iPAPR} constraining the total energy of 36 CCs for TR2 within 5\% of total energy for useful signal. Additionally, we cannot use many unused tones due to the strict compliance with the regulatory and \ac{3GPP} standard spectral mask requirements. For high numerical accuracy of solutions of TR1 and TR2 schemes, we employ CVX since TR1/TR2 does not offer closed-form solutions due to the minimization of the infinite norm. Noticeably, although TR1 and TR2 schemes offer relatively low complexity compared to our proposed \ac{TOP-ADMM} method, TR1 and TR2, unfortunately, cannot meet desired \ac{PAPR} targets under practical constraints---see Fig.~\ref{fig:cmp__tr_unconstrained__constrained__icf__topadmm}\subref{fig:1x1__TDL-A300ns10Hzlow__tr_unconstrained__constrained__icf__topadmm__ccdf_iPAPR}. Moreover, recall that the guard bands of NR/LTE can be used to deploy NB-IoT carriers, among other practical challenges for the \ac{PAPR} reduction in NR/LTE. } 

\begin{figure}[!t]
    \begin{minipage}[t]{0.99\linewidth}
    \centering
    \includegraphics[width=\textwidth,trim=32.5mm 92mm 31mm 92mm,clip]{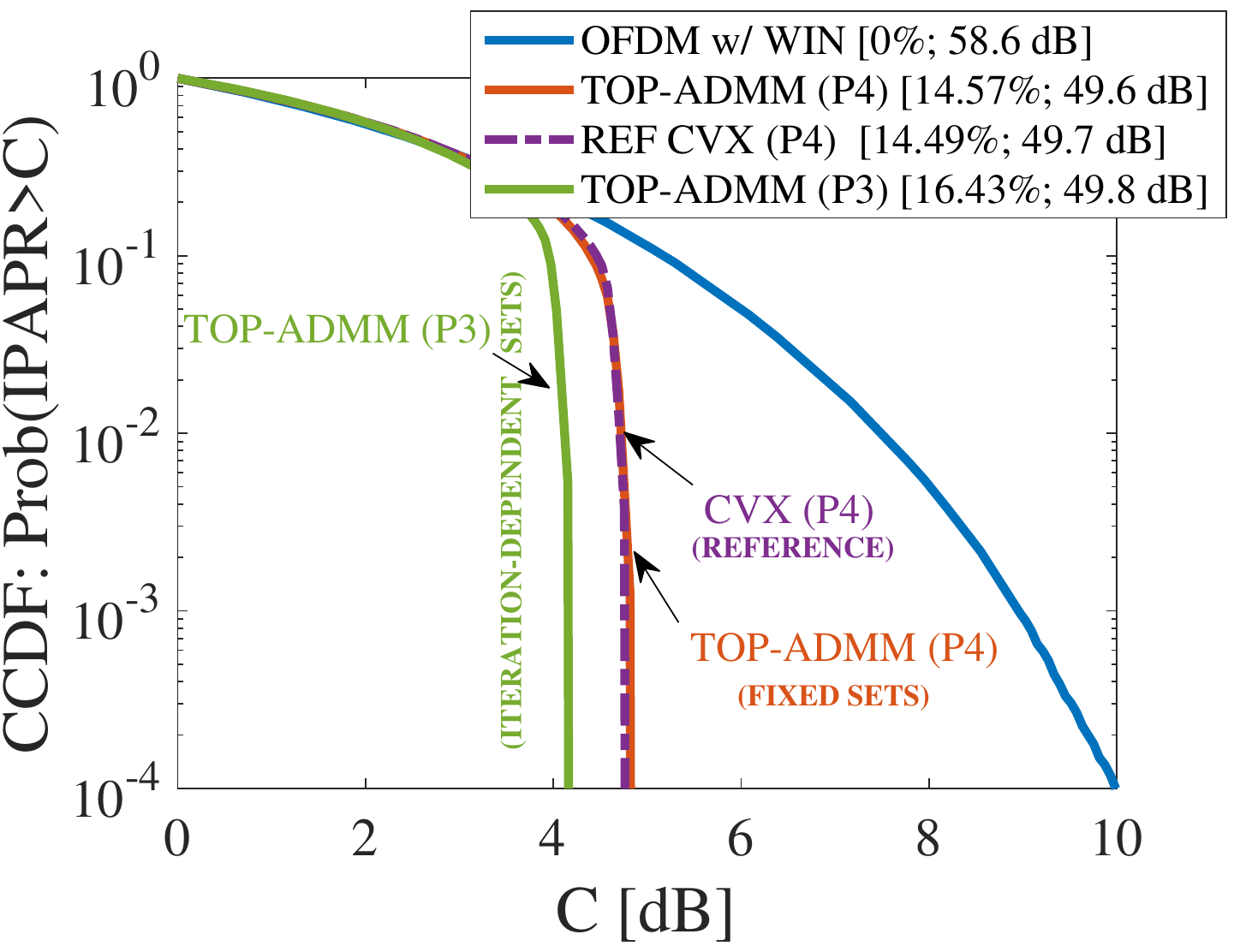}
    \caption{\skblue{Broadcast channel: \ac{IPAPR} distribution comparison between \texttt{P3} (with iteration-dependent sets) and \texttt{P4} (with fixed sets). }}\label{fig:cmp_ccdf__P3__P4__non_csi_aware}
    \end{minipage}
    \vspace{-1.5mm}
\end{figure}

\skblue{
Figure~\ref{fig:cmp_ccdf__P3__P4__non_csi_aware} illustrates the \ac{CCDF} of \ac{IPAPR} considering cell-specific broadcast/control channel or non-\ac{CSI}-aware, \ie, the base station has no downlink channel knowledge, under Test~2 (see Table~\ref{table:sim_assumptions}) but with a single antenna at the transmitter and the receiver. By the formulation of our proposed problem \texttt{P3}/\texttt{P4}, it can fall back and support the case where the base station has no downlink channel knowledge, which can cover the cell-specific broadcast/control channel case, \eg, by setting $\left\{ \mat{Q}[k]\!=\!\nu \I\right\}$ with $\nu\!\geq\!0$---see~\eqref{eqn:def_Qk_matrix}---and $\zeta\!\geq 0\!$. For brevity, we have set the following parameters to evaluate \texttt{P3}/\texttt{P4} for the non-\ac{CSI}-aware case: $\nu\!=\!0$, \ie, $\mat{Q}[k]\!=\!\mat{0}$ and $\zeta\!=\!0.05$. For \ac{TOP-ADMM}, we have set the step-size $\tau\!=\!1.945$ targeting either \texttt{P3} or \texttt{P4}. As expected by our conjecture, \ac{TOP-ADMM} targeting \texttt{P3}, \ie, with iteration-dependent sets, yields a better solution---at least not worse---than \ac{TOP-ADMM} targeting \texttt{P4} in terms of \ac{PAPR}. Moreover, our proposed computationally efficient \ac{TOP-ADMM}-based solution targeting \texttt{P4} converges to a similar solution using, high numerical accuracy, CVX \skblue{wrapper (with SDPT3 solver)~\cite{Grant2014CVX:Beta}}. However, CVX is very slow to run on a personal computer even for single antennas.  Therefore, we could not obtain solutions using CVX for \texttt{P4} with more than one antenna. Under the non-\ac{CSI}-aware case, the proposed problem~\texttt{P3} or~\texttt{P4} minimizes the transmit \ac{EVM} subject to the \ac{PAPR} and \ac{ACLR} constraints. In the sequel, we show numerically that when the base station with many antennas has (practically even imperfect) downlink channel estimates, the proposed problem~\texttt{P3} or~\texttt{P4}---solved by using our implementation-friendly \ac{TOP-ADMM}-based algorithms---mitigates the signal distortion seen at the receiver by capitalizing on the excess spatial degrees of freedom. 
}

\iftrue
\begin{figure*}[tp!]
  \begin{minipage}[t]{.99\linewidth}
  \begin{subfigure}[t]{0.49\textwidth}
    \centering
    \includegraphics[width=\textwidth,trim=32.5mm 92mm 31mm 92mm,clip]{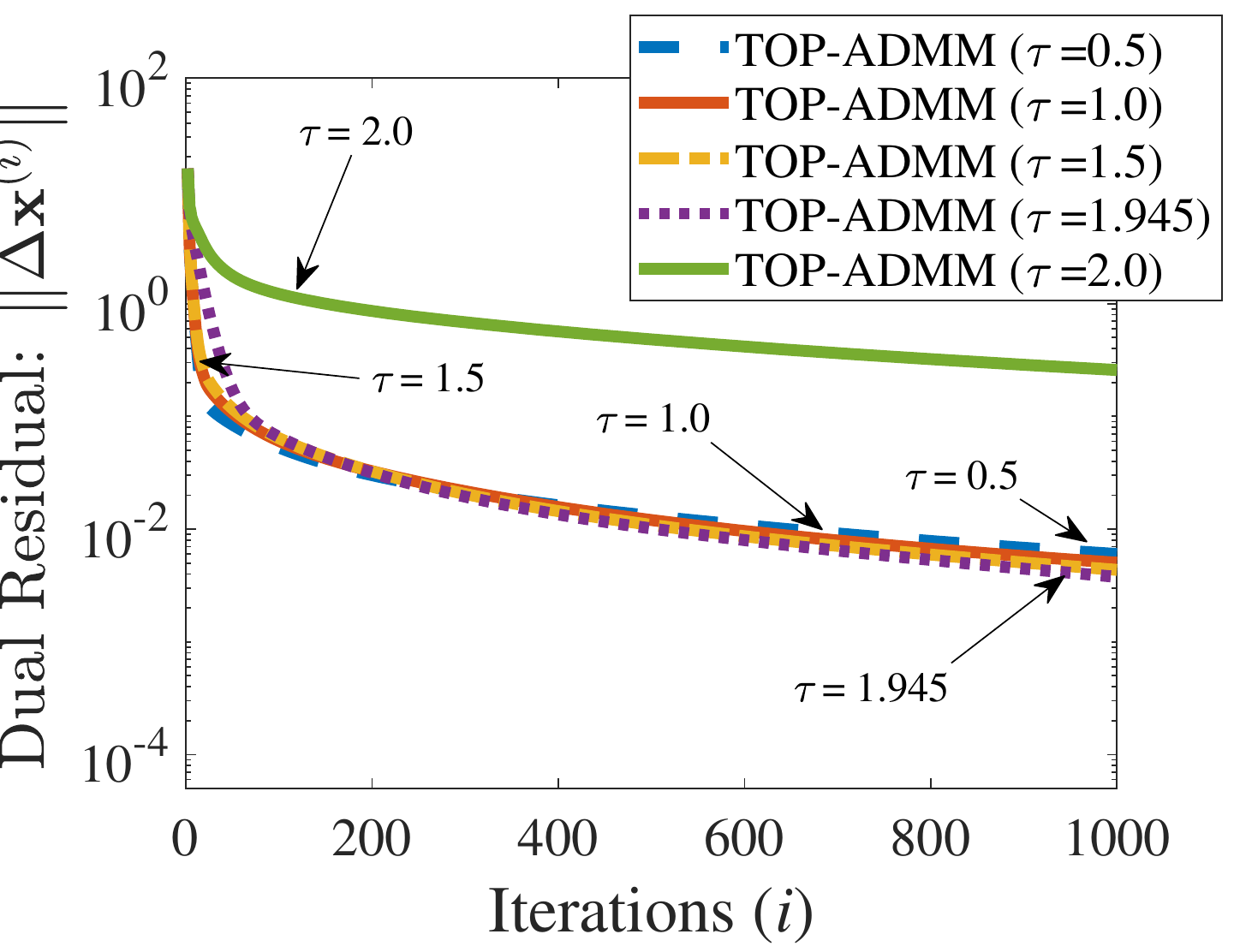}
    \caption{\ac{TOP-ADMM} ($\tau$): Dual residual vs. Iterations}\label{fig:FINAL__dual_residual_vs_iter__4x2_TDLD__TOPADMM}
  \end{subfigure}
  \begin{subfigure}[t]{0.49\textwidth}
    \centering
    \includegraphics[width=\textwidth,trim=32.5mm 92mm 31mm 92mm,clip]{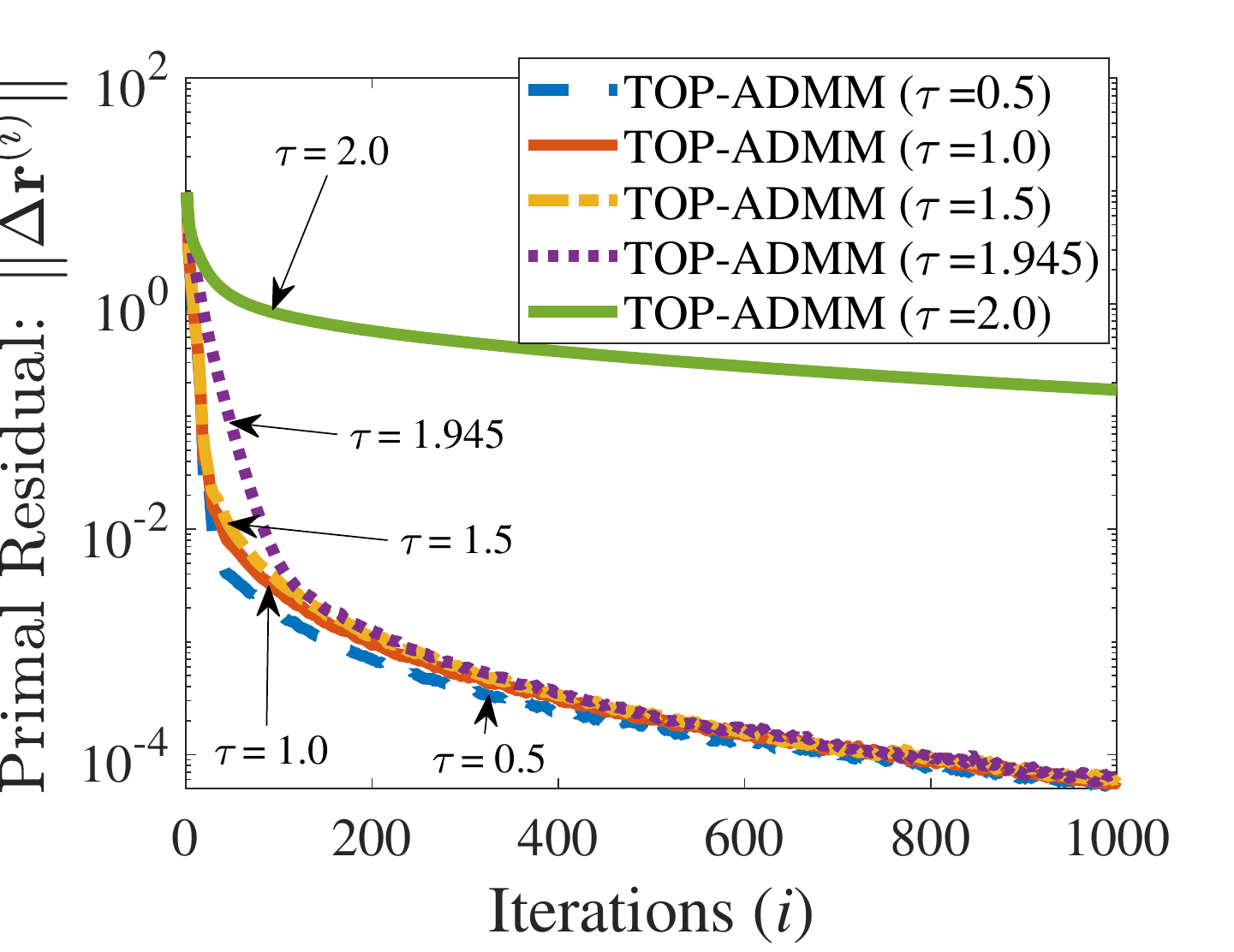}
    \caption{\ac{TOP-ADMM} ($\tau$): Primal residual vs. Iterations}\label{fig:FINAL__primal_residual_vs_iter__4x2_TDLD__TOPADMM}
  \end{subfigure}
  \caption{{Algorithm~\ref{algo:generalized_top_admm_algorithm_iterates} convergence behaviour}}\label{fig:convergence_behaviour__topadmm}
  \end{minipage}
  \vspace{-2.5mm}
\end{figure*}

\begin{figure*}[tp!]
  \begin{minipage}[t]{.99\linewidth}
   \begin{subfigure}[t]{0.49\textwidth}
    \centering
    \includegraphics[width=\textwidth,trim=32.5mm 92mm 31mm 92mm,clip]{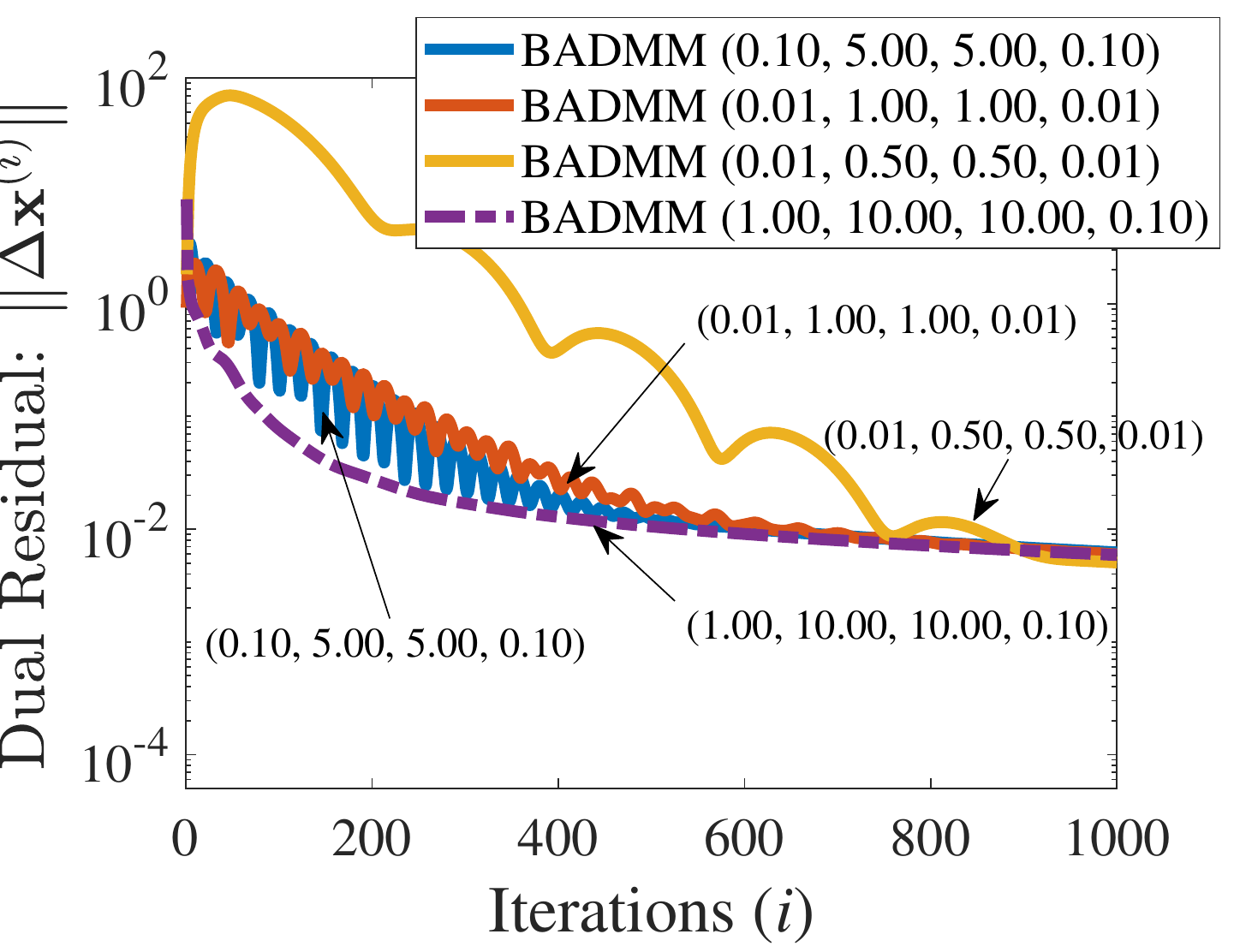}
    \caption{\ac{BADMM} ($\rho\!, \rho_x\!, \rho_z\!, \tau$): Dual residual vs. Iterations}\label{fig:FINAL__dual_residual_vs_iter__4x2_TDLD__BADMM}
  \end{subfigure}
  \begin{subfigure}[t]{0.49\textwidth}
    \centering
    \includegraphics[width=\textwidth,trim=32.5mm 92mm 31mm 92mm,clip]{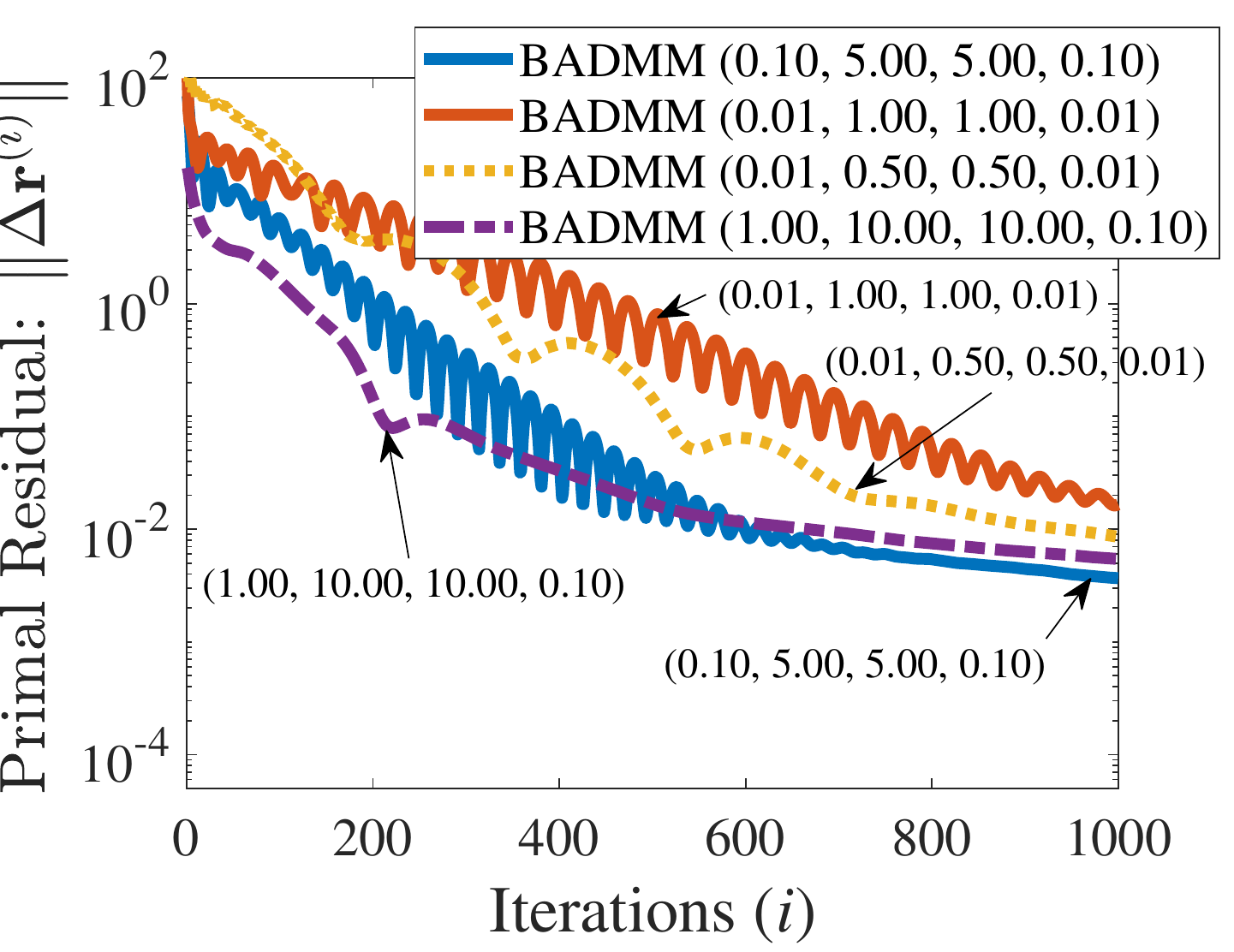}
    \caption{\ac{BADMM} ($\rho, \!\rho_x, \!\rho_z, \!\tau$): Primal residual vs. Iterations}\label{fig:FINAL__primal_residual_vs_iter__4x2_TDLD__BADMM}
  \end{subfigure}
   \caption{{Algorithm~\ref{algo:generalized_badmm_algorithm_iterates} convergence behaviour.}}\label{fig:convergence_behaviour__badmm}
  \end{minipage}
  \vspace{-2.5mm}
\end{figure*}

\begin{figure}[!htp]
    \vspace{-2.5mm}
    \begin{minipage}[t]{0.99\linewidth}
    \centering
    \includegraphics[width=\textwidth,trim=32.5mm 92mm 31mm 92mm,clip]{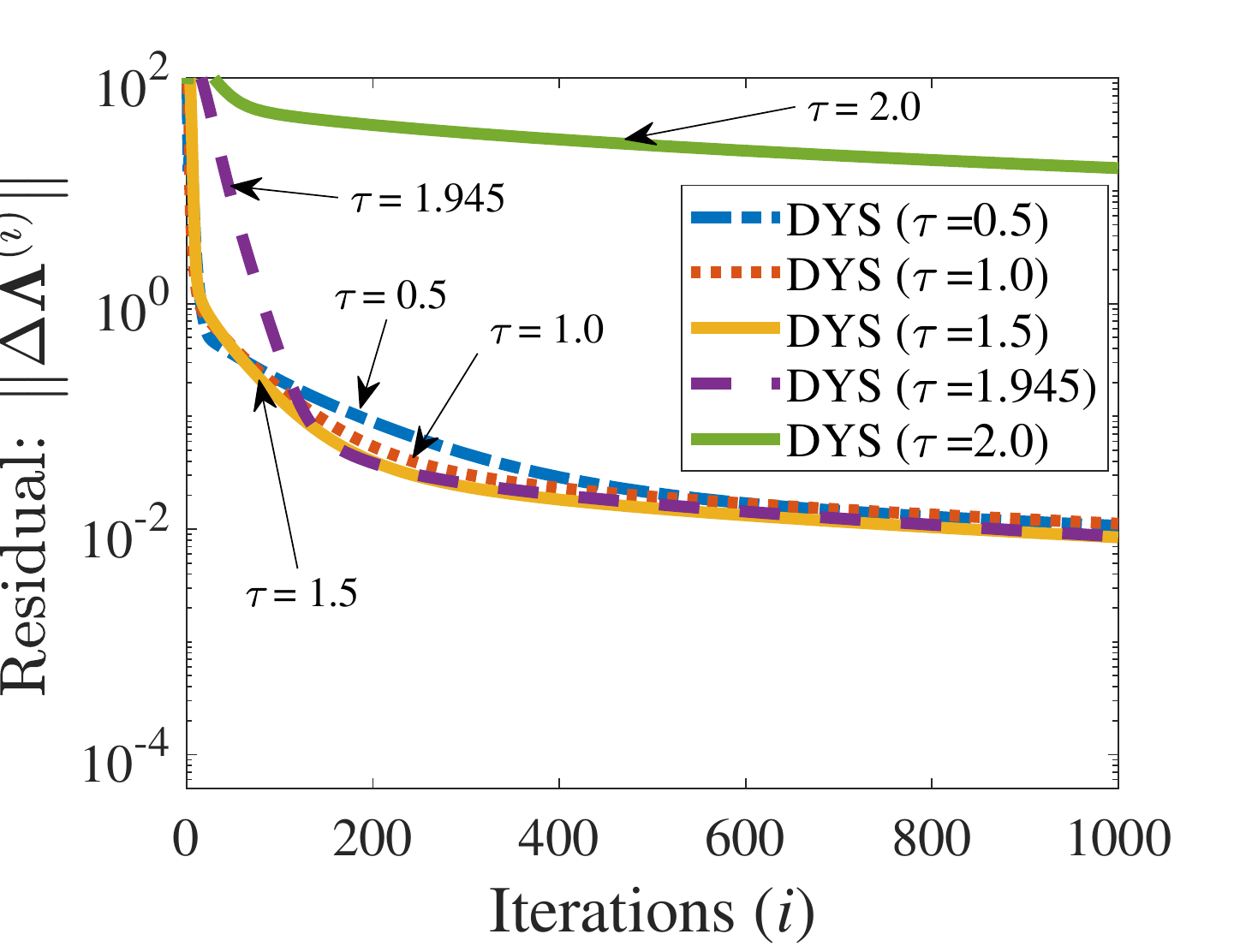}
    \caption{\ac{DYS} ($\tau$), Algorithm~\ref{algo:generalized_dys_algorithm_iterates}, convergence behaviour}\label{fig:convergence_behaviour__dys}
    \end{minipage}
    \vspace{-1.5mm}
\end{figure}

\begin{figure}[!htp]
    \vspace{-2.5mm}
    \begin{minipage}[t]{0.99\linewidth}
    \centering
    \includegraphics[width=\textwidth,trim=32.5mm 92mm 31mm 92mm,clip]{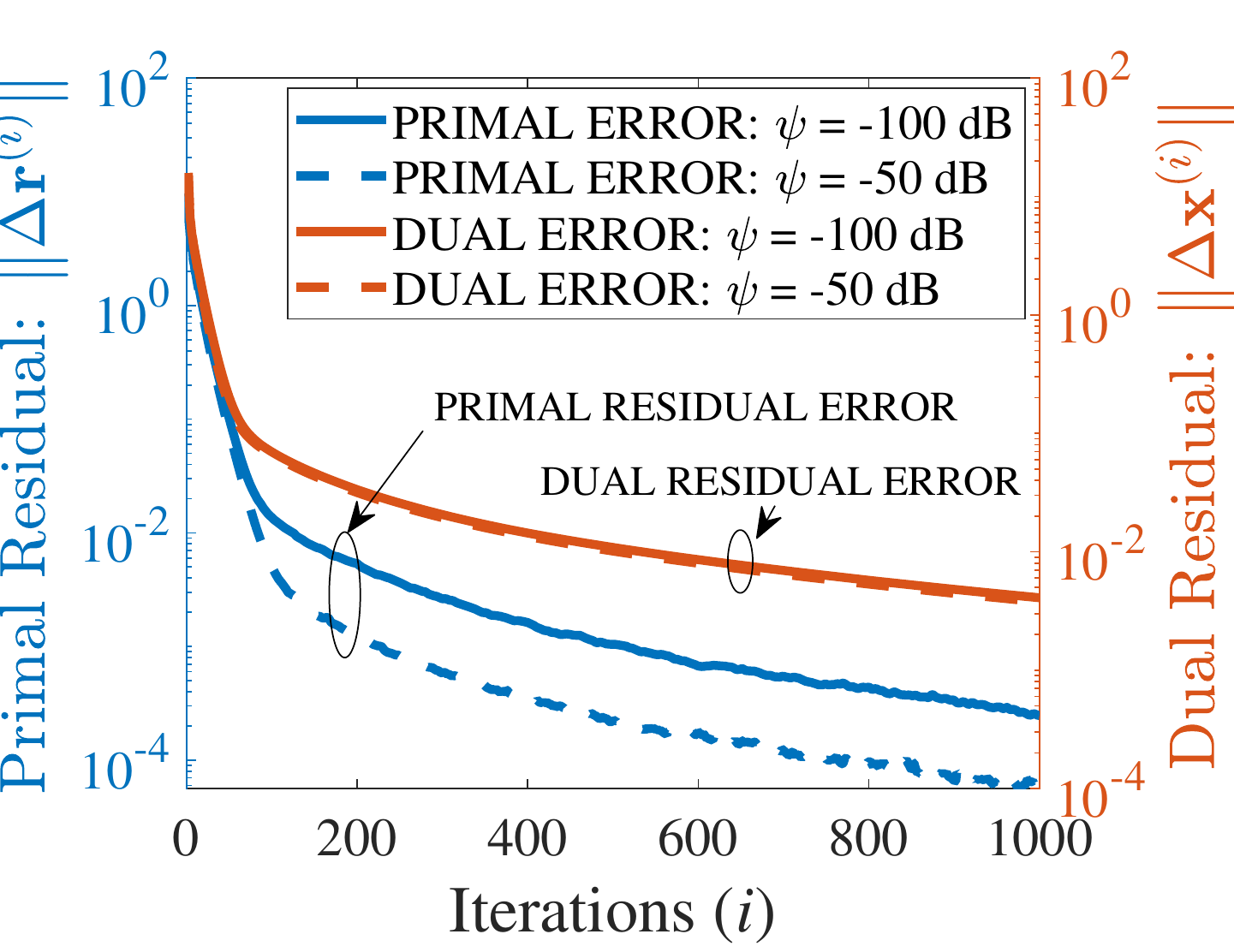}
    \caption{{Convergence behaviour of \ac{TOP-ADMM}($\tau\!=\!1.945$) with $\bm{\psi}_{\rm aclr} \!\in\!\left\{-50, -100 \right\} {\rm dB}$}}\label{fig:convergence_behaviour__topadmm_two_different_psi}
    \end{minipage}
    \vspace{-2.5mm}
\end{figure}

\else
\begin{figure*}[tp!]
  \begin{minipage}[t]{.315\linewidth}
  \begin{subfigure}[t]{\textwidth}
    \centering
    \includegraphics[width=\textwidth,trim=32.5mm 92mm 31mm 92mm,clip]{figures/FINAL__dual_residual_vs_iter__4x2_TDLD__TOPADMM.pdf}
    \caption{\ac{TOP-ADMM} ($\tau$): Dual residual vs. Iterations}\label{fig:FINAL__dual_residual_vs_iter__4x2_TDLD__TOPADMM}
  \end{subfigure}
  
  \begin{subfigure}[t]{\textwidth}
    \centering
    \includegraphics[width=\textwidth,trim=32.5mm 92mm 31mm 92mm,clip]{figures/FINAL__primal_residual_vs_iter__4x2_TDLD__TOPADMM.pdf}
    \caption{\ac{TOP-ADMM} ($\tau$): Primal residual vs. Iterations}\label{fig:FINAL__primal_residual_vs_iter__4x2_TDLD__TOPADMM}
  \end{subfigure}
    \caption{{Algorithm~\ref{algo:generalized_top_admm_algorithm_iterates} convergence behaviour}}\label{fig:convergence_behaviour__topadmm}
  \end{minipage}\hfill
  \begin{minipage}[t]{.315\linewidth}
  \begin{subfigure}[t]{\textwidth}
    \centering
    \includegraphics[width=\textwidth,trim=32.5mm 92mm 31mm 92mm,clip]{figures/FINAL__dual_residual_vs_iter__4x2_TDLD__BADMM.pdf}
    \caption{\ac{BADMM} ($\rho\!, \rho_x\!, \rho_z\!, \tau$): Dual residual vs. Iterations}\label{fig:FINAL__dual_residual_vs_iter__4x2_TDLD__BADMM}
  \end{subfigure}
  
  \begin{subfigure}[t]{\textwidth}
    \centering
    \includegraphics[width=\textwidth,trim=32.5mm 92mm 31mm 92mm,clip]{figures/FINAL__primal_residual_vs_iter__4x2_TDLD__BADMM.pdf}
    \caption{\ac{BADMM} ($\rho, \!\rho_x, \!\rho_z, \!\tau$): Primal residual vs. Iterations}\label{fig:FINAL__primal_residual_vs_iter__4x2_TDLD__BADMM}
  \end{subfigure}
    \caption{{Algorithm~\ref{algo:generalized_badmm_algorithm_iterates} convergence behaviour.}}\label{fig:convergence_behaviour__badmm}
  \end{minipage}\hfill
  \begin{minipage}[t]{.315\linewidth}
  \begin{subfigure}[t]{\textwidth}
    \centering
    \includegraphics[width=\textwidth,trim=32.5mm 92mm 31mm 92mm,clip]{figures/FINAL_fixed_point_residual_vs_iter__4x2_TDLD__DYS.pdf}
  \end{subfigure}
  \caption{\ac{DYS} ($\tau$), Algorithm~\ref{algo:generalized_dys_algorithm_iterates}, convergence behaviour}\label{fig:convergence_behaviour__dys}
  
  \begin{subfigure}[t]{\textwidth}
    \centering
    \includegraphics[width=\textwidth,trim=32.5mm 92mm 31mm 92mm,clip]{figures/FINAL_4x2_TDLD__residual_errors__fixed_tau_Psi_-50dB_and_-100dB__TOPADMM.pdf}
  \end{subfigure}
    \caption{{Convergence behaviour of \ac{TOP-ADMM}($\tau\!=\!1.945$) with $\bm{\psi}_{\rm aclr} \!\in\!\left\{-50, -100 \right\} {\rm dB}$}}\label{fig:convergence_behaviour__topadmm_two_different_psi}
  \end{minipage}
  \vspace{-2.5mm}
\end{figure*}
\fi

For benchmarking purposes, we evaluate \ac{ICF}~\cite{Armstrong__icf__2002} as a baseline \skblue{for distortion-based \ac{PAPR} reduction techniques}.  Additionally, we evaluate the prior art, in particular the method referred to as {PROXINF-ADMM1}, which essentially targets to solve convex programming problem~\cite[Equation~(14)/(15)]{Bao_papr_admm__2018}. Observe that the authors in~\cite{Bao_papr_admm__2018} employed double loops, which can be avoided by directly employing our (proposed) three-operator \ac{ADMM}-like approaches.

Figures~\ref{fig:convergence_behaviour__topadmm}--\ref{fig:convergence_behaviour__dys} exhibit the convergence behaviour of the (proposed) \ac{TOP-ADMM}-type algorithms, namely, \ac{TOP-ADMM} (Algorithm~\ref{algo:generalized_top_admm_algorithm_iterates}), \ac{BADMM} (Algorithm~\ref{algo:generalized_badmm_algorithm_iterates}), and \ac{DYS} (Algorithm~\ref{algo:generalized_dys_algorithm_iterates}). \sk{We use} a set of tunable parameters to solve the optimization \skblue{problem~\texttt{P3}} considering Test~1 with four transmit and two receive antennas---see Table~\ref{table:sim_assumptions}---for \ac{PAPR} constraint with $\bm{\gamma}_{\rm par}\!=\! 4$~dB and \ac{ACLR} constraint with $\bm{\psi}_{\rm aclr}\!=\! -50$~dB. We do not know an optimal solution to \skblue{problem~\texttt{P3}} a priori (or through any black-box global optimization solvers). Thus, as remarked in \cite{Boyd2011}, we characterize the convergence of the (proposed) \ac{TOP-ADMM}-type algorithms, see Figure~\ref{fig:convergence_behaviour__topadmm}, and \ac{BADMM}, see Figure~\ref{fig:convergence_behaviour__badmm}, in terms of the primal and dual residual errors, \ie, $\left\|\Delta \mathbf{r}^{\left(i\right)}\right\| \!\coloneqq\! \left\| \overline{\mat{Y}}^{\left(i\right)} \!-\! \overline{\mat{X}}^{\left(i\right)} \right\|_F$ and $\left\|\Delta \mathbf{x}^{\left(i\right)}\right\| \!\coloneqq\! \left\| \overline{\mat{X}}^{\left(i\right)} \!-\! \overline{\mat{X}}^{\left(i-1\right)} \right\|_F$, respectively. For the convergence analysis of \ac{DYS} (Algorithm~\ref{algo:generalized_dys_algorithm_iterates}), the (fixed-point) residual error, cf.~\cite{Davis2017}, $\left\|\Delta \Lambda^{\left(i\right)}\right\| \!\coloneqq\! \left\| {\bm{\Lambda}}^{\left(i\right)} \!-\! {\bm{\Lambda}}^{\left(i-1\right)} \right\|_F$ \sk{plays a similar role as the} primal and dual residual errors---see Figure~\ref{fig:convergence_behaviour__dys}. 

In \ac{TOP-ADMM} (Algorithm~\ref{algo:generalized_top_admm_algorithm_iterates}), the scaled dual variable $\nicefrac{{\bm{\Lambda}}^{\left(i+1\right)}}{\rho}$ \sk{could be used} such that Algorithm~\ref{algo:generalized_top_admm_algorithm_iterates} becomes free from the $\rho$ parameter. Consequently, Algorithm~\ref{algo:generalized_top_admm_algorithm_iterates} has only one tunable step-size $\tau$ parameter. In Fig.~\ref{fig:convergence_behaviour__topadmm}, we show the residual errors for the step-size $\tau \!\in\! \left\{0.500, 1.000, 1.500, 1.945, 2.000 \right\}$. Similar to \ac{TOP-ADMM} with a scaled dual variable, the \ac{DYS}-based Algorithm~\ref{algo:generalized_dys_algorithm_iterates}  also has only one parameter, where Fig.~\ref{fig:convergence_behaviour__dys} illustrates the primal and dual residual errors against iterations for various step-sizes $\tau \!\in\! \left\{0.500, 1.000, 1.500, 1.945, 2.000 \right\}$. In contrast to \ac{TOP-ADMM} and \ac{DYS}, the \ac{BADMM} algorithm is not blessed with a single parameter but with four parameters $\left(\rho, \rho_x, \rho_z, \tau \right)$. In Fig.~\ref{fig:convergence_behaviour__badmm}, the primal and dual residual errors are shown for some choices of parameters within a bracket. Conspicuously, the primal residual error of the sequences generated by \ac{TOP-ADMM} is nearly 10 times smaller than the \ac{BADMM}. Although these metrics, \ie, primal and residual errors, are commonly used in the (numerical) optimization literature, \sk{subsequently,} we focus directly on the performance metrics of interest against iteration. Nevertheless, from these numerical convergence figures, we fix the tunable parameters of the respective algorithms for the subsequent performance figures. Therefore, for \ac{TOP-ADMM}, \ac{BADMM}, and \ac{DYS}, we use $\tau\!=\!1.945$, $\left(\rho\!=\!0.01, \rho_x\!=\!1, \rho_z\!=\!1, \tau\!=\!0.01 \right)$, and $\tau\!=\!1.945$, respectively, unless mentioned otherwise. Additionally, in Fig.~\ref{fig:convergence_behaviour__topadmm_two_different_psi}, we show the convergence behaviour of \ac{TOP-ADMM} for two different \ac{ACLR} constraints having $\bm{\psi}_{\rm aclr} \!\in\!\left\{-50, -100 \right\}$~dB. In terms of primal residual error, \sk{the convergence noticeably} improves with the relaxed \ac{ACLR} constraint having $\bm{\psi}_{\rm aclr} \!=\!-50$~dB. Similar behaviour is observed with \ac{BADMM} and \ac{DYS}. To this end, we fix both \ac{ACLR} and \ac{PAPR} constraints with $\bm{\psi}_{\rm aclr}\!=\!-50$~dB and $\bm{\gamma}_{\rm par}\!=\!4$~dB, respectively, \skblue{unless stated otherwise}.

\iftrue
\begin{figure*}[tp!]
  \begin{minipage}[t]{.485\linewidth}
    \centering
    \includegraphics[width=\textwidth,trim=32.5mm 92mm 31mm 92mm,clip]{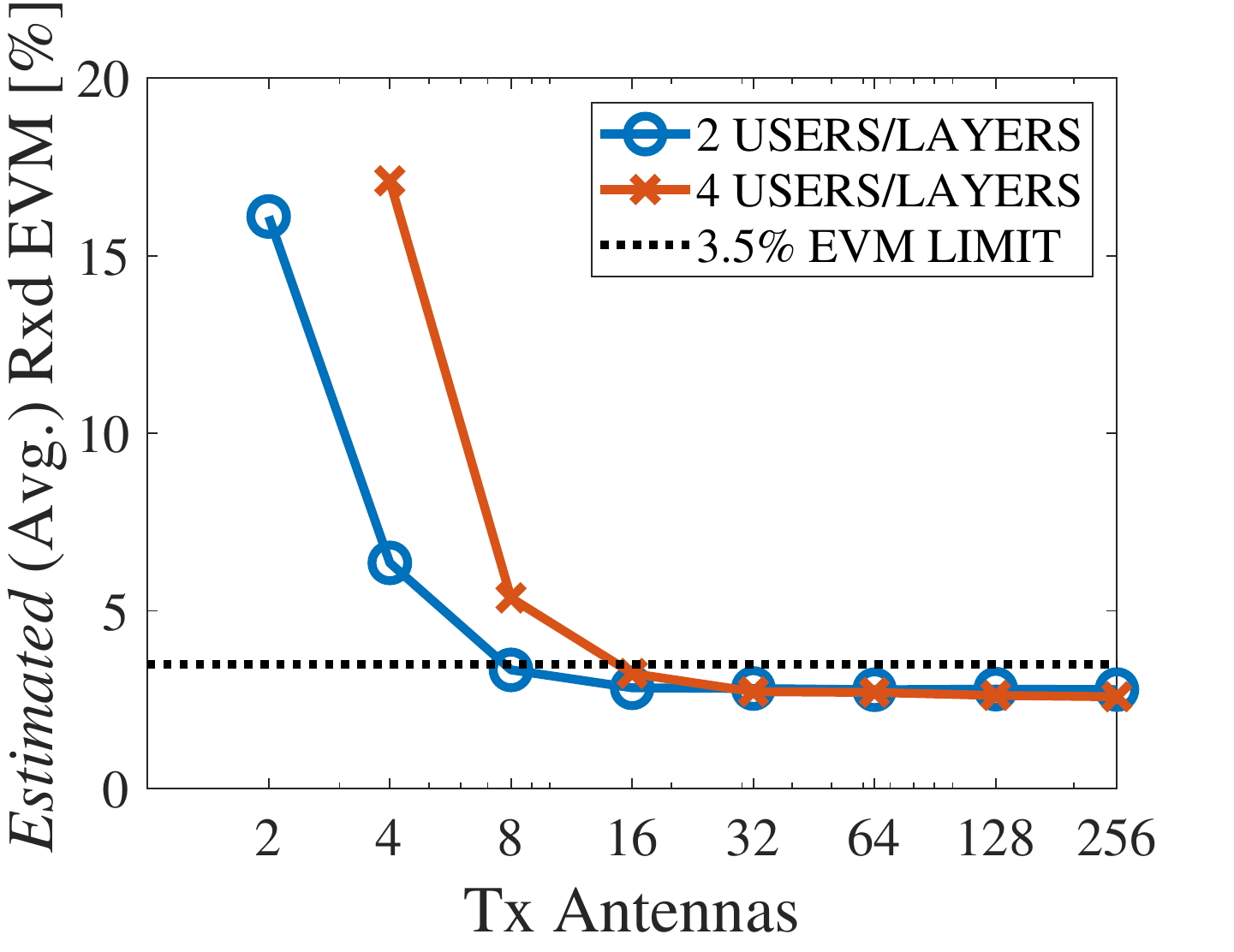}
    \caption{\skblue{Estimated average received \ac{EVM} against increasing excess spatial degrees of freedom for $\NT \!\times\! 2$ and $\NT \!\times\! 4$ DL \ac{MU-MIMO}/\ac{SU-MIMO}  setup under Test~2.}}\label{fig:Average_Estimated_Rxd_EVM_over_Mantennas__Mx2_or_Mx4}
  \end{minipage}\hfill
  \begin{minipage}[t]{.485\linewidth}
    \centering
    \includegraphics[width=\textwidth,trim=32.5mm 92mm 31mm 92mm,clip]{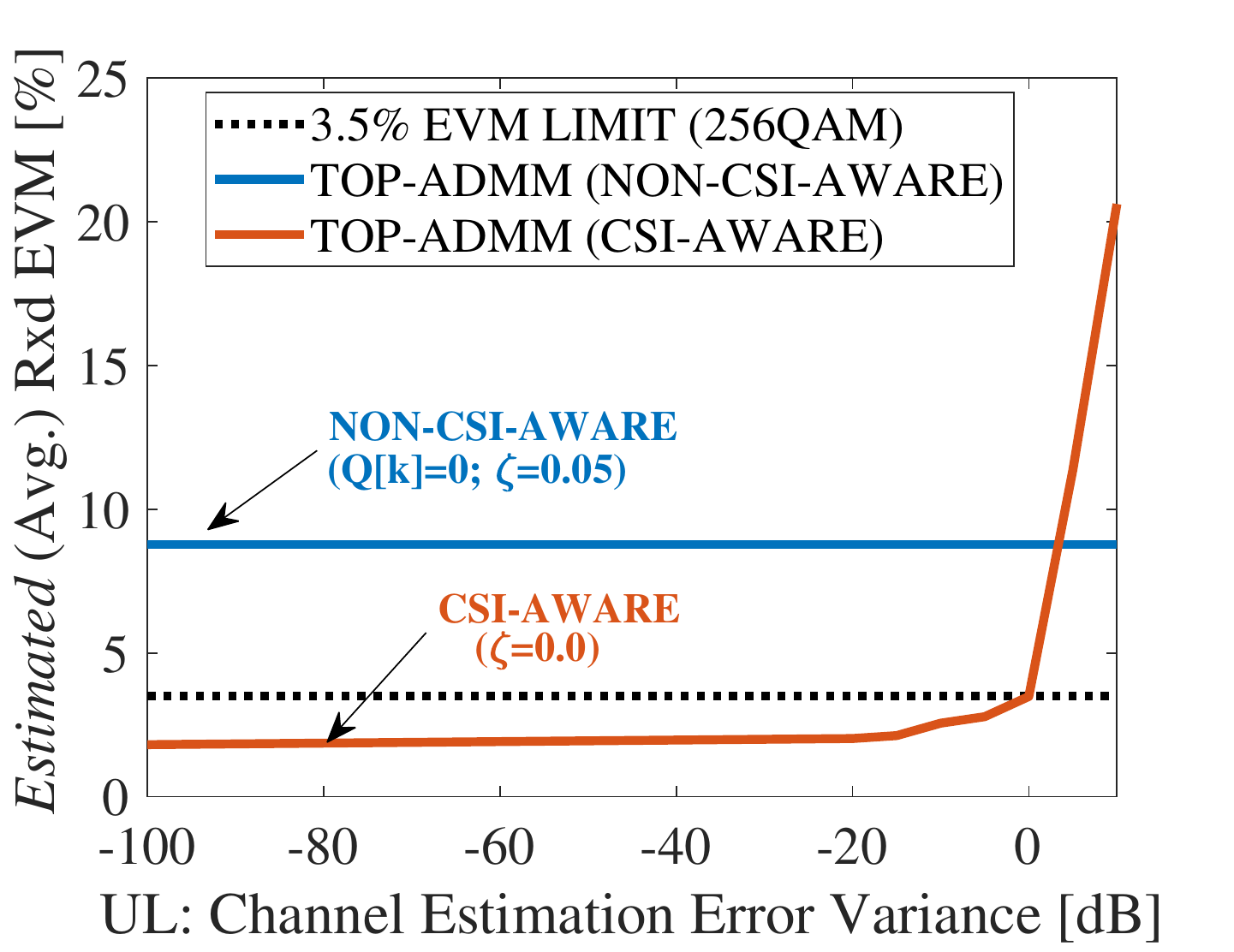}
    \caption{\skblue{Estimated average received \ac{EVM} against increasing UL channel estimation error variance with a fixed regularizer $\nu\!=\!0.001$ for $\left\{ \mat{Q}[k] \right\}$---cf.~\eqref{eqn:def_Qk_matrix}---under Test~2.}}\label{fig:64x2__Avg_EstRxdEVM__vs__channel_est_error_var__TOPADMM}
  \end{minipage}
  \vspace{-2.5mm}
\end{figure*}

\begin{figure}[!htp]
    \begin{minipage}[t]{0.99\linewidth}
    \centering
    \includegraphics[width=\textwidth,trim=32.5mm 92mm 31mm 92mm,clip]{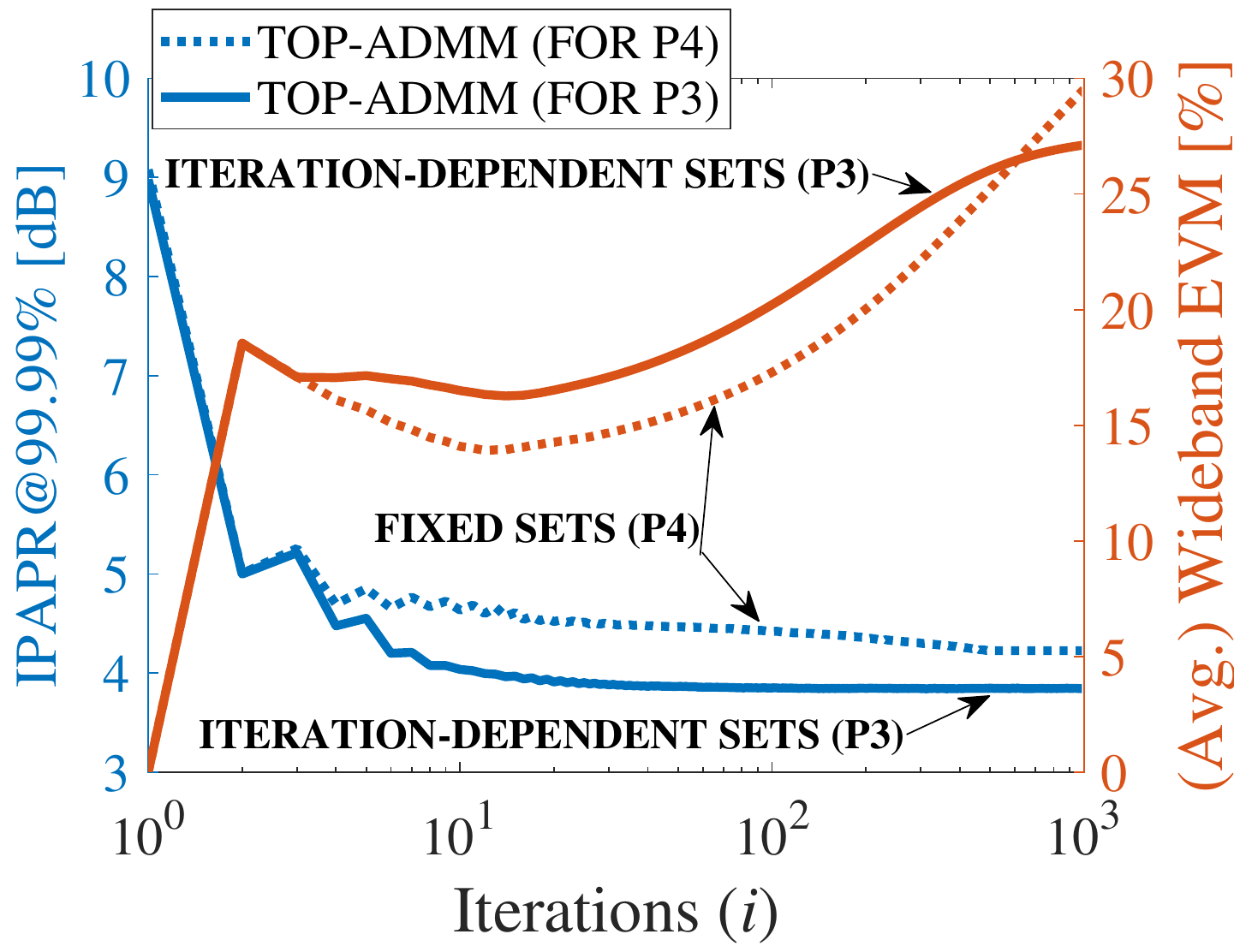}
    \caption{\skblue{CSI-aware: Comparison between \ac{TOP-ADMM} (for \texttt{P3} with iteration-dependent sets) and  \ac{TOP-ADMM} (for \texttt{P4} with fixed sets) under Test~1.}}\label{fig:imperfect_csi_aware__cmp_p3_and_p4}
    \end{minipage}
    \vspace{-1.5mm}
\end{figure}

\else

\begin{figure*}[tp!]
  \begin{minipage}[t]{.315\linewidth}
    \centering
    \includegraphics[width=\textwidth,trim=32.5mm 92mm 31mm 92mm,clip]{figures/Average_Estimated_Rxd_EVM_over_Mantennas__Mx2_or_Mx4.pdf}
    \caption{\skblue{Estimated average received \ac{EVM} against increasing excess spatial degrees of freedom for $\NT \!\times\! 2$ and $\NT \!\times\! 4$ DL \ac{MU-MIMO}/\ac{SU-MIMO}  setup under Test~2.}}\label{fig:Average_Estimated_Rxd_EVM_over_Mantennas__Mx2_or_Mx4}
  \end{minipage}\hfill
  \begin{minipage}[t]{.315\linewidth}
    \centering
    \includegraphics[width=\textwidth,trim=32.5mm 92mm 31mm 92mm,clip]{figures/64x2__Avg_EstRxdEVM__vs__channel_est_error_var__TOPADMM.pdf}
    \caption{\skblue{Estimated average received \ac{EVM} against increasing UL channel estimation error variance with a fixed regularizer $\nu\!=\!0.001$ for $\left\{ \mat{Q}[k] \right\}$---cf.~\eqref{eqn:def_Qk_matrix}---under Test~2.}}\label{fig:64x2__Avg_EstRxdEVM__vs__channel_est_error_var__TOPADMM}
  \end{minipage}\hfill
  \begin{minipage}[t]{.315\linewidth}
  \centering
    \includegraphics[width=\textwidth,trim=32.5mm 92mm 31mm 92mm,clip]{figures/64x2_TDL-D30ns10Hzlow_prac_prg2_par4dB_-50dBc__top_admm__P3_vs_P4__csi_aware__iapr_and_wbtxevm_vs_iter.pdf}
    \caption{\skblue{CSI-aware: Comparison between \ac{TOP-ADMM} (for \texttt{P3} with iteration-dependent sets) and  \ac{TOP-ADMM} (for \texttt{P4} with fixed sets) under Test~1.}}\label{fig:imperfect_csi_aware__cmp_p3_and_p4}
  \end{minipage}
  \vspace{-2.5mm}
\end{figure*}
\fi

\skblue{
Figure~\ref{fig:Average_Estimated_Rxd_EVM_over_Mantennas__Mx2_or_Mx4} shows the inband performance with increasing excess spatial degrees of freedom, \ie, number of transmit antennas under Test~2 using \ac{TOP-ADMM} (targeting \texttt{P3}) with the above discussed fixed parameters. More specifically, this figure depicts the estimated (averaged) received \ac{EVM} performance of \ac{MU-MIMO} or \ac{SU-MIMO} with 2/4 users or spatial layers against increasing excess spatial degrees of freedom. As expected, the proposed problem~\texttt{P3} (or \texttt{P4}) mitigates the incurred signal distortion when there are good enough excess spatial degrees of freedom. However, the estimated received \ac{EVM} may not necessarily vanish entirely or become zero with very high excess spatial degrees of freedom due to many practical reasons such as imperfect channel estimates and channel aging. 
}

\skblue{
Figure~\ref{fig:64x2__Avg_EstRxdEVM__vs__channel_est_error_var__TOPADMM} illustrates the robustness against the usage of erroneous channel estimates with a fixed regularizer $\nu\!=\!0.001$ in the computation of distortion mitigation $\left\{ \mat{Q}[k] \right\}$---see.~\eqref{eqn:def_Qk_matrix}---under Test~2. The channel estimation error variance equal to -100~dB represents perfect channel estimates at the transmitter. Therefore, as expected, the received \ac{EVM} increases if the channel estimation error variance increases. Hence, if channel estimation error variance is high, then one can instead make the proposed \ac{PAPR} reduction without capitalizing on the \ac{CSI} knowledge if any, \eg, by simply setting $\left\{ \mat{Q}[k] \!=\! \mat{0}\right\}$ and $\zeta \! = \! 0.05$, \ie, the fall back mode, where \texttt{P3}/\texttt{P4} minimizes the transmit \ac{EVM} as required for the broadcast/control channel (or non-\ac{CSI}-aware case). Additionally, the practical system employs a link adaptation, where the modulation and coding scheme for the downlink transmission can be adapted according to the channel conditions (via \ac{CSI} report feedback).
In Fig.~\ref{fig:imperfect_csi_aware__cmp_p3_and_p4}, we compare the \ac{TOP-ADMM} solution achieved by solving \texttt{P3} with \texttt{P4} considering Test~1 ($64 \! \times \! 2$ \ac{MIMO} setup). As expected, the solution of \texttt{P3} (utilizing iteration-dependent sets) renders better performance than the solution of \texttt{P4} (using fixed constraint sets).
}

\iftrue
\begin{figure*}[tp!]
  \begin{minipage}[t]{.99\linewidth}
  \begin{subfigure}[t]{0.49\textwidth}
    \centering
    \includegraphics[width=\textwidth,trim=32.5mm 92mm 31mm 92mm,clip]{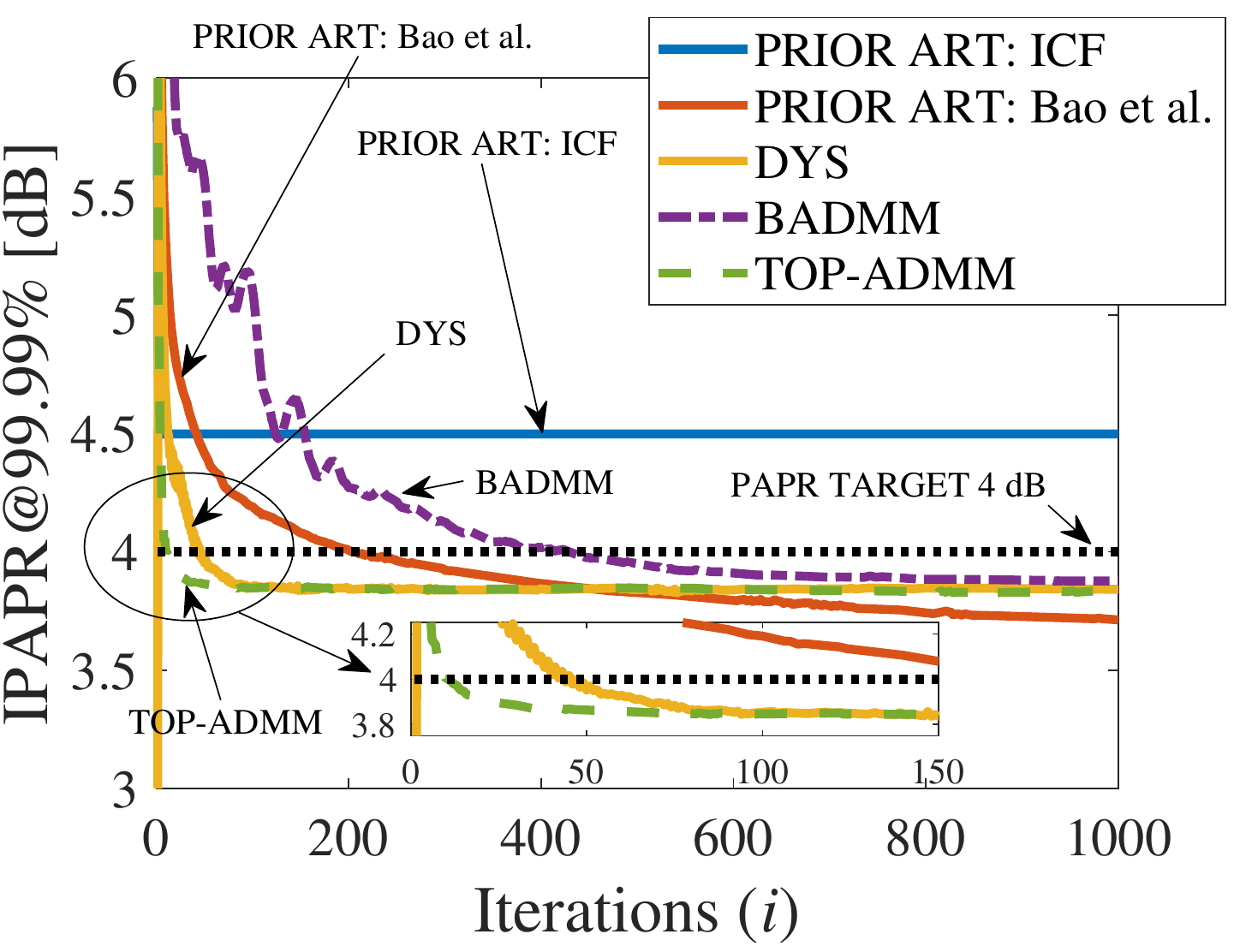}
    \caption{\ac{IPAPR} vs. {Iterations}; Test~1 (64Tx)}\label{fig:FINAL__64x2_TDLD10Hz30ns__target4dB__Tx1_iaprVsIter__icf_bao_topadmm}
  \end{subfigure}
  \begin{subfigure}[t]{0.49\textwidth}
    \centering
    \includegraphics[width=\textwidth,trim=32.5mm 92mm 31mm 92mm,clip]{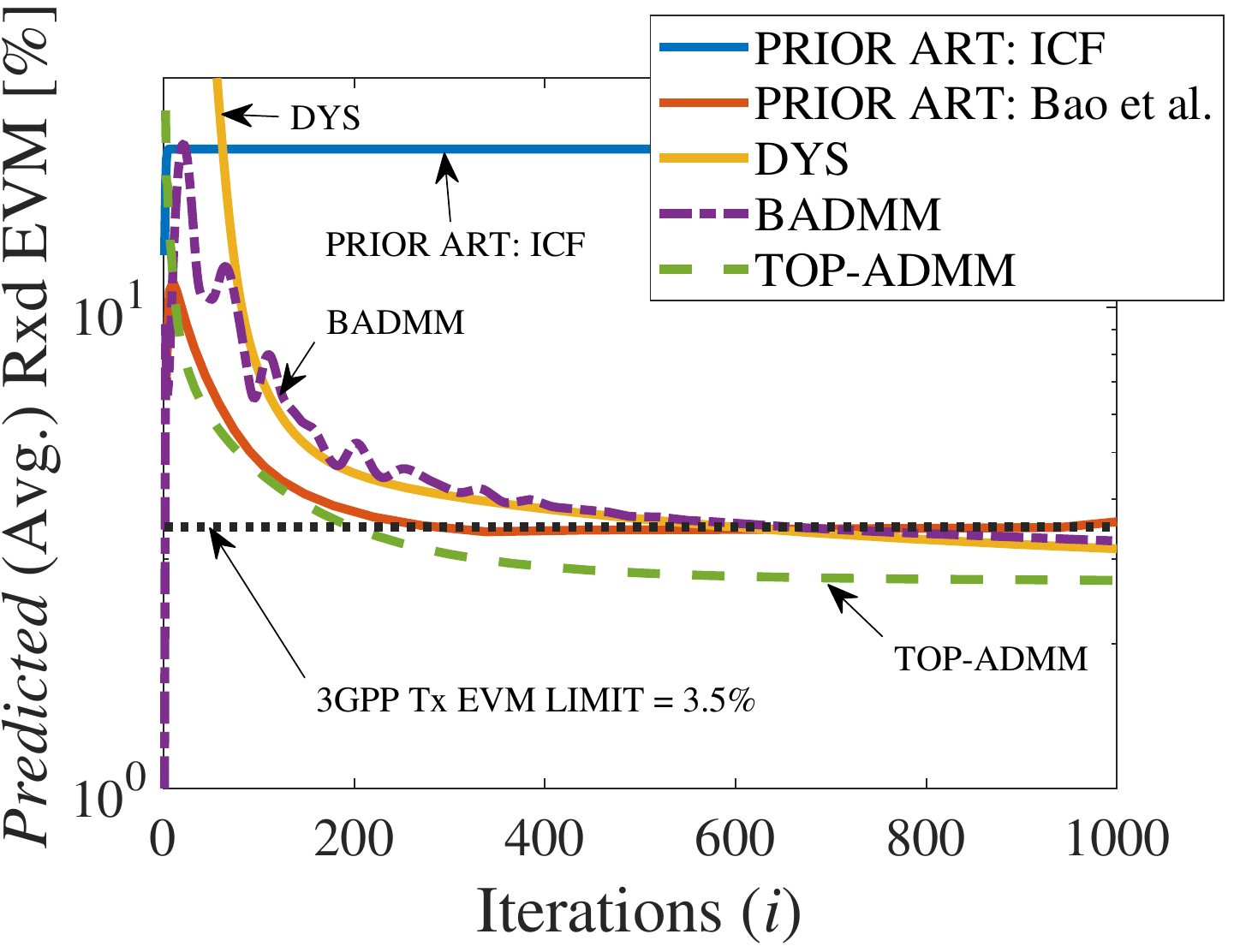}
    \caption{Predicted \ac{EVM} vs. {Iterations}; Test~1 (64Tx)}\label{fig:FINAL__64x2_TDLD10Hz30ns__target4dB__predictedEVMvsIter__icf_bao_topadmm}
  \end{subfigure}
  
    \begin{subfigure}[t]{0.49\textwidth}
    \centering
    \includegraphics[width=\textwidth,trim=32.5mm 92mm 31mm 92mm,clip]{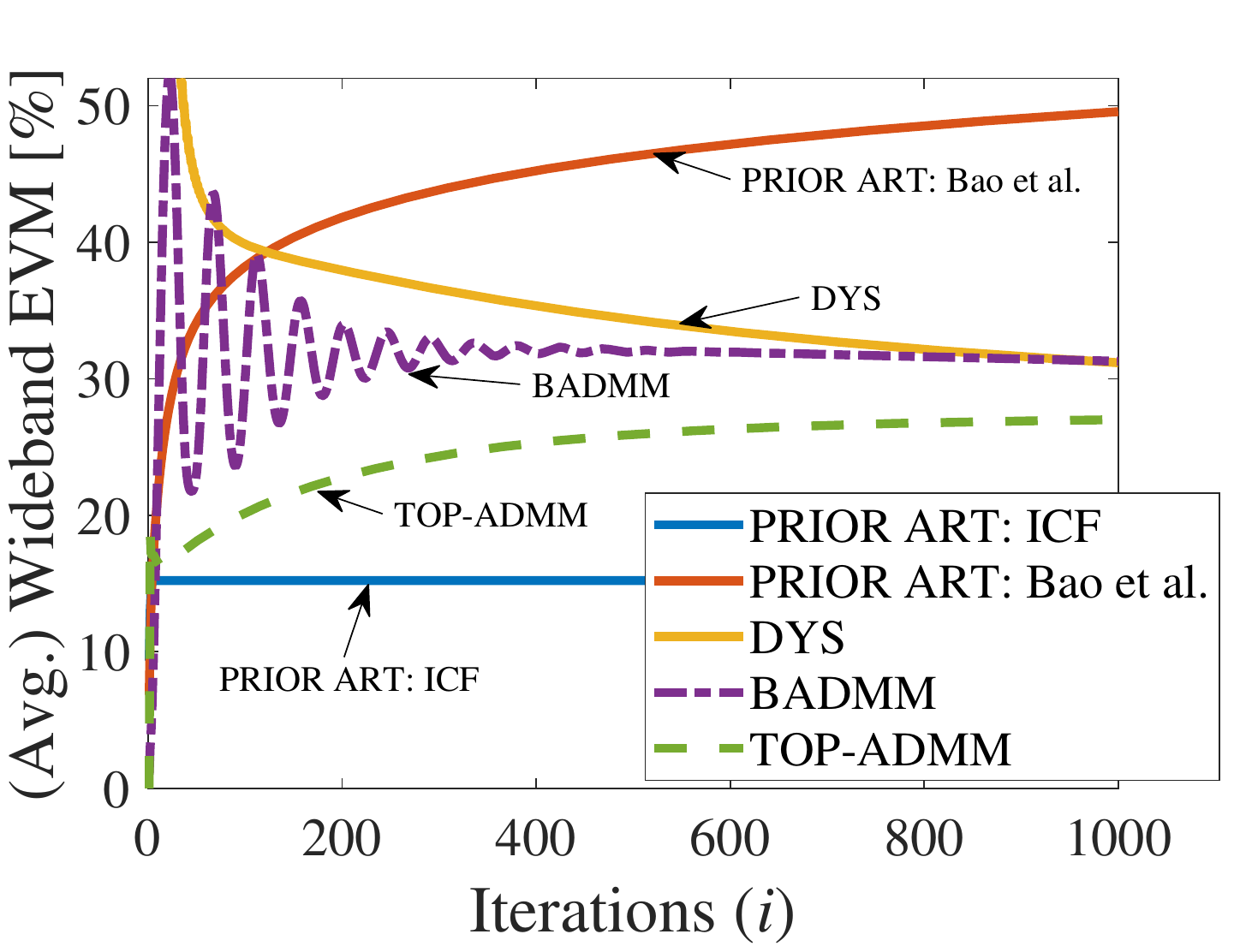}
    \caption{Transmit \ac{EVM} vs. {Iterations}; Test~1 (64Tx)}\label{fig:FINAL__64x2_TDLD10Hz30ns__target4dB__wbTxEVMvsIter__icf_bao_topadmm}
  \end{subfigure}
  \begin{subfigure}[t]{0.49\textwidth}
    \centering
    \includegraphics[width=\textwidth,trim=32.5mm 92mm 31mm 92mm,clip]{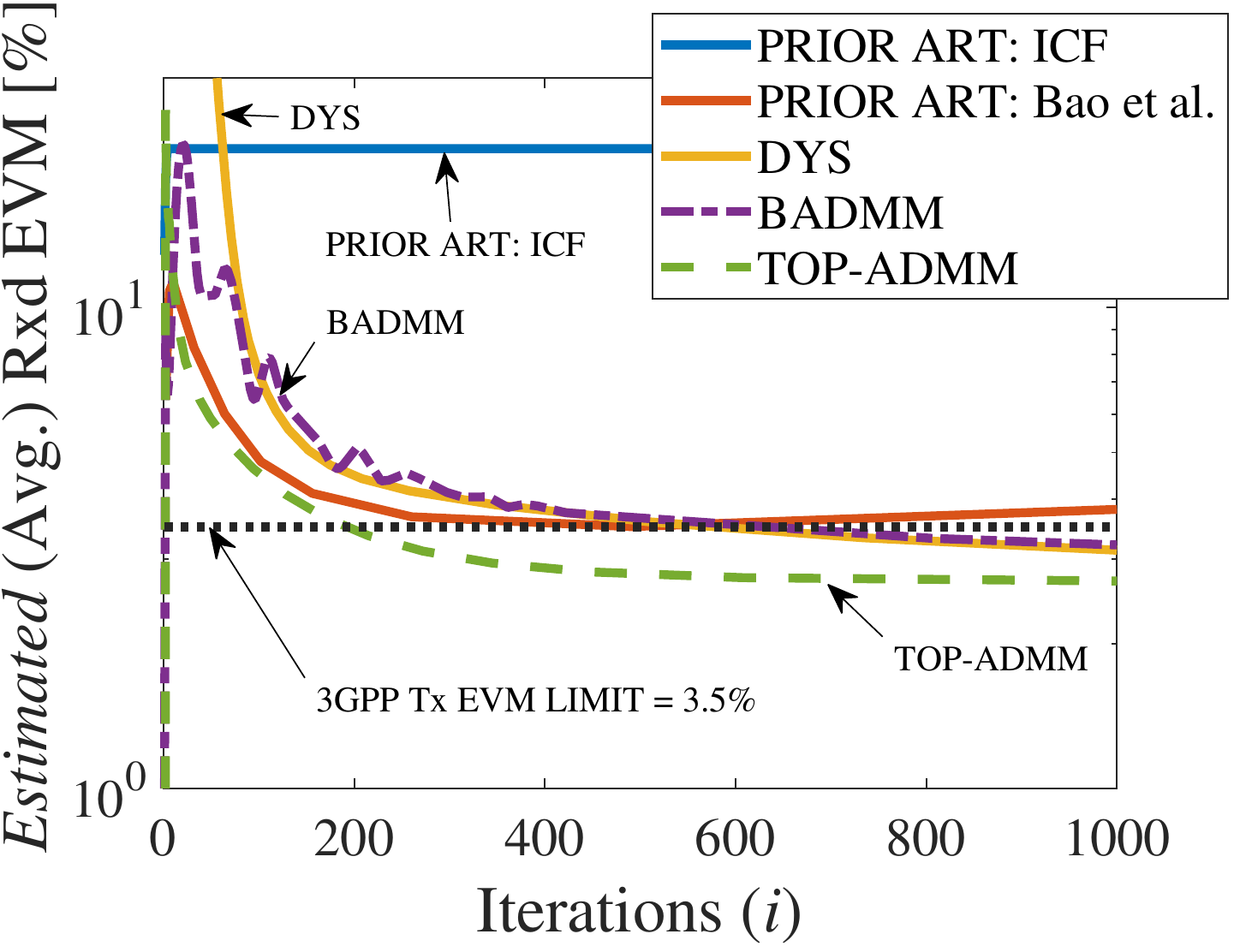}
    \caption{Estimated \ac{EVM} vs. {Iterations}; Test~1 (64Tx) }\label{fig:FINAL__64x2_TDLD10Hz30ns__target4dB__estimatedEVMvsIter__icf_bao_topadmm}
  \end{subfigure}
  \caption{Performance comparison of proposed \ac{TOP-ADMM}-type algorithms (at \ac{PAPR} target of 4~dB) with two prior arts, namely the non-CSI-aware \ac{ICF}~\cite{Armstrong__icf__2002} and CSI-aware Bao et al. algorithm~\cite{Bao_papr_admm__2018} in Test~1.}\label{fig:performance_comparison_proposed_three_operator_admm_types_with_prior_arts}
  \end{minipage}
  \vspace{-3.5mm}
\end{figure*}
\begin{figure*}[!htp]
  \begin{minipage}[t]{.99\linewidth}
  \begin{subfigure}[t]{0.49\textwidth}
    \centering
    \includegraphics[width=\textwidth,trim=32.5mm 92mm 31mm 92mm,clip]{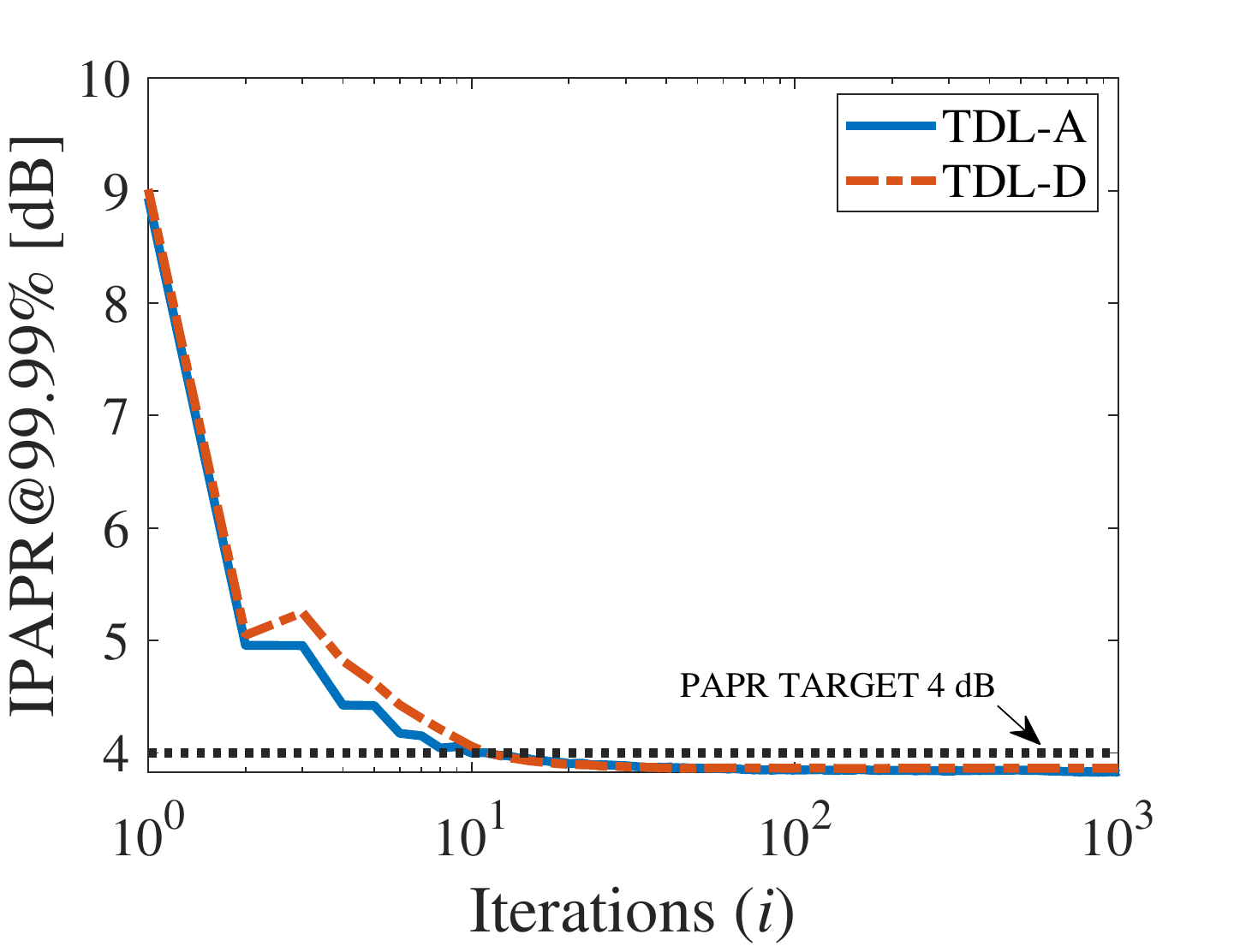}
    \caption{\ac{IPAPR} vs. Iterations}\label{fig:FINAL_64x2_TDLD10Hz30ns_vs_TDLA10Hz300ns__target4dB__Tx1_iaprVsIter_topadmm}
  \end{subfigure}
  \begin{subfigure}[t]{0.49\textwidth}
    \centering
    \includegraphics[width=\textwidth,trim=32.5mm 92mm 31mm 92mm,clip]{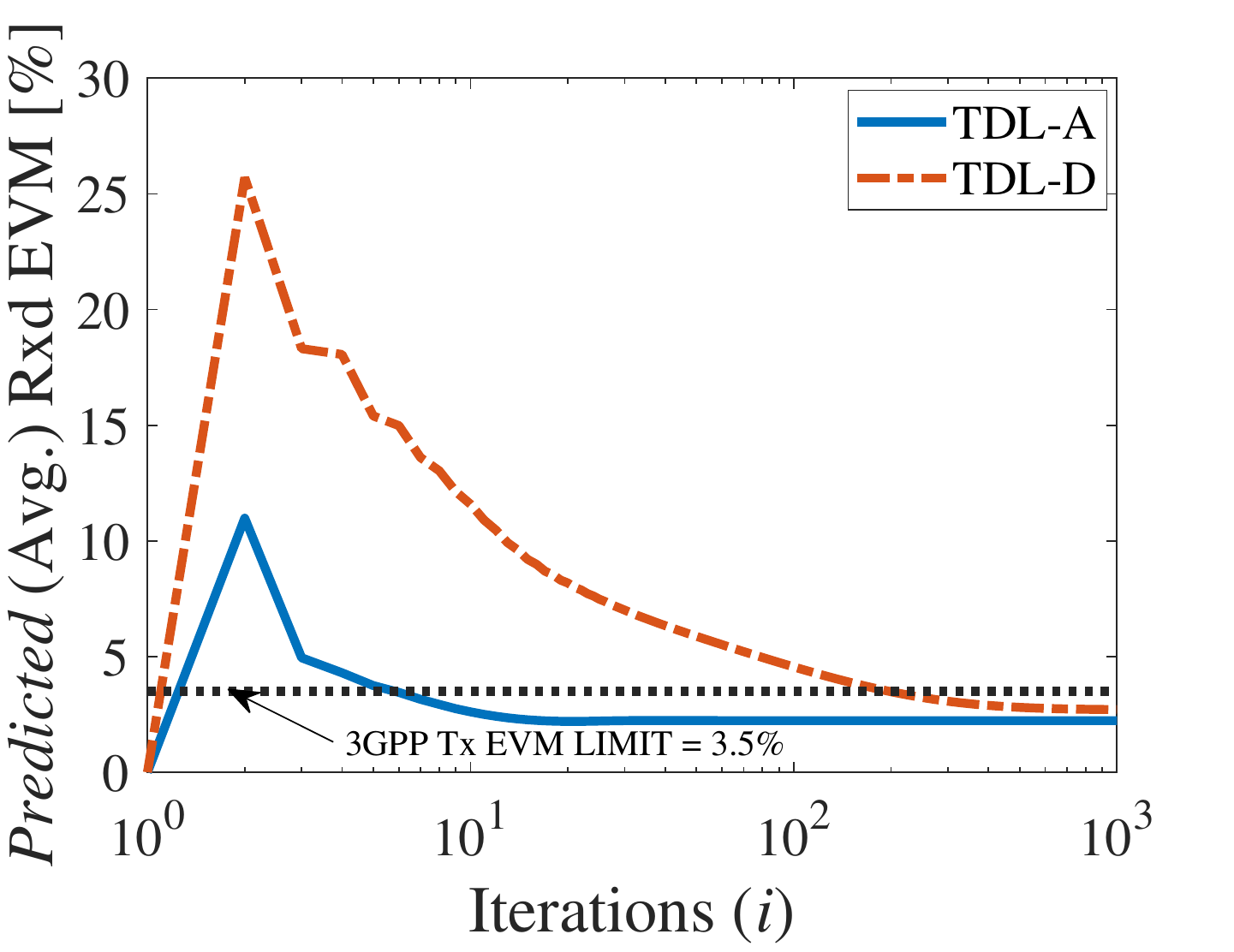}
    \caption{Predicted \ac{EVM} vs. Iterations}\label{fig:FINAL_64x2_TDLD10Hz30ns_vs_TDLA10Hz300ns__target4dB__predictedEVMvsIter__topadmm}
  \end{subfigure}
    \caption{{TOP-ADMM:Test~2 (\mbox{TDL-A}; Non-LOS, low spatial corr.) vs. Test~1 (TDL-D; LOS+Non-LOS, high spatial corr.). }}\label{fig:top_admm_perf__tdld_vs_tdla}
  \end{minipage}
  \vspace{-1.5mm}
\end{figure*}

\else
\begin{figure*}[tp!]
  \begin{minipage}[t]{.63\linewidth}
  \begin{subfigure}[t]{0.49\textwidth}
    \centering
    \includegraphics[width=\textwidth,trim=32.5mm 92mm 31mm 92mm,clip]{figures/FINAL__64x2_TDLD10Hz30ns__target4dB__Tx1_iaprVsIter__icf_bao_topadmm.pdf}
    \caption{\ac{IPAPR} vs. {Iterations}; Test~1 (64Tx)}\label{fig:FINAL__64x2_TDLD10Hz30ns__target4dB__Tx1_iaprVsIter__icf_bao_topadmm}
  \end{subfigure}
  \begin{subfigure}[t]{0.49\textwidth}
    \centering
    \includegraphics[width=\textwidth,trim=32.5mm 92mm 31mm 92mm,clip]{figures/FINAL__64x2_TDLD10Hz30ns__target4dB__predictedEVMvsIter__icf_bao_topadmm.pdf}
    \caption{Predicted \ac{EVM} vs. {Iterations}; Test~1 (64Tx)}\label{fig:FINAL__64x2_TDLD10Hz30ns__target4dB__predictedEVMvsIter__icf_bao_topadmm}
  \end{subfigure}
  
    \begin{subfigure}[t]{0.49\textwidth}
    \centering
    \includegraphics[width=\textwidth,trim=32.5mm 92mm 31mm 92mm,clip]{figures/FINAL__64x2_TDLD10Hz30ns__target4dB__wbTxEVMvsIter__icf_bao_topadmm.pdf}
    \caption{Transmit \ac{EVM} vs. {Iterations}; Test~1 (64Tx)}\label{fig:FINAL__64x2_TDLD10Hz30ns__target4dB__wbTxEVMvsIter__icf_bao_topadmm}
  \end{subfigure}
  \begin{subfigure}[t]{0.49\textwidth}
    \centering
    \includegraphics[width=\textwidth,trim=32.5mm 92mm 31mm 92mm,clip]{figures/FINAL__64x2_TDLD10Hz30ns__target4dB__estimatedEVMvsIter__icf_bao_topadmm.pdf}
    \caption{Estimated \ac{EVM} vs. {Iterations}; Test~1 (64Tx) }\label{fig:FINAL__64x2_TDLD10Hz30ns__target4dB__estimatedEVMvsIter__icf_bao_topadmm}
  \end{subfigure}
  \caption{Performance comparison of proposed \ac{TOP-ADMM}-type algorithms (at \ac{PAPR} target of 4~dB) with two prior arts, namely the non-CSI-aware \ac{ICF}~\cite{Armstrong__icf__2002} and CSI-aware Bao et al. algorithm~\cite{Bao_papr_admm__2018} in Test~1.}\label{fig:performance_comparison_proposed_three_operator_admm_types_with_prior_arts}
  \end{minipage}\hfill
  \begin{minipage}[t]{.315\linewidth}
  \begin{subfigure}[t]{\textwidth}
    \centering
    \includegraphics[width=\textwidth,trim=32.5mm 92mm 31mm 92mm,clip]{figures/FINAL_64x2_TDLD10Hz30ns_vs_TDLA10Hz300ns__target4dB__Tx1_iaprVsIter_topadmm.pdf}
    \caption{\ac{IPAPR} vs. Iterations}\label{fig:FINAL_64x2_TDLD10Hz30ns_vs_TDLA10Hz300ns__target4dB__Tx1_iaprVsIter_topadmm}
  \end{subfigure}
  
  \begin{subfigure}[t]{\textwidth}
    \centering
    \includegraphics[width=\textwidth,trim=32.5mm 92mm 31mm 92mm,clip]{figures/FINAL_64x2_TDLD10Hz30ns_vs_TDLA10Hz300ns__target4dB__predictedEVMvsIter__topadmm.pdf}
    \caption{Predicted \ac{EVM} vs. Iterations}\label{fig:FINAL_64x2_TDLD10Hz30ns_vs_TDLA10Hz300ns__target4dB__predictedEVMvsIter__topadmm}
  \end{subfigure}
    \caption{{TOP-ADMM:Test~2 (\mbox{TDL-A}; Non-LOS, low spatial corr.) vs. Test~1 (TDL-D; LOS+Non-LOS, high spatial corr.). }}\label{fig:top_admm_perf__tdld_vs_tdla}
  \end{minipage}
  \vspace{-3.5mm}
\end{figure*}

\fi

\iftrue 

\begin{figure*}[!ht]
\centering 
\begin{minipage}[t]{.48\textwidth}
\begin{subfigure}{\textwidth}
    \centering
    \includegraphics[width=\textwidth,trim=32.5mm 92mm 31mm 92mm,clip]{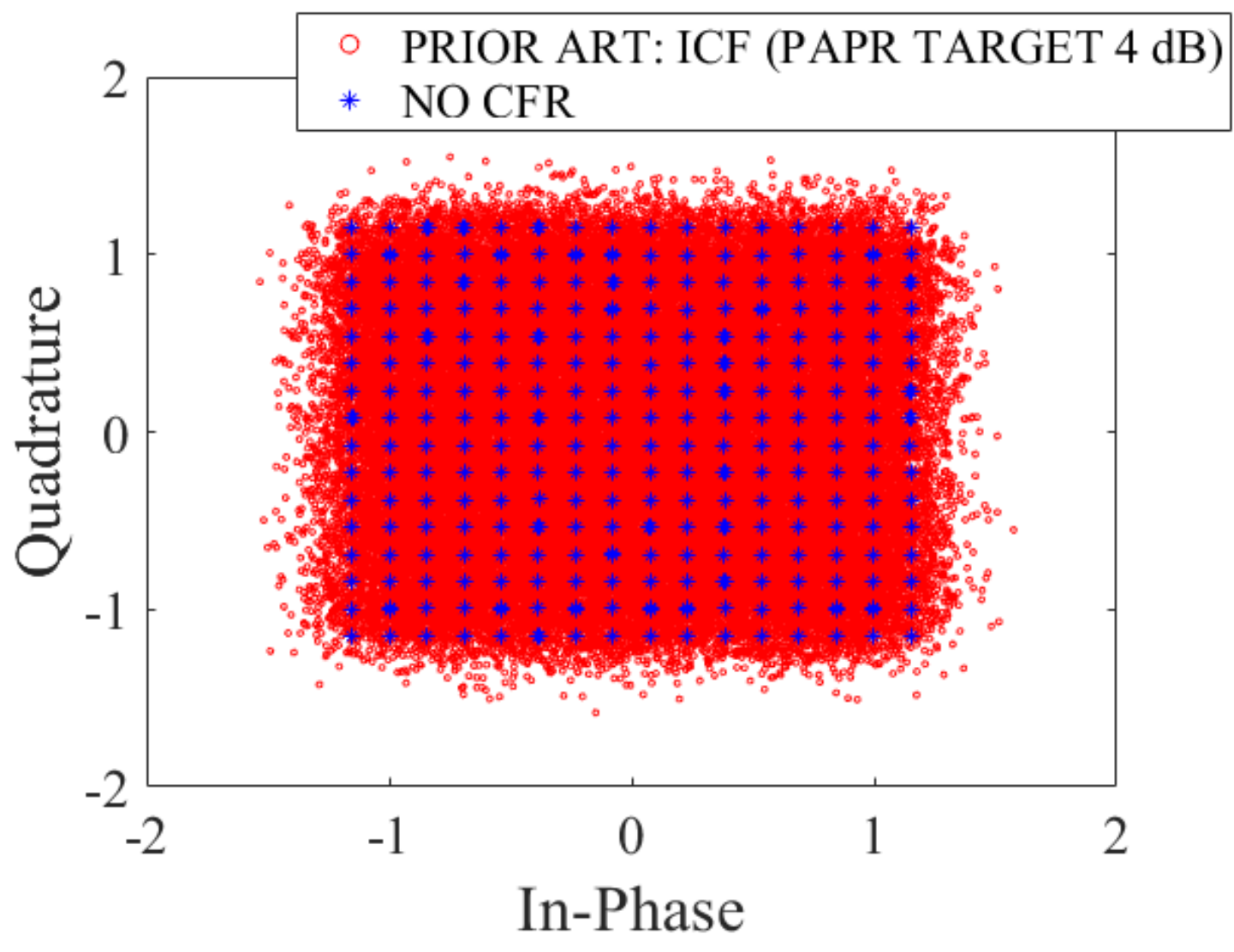}
    \caption{Estimated \sk{scatter plot} of \ac{ICF}~\cite{Armstrong__icf__2002}}\label{fig:FINAL__64x2_TDLD10Hz30ns__target4dB__estimatedScatterplot__icf}
\end{subfigure}\hfil 
\begin{subfigure}{\textwidth}
    \centering
    \includegraphics[width=\textwidth,trim=32.5mm 92mm 31mm 92mm,clip]{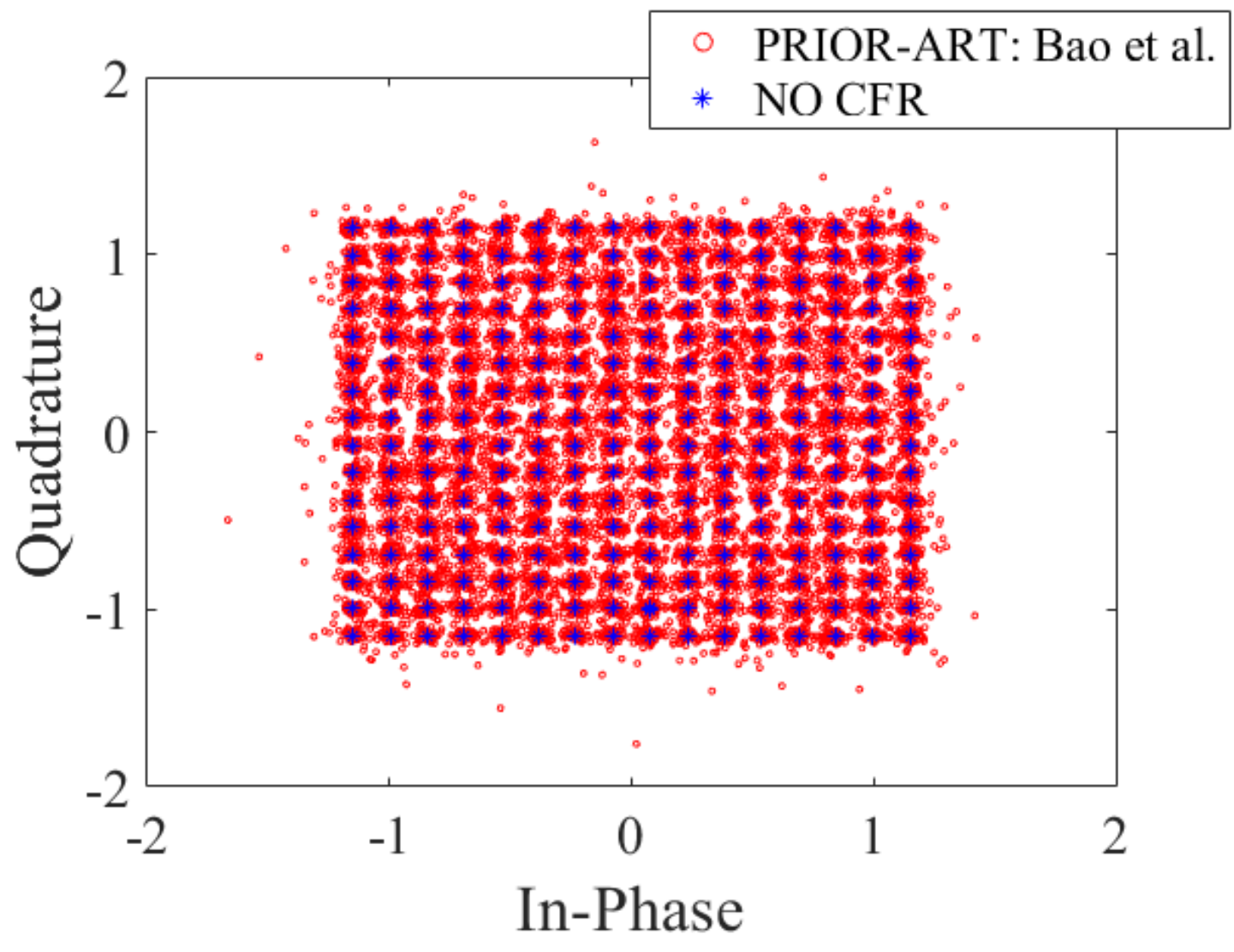}
    \caption{Estimated \sk{scatter plot} of \texttt{Bao et al.}~\cite{Bao_papr_admm__2018}}\label{fig:FINAL__64x2_TDLD10Hz30ns__target4dB__estimatedScatterplot__bao_speed}
  \end{subfigure}\hfil 
\begin{subfigure}{\textwidth}
  \centering
    \includegraphics[width=\textwidth,trim=32.5mm 92mm 31mm 92mm,clip]{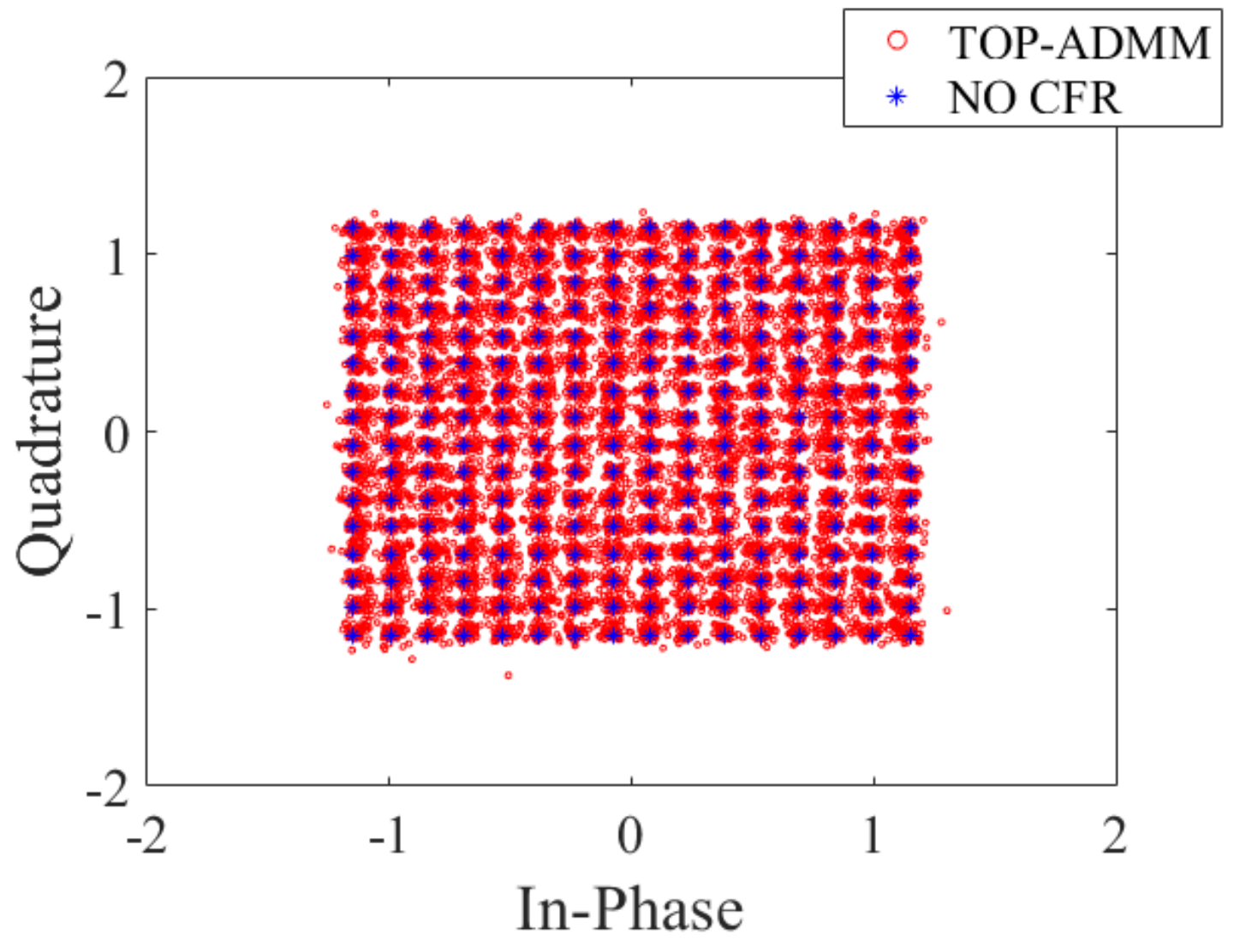}
    \caption{Estimated \sk{scatter plot} of \texttt{TOP-ADMM}}\label{fig:FINAL__64x2_TDLD10Hz30ns__target4dB__estimatedScatterplot__topadmm}
\end{subfigure}
\end{minipage}\hfil
\begin{minipage}[t]{.48\textwidth}
\begin{subfigure}{\textwidth}
    \centering
    \includegraphics[width=\textwidth,trim=32.5mm 92mm 31mm 92mm,clip]{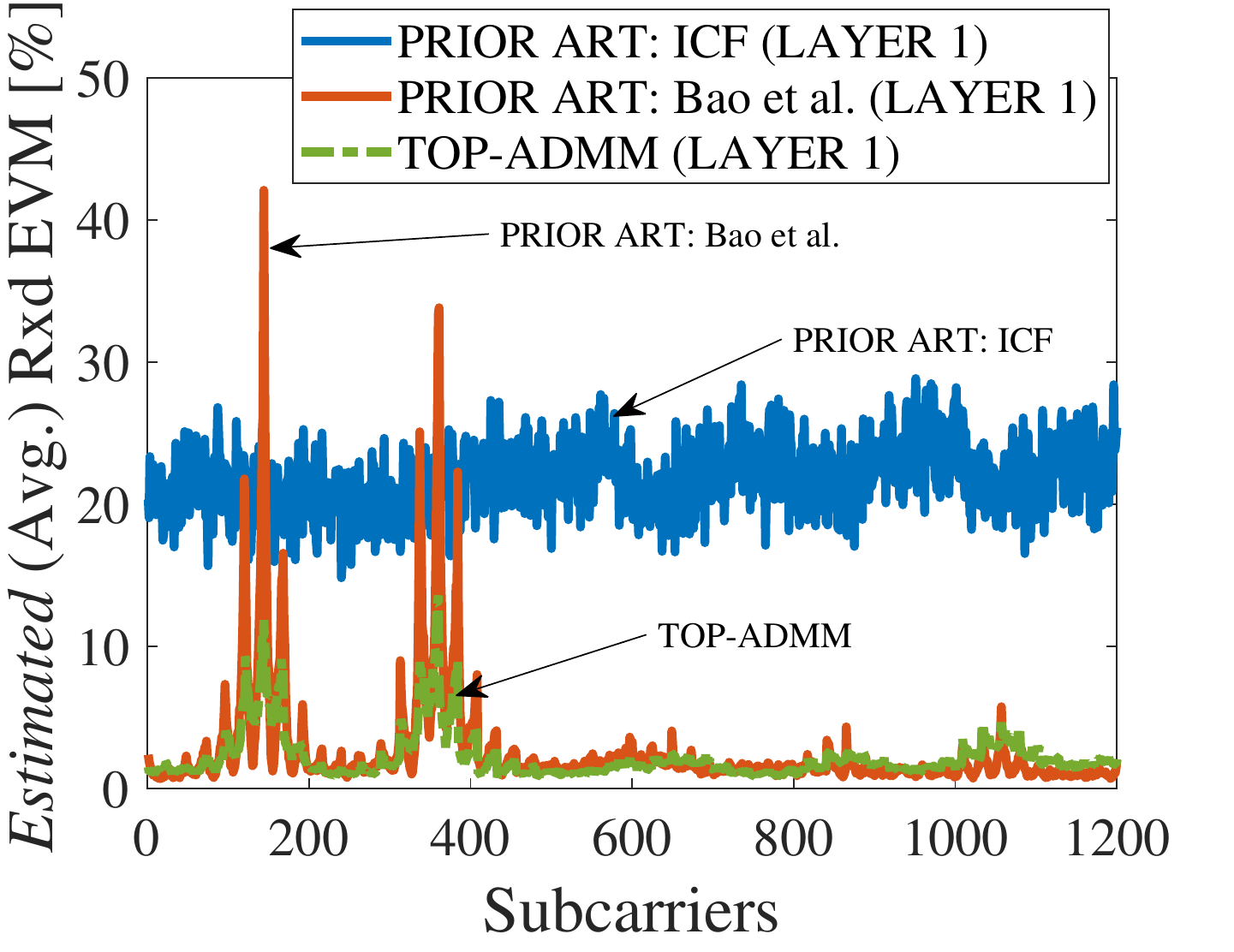}
    \caption{Estimated received \ac{EVM} distribution \sk{in the subcarrier domain} for spatial layer~1}\label{fig:FINAL__64x2_TDLD10Hz30ns__target4dB___Estimated_rxTxEVM_vs_PRB__LAYER1__icf_bao_topadmm}
\end{subfigure}\hfil 
\begin{subfigure}{\textwidth}
  \centering
    \includegraphics[width=\textwidth,trim=32.5mm 92mm 31mm 92mm,clip]{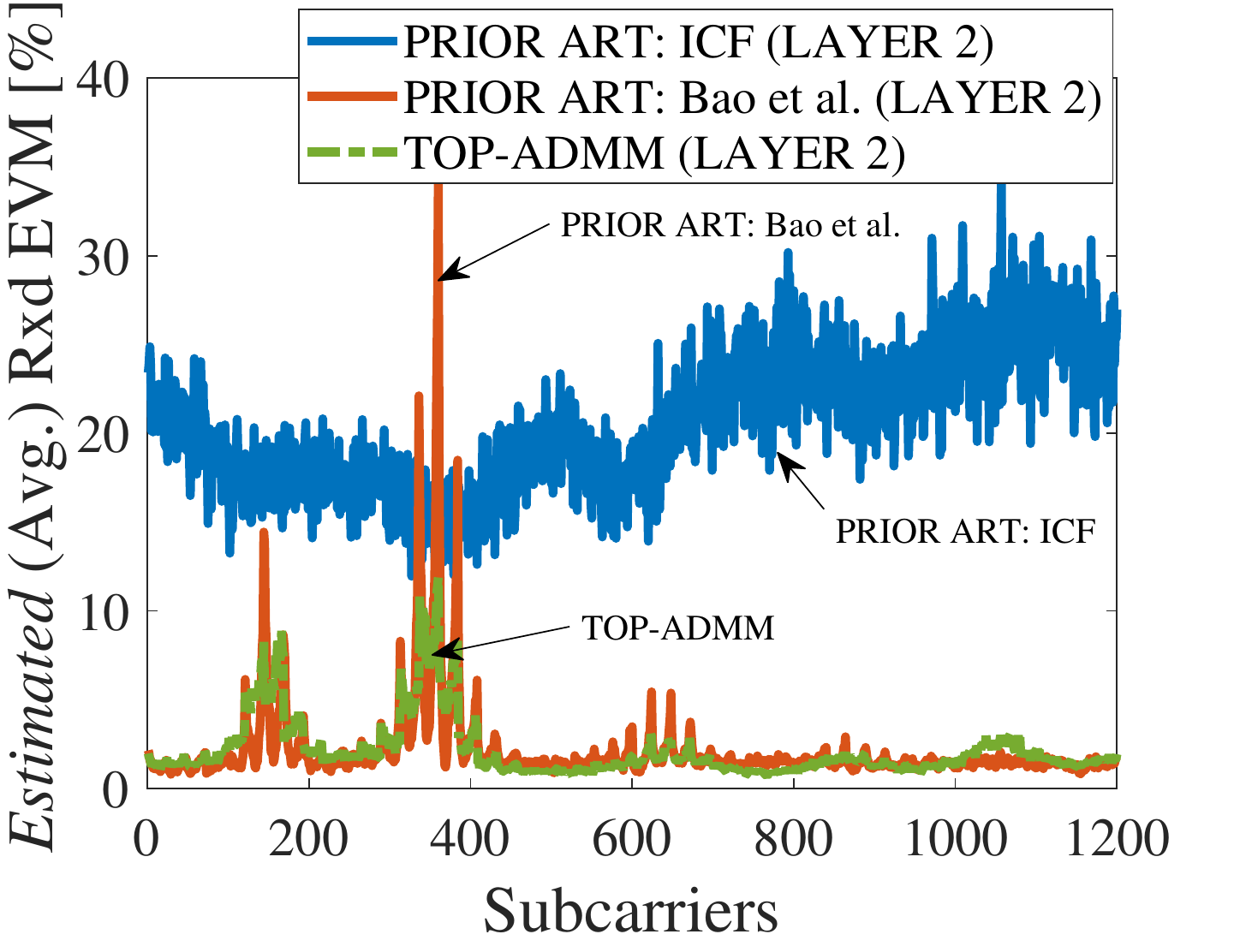}
    \caption{Estimated received \ac{EVM} distribution \sk{in the subcarrier domain} for spatial layer~2}\label{fig:FINAL__64x2_TDLD10Hz30ns__target4dB___Estimated_rxTxEVM_vs_PRB__LAYER2__icf_bao_topadmm}
\end{subfigure}\hfil 
\begin{subfigure}{\textwidth}
    \centering
    \includegraphics[width=\textwidth,trim=32.5mm 92mm 31mm 92mm,clip]{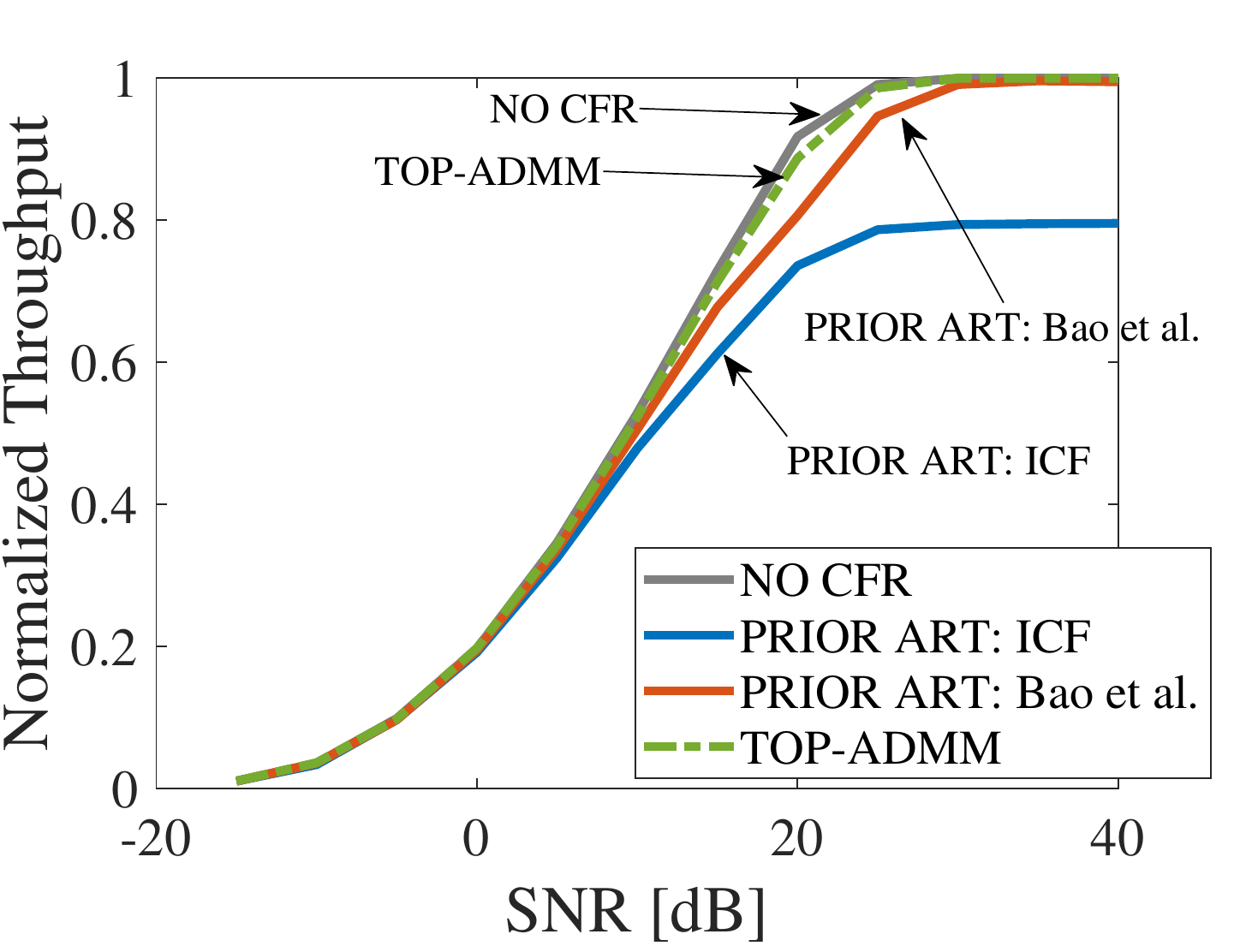}
    \caption{\skblue{Throughput versus averaged SNR per receive antenna with link adaptation enabled} }\label{fig:FINAL_16x2_Throughput_vs_SNR__LA__TDLA300ns10Hz}
\end{subfigure}
\end{minipage}
\caption{\sk{Scatter plot}, EVM, ACLR, \skblue{and Throughput} performance comparison of TOP-ADMM with prior art. }\label{fig:scatterplot_evm_distribution_top_admm_prior_art}
\vspace{-3.5mm}
\end{figure*}

\else

\begin{figure*}[tp!]
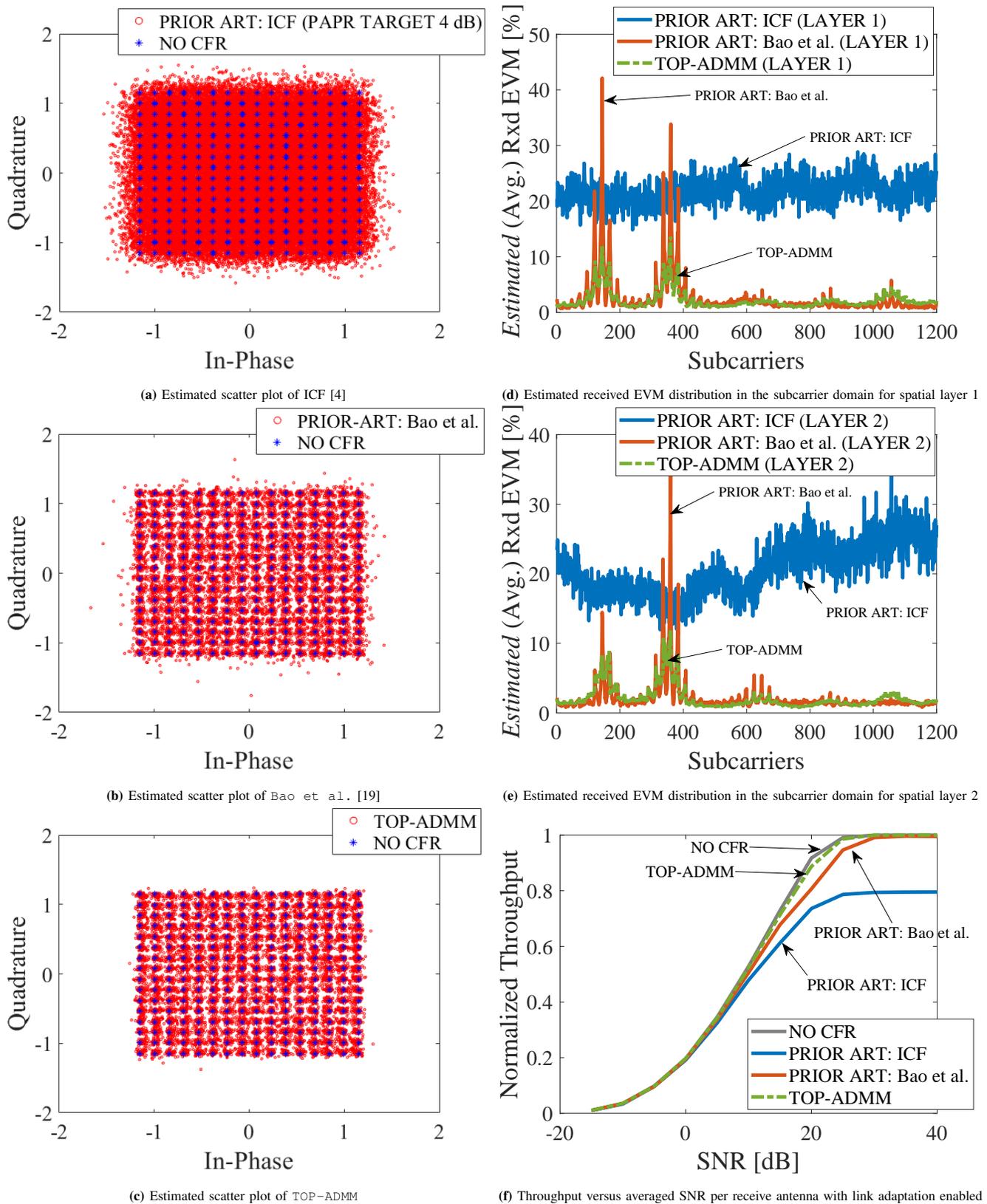

  \begin{minipage}[t]{.315\linewidth}
  \begin{subfigure}[t]{\textwidth}
    \centering
    \includegraphics[width=\textwidth,trim=32.5mm 92mm 31mm 92mm,clip]{figures/FINAL__64x2_TDLD10Hz30ns__target4dB__estimatedScatterplot__icf.pdf}
    \caption{Estimated \sk{scatter plot} of \ac{ICF}~\cite{Armstrong__icf__2002}}\label{fig:FINAL__64x2_TDLD10Hz30ns__target4dB__estimatedScatterplot__icf}
  \end{subfigure}
  \end{minipage}\hfill
  \begin{minipage}[t]{.315\linewidth}
  \begin{subfigure}[t]{\textwidth}
    \centering
    \includegraphics[width=\textwidth,trim=32.5mm 92mm 31mm 92mm,clip]{figures/FINAL__64x2_TDLD10Hz30ns__target4dB__estimatedScatterplot__bao_speed.pdf}
    \caption{Estimated \sk{scatter plot} of \texttt{Bao et al.}~\cite{Bao_papr_admm__2018}}\label{fig:FINAL__64x2_TDLD10Hz30ns__target4dB__estimatedScatterplot__bao_speed}
  \end{subfigure}
  \end{minipage}\hfill
  \begin{minipage}[t]{.315\linewidth}
  \begin{subfigure}[t]{\textwidth}
  \centering
    \includegraphics[width=\textwidth,trim=32.5mm 92mm 31mm 92mm,clip]{figures/FINAL__64x2_TDLD10Hz30ns__target4dB__estimatedScatterplot__topadmm.pdf}
    \caption{Estimated \sk{scatter plot} of \texttt{TOP-ADMM}}\label{fig:FINAL__64x2_TDLD10Hz30ns__target4dB__estimatedScatterplot__topadmm}
    \end{subfigure}
  \end{minipage}
  \vfill
  \begin{minipage}[t]{.315\linewidth}
  \begin{subfigure}[t]{\textwidth}
    \centering
    \includegraphics[width=\textwidth,trim=32.5mm 92mm 31mm 92mm,clip]{figures/FINAL__64x2_TDLD10Hz30ns__target4dB___Estimated_rxTxEVM_vs_subcarrier__LAYER1__icf_bao_topadmm.pdf}
    \caption{Estimated received \ac{EVM} distribution \sk{in the subcarrier domain} for spatial layer~1}\label{fig:FINAL__64x2_TDLD10Hz30ns__target4dB___Estimated_rxTxEVM_vs_PRB__LAYER1__icf_bao_topadmm}
  \end{subfigure}
  \end{minipage}\hfill
  \begin{minipage}[t]{.315\linewidth}
  \begin{subfigure}[t]{\textwidth}
  \centering
    \includegraphics[width=\textwidth,trim=32.5mm 92mm 31mm 92mm,clip]{figures/FINAL__64x2_TDLD10Hz30ns__target4dB___Estimated_rxTxEVM_vs_subcarrier__LAYER2__icf_bao_topadmm.pdf}
    \caption{Estimated received \ac{EVM} distribution \sk{in the subcarrier domain} for spatial layer~2}\label{fig:FINAL__64x2_TDLD10Hz30ns__target4dB___Estimated_rxTxEVM_vs_PRB__LAYER2__icf_bao_topadmm}
    \end{subfigure}
  \end{minipage}\hfill
  \begin{minipage}[t]{.315\linewidth}
  \begin{subfigure}[t]{\textwidth}
    \centering
    \includegraphics[width=\textwidth,trim=32.5mm 92mm 31mm 92mm,clip]{figures/FINAL_16x2_Throughput_vs_SNR__LA__TDLA300ns10Hz.pdf}
    \caption{\skblue{Throughput versus averaged SNR per receive antenna with link adaptation enabled} }\label{fig:FINAL_16x2_Throughput_vs_SNR__LA__TDLA300ns10Hz}
  \end{subfigure}
  \end{minipage}
  \caption{\sk{Scatter plot}, EVM, ACLR, \skblue{and Throughput} performance comparison of TOP-ADMM with prior art. }\label{fig:scatterplot_evm_distribution_top_admm_prior_art}
  \vspace{-3.5mm}
\end{figure*}

\fi

Figure~\ref{fig:performance_comparison_proposed_three_operator_admm_types_with_prior_arts} compares the performance metrics against iterations of proposed \ac{TOP-ADMM}-type methods with the two considered prior arts. Note that the prior art by Bao et al.~\cite{Bao_papr_admm__2018} has two inner iterations for each outer iteration, which are included in these plots for a fair comparison. Figure~\ref{fig:performance_comparison_proposed_three_operator_admm_types_with_prior_arts}\subref{fig:FINAL__64x2_TDLD10Hz30ns__target4dB__Tx1_iaprVsIter__icf_bao_topadmm} shows the achieved \ac{IPAPR} at 99.99\% probability within each \ac{OFDM} symbol. Clearly, \ac{TOP-ADMM} \sk{followed by \ac{DYS}} has a relatively faster convergence compared to \ac{BADMM}. One of the reasons for the slower convergence  of \ac{BADMM} among the proposed methods could be due to the curse of the number of tuning parameters. In other words, a better and finer grid search of tunable parameters of \ac{BADMM} may be required to be competitive. Nevertheless, \sk{we iterate that the target optimization problem in~\cite{Bao_papr_admm__2018} minimizes signal peaks} subject to equality constraints on \ac{EVM} mitigation and \ac{ACLR}. However, in practice, we need to adjust at least one of the performance metrics, \ie, \ac{PAPR} or \ac{EVM}, according to the traffic and channel conditions. Therefore, \sk{the method in \cite{Bao_papr_admm__2018}} achieves the \midtilde0.25 dB smaller \ac{IPAPR} than the proposed methods but at the cost of a nearly two-fold increase in transmit \ac{EVM}, see Fig.~\ref{fig:performance_comparison_proposed_three_operator_admm_types_with_prior_arts}\subref{fig:FINAL__64x2_TDLD10Hz30ns__target4dB__wbTxEVMvsIter__icf_bao_topadmm}. \sk{This two times increase in the transmit \ac{EVM}, \ie, \midtilde6~dB increase in the distortion energy, consequently penalizes the energy of the useful transmit signal by \midtilde6~dB since the total transmit power is fixed and shared by both the signal and distortion.}  
It can clearly be argued that this \ac{IPAPR} gain is worthwhile in practice. Moreover, for a larger modulation alphabet, such as 1024QAM in the upcoming NR releases, \sk{a} lower estimated received \ac{EVM} \sk{would be required} compared to 256QAM. Thus, one might need to lower the target \ac{PAPR} (or back off). \sk{Hence, the optimization problem (and method) proposed in~\cite{Bao_papr_admm__2018} would be unfortunately and arguably impractical for realistic radio systems.} Furthermore, Figs.~\ref{fig:performance_comparison_proposed_three_operator_admm_types_with_prior_arts}\subref{fig:FINAL__64x2_TDLD10Hz30ns__target4dB__predictedEVMvsIter__icf_bao_topadmm},
\subref{fig:FINAL__64x2_TDLD10Hz30ns__target4dB__estimatedEVMvsIter__icf_bao_topadmm} illustrate the predicted and estimated \ac{EVM}, respectively, over iterations. All the methods except \ac{ICF} meet the desired estimated \ac{EVM} target of 5.6\% (additionally, as a reference, we have also shown the 3GPP transmit EVM requirement of 3.5\% for 256QAM). However, 
\ac{TOP-ADMM} has slightly better estimated/predicted received \ac{EVM} performance compared to 
\ac{DYS} and \ac{BADMM}, and noticeably \sk{compared to} prior arts.

In Fig.~\ref{fig:top_admm_perf__tdld_vs_tdla}, we compare \ac{TOP-ADMM} performance in terms of \ac{IPAPR} and predicted received \ac{EVM} against the iterations achieved in Test~1, \ie, the TDL-D channel model with non-\ac{LOS} and \ac{LOS} components and high-like spatial correlation,  and in Test~2, \ie, the TDL-A channel model with only non-\ac{LOS} components and low-like spatial correlation. We construe the following key observation from this plot.
\begin{observation}
The convergence, in terms of the predicted/estimated received \ac{EVM} over iterations, slows down for a channel with increasing Rician $K$-factor, \ie, \ac{LOS} component, and spatial correlation. 
\end{observation}
The slower convergence under such channel conditions could be due to more active constraints.

For completeness, Fig.~\ref{fig:scatterplot_evm_distribution_top_admm_prior_art} presents the \sk{scatter plot, \ac{PSD}, and \ac{EVM} distribution after the last considered iteration (simply fixed to 1000 for all the algorithms) of \ac{ICF}, the Bao et al. method, and \ac{TOP-ADMM}.} \sk{Note that we average the \ac{EVM} and \ac{PSD} metrics over 28 \ac{OFDM} symbols to analyze instantaneous-like behaviour.} In Figs.~\ref{fig:scatterplot_evm_distribution_top_admm_prior_art}\subref{fig:FINAL__64x2_TDLD10Hz30ns__target4dB__estimatedScatterplot__icf}--\subref{fig:FINAL__64x2_TDLD10Hz30ns__target4dB__estimatedScatterplot__topadmm}, we illustrate the \sk{scatter plot} of estimated/equalized received symbols for \ac{ICF}, the Bao et al. method, and \ac{TOP-ADMM}, respectively. Incontestably, \sk{the scatter plot} of \ac{TOP-ADMM} shows qualitatively the lowest estimated received \ac{EVM} compared to prior arts (particularly, the \ac{CSI}-aware Bao et al. method). We show the distribution of the average (over time domain \ac{OFDM} symbols) estimated received \ac{EVM} for Spatial Layers~1~and~2 across \sk{subcarriers} in Fig.~\ref{fig:scatterplot_evm_distribution_top_admm_prior_art}\subref{fig:FINAL__64x2_TDLD10Hz30ns__target4dB___Estimated_rxTxEVM_vs_PRB__LAYER1__icf_bao_topadmm},
\subref{fig:FINAL__64x2_TDLD10Hz30ns__target4dB___Estimated_rxTxEVM_vs_PRB__LAYER2__icf_bao_topadmm}, respectively. Clearly, for the non-robust Bao et al. method, in some sets of subcarriers, we observe high amounts of estimated received \ac{EVM} compared to not only \ac{TOP-ADMM} but also classical \ac{ICF}. The estimated received \ac{EVM} in some resources can be large due to the composite mismatch effect under channel uncertainty. \skblue{In Fig.~\ref{fig:scatterplot_evm_distribution_top_admm_prior_art}\subref{fig:FINAL_16x2_Throughput_vs_SNR__LA__TDLA300ns10Hz}, we also depict the (normalized) throughput against the average \ac{SNR} per receive antenna after the respective \ac{PAPR} reduction algorithms with link adaptation enabled under Test~1. Around 20~dB SNR, we see that prior art~\cite{Bao_papr_admm__2018} performs poorer than our method. The reason is that the method in~\cite{Bao_papr_admm__2018} renders high transmit signal distortion energy, see~Fig.~\ref{fig:performance_comparison_proposed_three_operator_admm_types_with_prior_arts}\subref{fig:FINAL__64x2_TDLD10Hz30ns__target4dB__wbTxEVMvsIter__icf_bao_topadmm}, which consequently penalizes the useful signal's transmit energy. However, as expected, at high enough SNR, our proposed TOP-ADMM and the method in~\cite{Bao_papr_admm__2018} will have a negligible impact on the throughput performance. }

\vspace{-2.5mm}
\section{Conclusion} \label{sec:conclusion_future_work}

In this paper, \skblue{we sought a principled approach to mitigate (CSI-aware catering to user-specific beamformed channel under channel uncertainty) or minimize (non-CSI-aware supporting cell-specific non-beamformed broadcast channel) flexibly signal distortion seen at the receivers subject to nonconvex \ac{PAPR} and \ac{ACLR} constraints for a large-scale \ac{MIMO}-\ac{OFDM}-based system.}
Unfortunately, a general-purpose solver for such a large-scale (nonconvex) optimization problem is practically infeasible. 
Hence, to tackle such a \sk{nonconvex} problem, \sk{we applied iteration-dependent ``convexification" to the nonconvex \ac{PAPR} and \ac{ACLR} constraints. We conjecture that such iteration-dependent ``convexification" approaches the nonconvex constraint set asymptotically. Furthermore, we capitalized on 
a divide-and-conquer approach to decompose the large-scale problem into smaller easy-to-solve subproblems. More concretely, we proposed a powerful and simple  
optimization solution using our recently proposed \ac{TOP-ADMM} algorithm to tackle such a problem and compared it to 
two relatively new algorithms popular within machine learning and (numerical) optimization, the so-called \ac{BADMM} and \ac{DYS} algorithms.} 
Consequently, we developed three \sk{``implementation-friendly"} algorithms to tackle the proposed large-scale (nonconvex) optimization problem.  Numerical results \sk{showed} that the proposed low-complexity algorithms can meet the estimated (received) \ac{EVM} target while meeting the desired \ac{PAPR} and \ac{ACLR} constraints. This is arguably the first work to propose computationally efficient algorithms for a robust \ac{EVM} mitigation problem under channel uncertainty with nonconvex \ac{PAPR} and \ac{ACLR} constraints.

\vspace{-1.5mm}
\begin{appendices}
\section{\skblue{Optimality Conditions of TOP-ADMM for \texttt{P3}/\texttt{P4}}} \label{appendix:optimality_condition_of_top_admm_for_p3_or_p4}

\skblue{
We first form the Lagrangian to~\eqref{eqn:problem_p3} (considering in~\eqref{eqn:general_consensus_top_admm__generic_form} form)---no convexity assumptions are required:
\begin{align*}
    \mathcal{L}\!\left(\overline{\mat{X}}\!,  \overline{\mat{Z}}\!, \bm{\Lambda} \!\right) \! 
    \coloneqq&  \!
    \Ind_{\mathbfcal{U}}\!\left( \overline{\mat{X}} \right) \! +\! \Ind_{\mathbfcal{P}}\!\left( \overline{\mat{Z}} \right)  \!+\! h\!\left( \overline{\mat{Z}}  \right) 
    \!+\! 2\Re\!\left\{ \! \trace\!\left[\!\bm{\Lambda}^{\!\herm} \!\left( \overline{\mat{X}} \! - \! \overline{\mat{Z}} \right)\!\right] \!\right\}\!.
\end{align*}
It is known that the \ac{KKT} conditions are first-order optimality conditions which are necessary conditions for the solution point of~\eqref{eqn:general_consensus_top_admm__generic_form} to be optimal under some suitable constraint qualifications, see, \eg, \cite{Boyd2004ConvexOptimization}. Then, we assume the subdifferential exist for the (nonconvex) indicator function. 
Let $\left[\overline{\mat{X}}^\star;  \overline{\mat{Z}}^\star; \bm{\Lambda}^\star\right]$  be a \ac{KKT} point, which fulfils the dual feasibility, \ie, $ \vec{0} \! \in\!  \partial \Ind_{\mathbfcal{U}}\left( \overline{\mat{X}}^\star \right)  \! +\! \bm{\Lambda}^{\star}$ and $\vec{0}
    \! \in \! \partial \Ind_{\mathbfcal{P}}\left( \overline{\mat{Z}}^\star \right)  \!+\! \nabla h\left( \overline{\mat{Z}}^\star  \right)  \!-\!  \bm{\Lambda}^\star$
and the primal feasibility, \ie, $\overline{\mat{X}}^\star \!-\! \overline{\mat{Z}}^\star \!=\! \vec{0}$. Following~\cite[Appendix~B]{Kant_journal_msp_top_admm:2020}, we can easily show that the sequences generated by \ac{TOP-ADMM} for either \texttt{P3} (under the conjecture that the residual errors vanish asymptotically) or \texttt{P4} (see~\cite{Kant_etal__intro_to_TOPADMM_2022}), at any limit point, converges to a \ac{KKT} point.
}

\end{appendices}

\vspace{-3mm}
\singlespacing
\bibliographystyle{IEEEtran}
\bibliography{IEEEabrv,ref}

\end{document}